\documentclass[12pt]{article}
\usepackage{eurosym}
\usepackage{amsmath}
\usepackage{amsfonts}
\usepackage{makeidx}
\usepackage{amssymb}
\usepackage{times}
\usepackage{anysize}
\usepackage{pst-node}
\usepackage{pst-tree}
\usepackage{lscape}
\usepackage{graphics,graphicx}
\usepackage{hyperref}
\usepackage{cite}

\makeatletter
\def\pstree@balancedfit#1#2{\edef\next{\noexpand\pstree@@balancedfit#1\noexpand\@nil#2\noexpand\@nil}\next
\ifnum\pst@cntg=\z@
\pstree@max{#1}\pst@cnth
\else
\pstree@max{#2}\pst@cnth
\fi
\advance\pst@cnth\pst@cnth
\advance\pst@cnth\psk@thistreesep\relax
\advance\pst@cnth\pstree@tspace\relax
\gdef\pstree@tspace{\z@}}
\def\pstree@@balancedfit#1,#2\@nil#3,#4\@nil{\ifnum#1=\pstree@stop
\let\next\relax
\pst@cntg=\@ne
\else
\ifnum#3=\pstree@stop
\let\next\relax
\pst@cntg=\z@
\else
\def\next{\pstree@@balancedfit#2\@nil#4\@nil}\fi
\fi
\next}
\makeatother
\psset{fillstyle=solid}

\newtheorem{theorem}{Theorem}[section]

\newtheorem{corollary}[theorem]{Corollary}

\newtheorem{lemma}[theorem]{Lemma}

\newtheorem{proposition}[theorem]{Proposition}
\newtheorem{remark}[theorem]{Remark}

\newenvironment{proof}[1][Proof]{\noindent\textbf{#1.} }{\ \rule{0.5em}{0.5em}}
\marginsize{2cm}{2cm}{1cm}{1cm}

\newtheorem{defn}{Definition}[section]

\input{tcilatex}
\begin{document}

\title{From Short-Range to Mean-Field Models in Quantum Lattices}
\author{J.-B. Bru, W. de Siqueira Pedra, K. Rodrigues Alves}
\date{\today }
\maketitle

\begin{abstract}
Realistic effective interparticle interactions of quantum many-body systems
are widely seen as being short-range. However, the rigorous mathematical
analysis of this type of model turns out to be extremely difficult, in
general, with many important fundamental questions remaining open still
nowadays. By contrast, mean-field models come from different approximations
or Ans\"{a}tze, and are thus less realistic, in a sense, but are technically
advantageous, by allowing explicit computations while capturing surprisingly
well many real physical phenomena. Here, we establish a precise mathematical
relation between mean-field and short-range models, by using the long-range
limit that is known in the literature as the Kac, or van der Waals,\textit{\ 
}limit. If both attractive and repulsive long-range forces are present then
it turns out that the limit mean-field model is not necessarily what one
traditionally guesses. One important innovation of our study, in contrast
with previous works on the subject, is the fact that we are able to show the
convergence of equilibrium states, i.e., of all correlation functions. This
paves the way for studying phase transitions, or at least important
fingerprints of them like strong correlations at long distances, for models
having interactions whose ranges are finite, but very large. It also sheds a
new light on mean-field models. Even on the level of pressures, our results
go considerably further than previous ones, by allowing, for instance, a
continuum of long-range interaction components, as well as very general
short-range Hamiltonians for the \textquotedblleft free\textquotedblright\
part of the model. The present results\ were made possible by the
variational approach of \cite{BruPedra2} for equilibrium states of
mean-field models, as well as the game theoretical characterization of these
states. Our results are obtained in an abstract, model-independent, way.
\end{abstract}

%TCIMACRO{\TeXButton{\tableofcontents }{\tableofcontents}}%
%BeginExpansion
\tableofcontents%
%EndExpansion

\newpage

\section{Introduction}

In theoretical physics, realistic effective interparticle interactions of
quantum many-body systems are widely seen as being short-range, like in the
celebrated Hubbard model or models with Yukawa-type potentials. However, the
rigorous mathematical analysis of this type of model turns out to be
extremely difficult, in general, with many important fundamental questions
(referring to the structure of phase diagrams, transport properties, etc.)
remaining open still nowadays. In contrast with short-range models,
mean-field models come from different approximations or Ans\"{a}tze, and are
thus less realistic, in a sense,\ but handy, by allowing explicit
computations while capturing surprisingly well many important real physical
phenomena, like for instance the superconductivity. They are thus essential
in condensed matter physics. Note that a mean-field interaction is, at least
in a formal sense, an extreme form of interaction that is weak, but acts
across very long distances. The aim of this paper is thus to establish a
precise mathematical relation between mean-field and short-range models, by
using the long-range limit known in the literature as the \emph{Kac\ limit},
or the \emph{van der Waals limit}. This is done in an abstract,
model-independent, way.

To understand our motivations and results, it is instructive to consider the
following example: \medskip

\noindent \textbf{Notation.} $\mathfrak{L}\doteq \mathbb{Z}^{d}$ represents
a cubic crystal of dimension $d\in \mathbb{N}$. In order to define the
thermodynamic limit, we use the cubic boxes $\Lambda _{L}\doteq \{\mathbb{Z}%
\cap \left[ -L,L\right] \}^{d}$ of volume $|\Lambda _{L}|$, where $L\in 
\mathbb{N}_{0}$. Let $\mathrm{S}$\ be some finite set representing an
orthonormal basis of spin modes. For instance, $\mathrm{S}=\{\uparrow
,\downarrow \}$, referring to fermions with spin $\uparrow $ and $\downarrow 
$, like electrons. Fermionic annihilation operators are denoted by $a_{x,%
\mathrm{s}}$ for $x\in \mathfrak{L}$ and $\mathrm{s}\in \mathrm{S}$. The
parameter $\beta \in \mathbb{R}^{+}$ fixes the inverse temperature. \medskip 

\noindent \textbf{Short-range model.} On the one hand, given two functions $%
h,f:\mathbb{R}^{d}\rightarrow \mathbb{R}$ and a parameter $\gamma \in (0,1)$%
, consider a translation-invariant lattice fermion system via box
Hamiltonians defined by%
\begin{equation}
H_{L}^{\mathrm{SR}}\doteq \sum\limits_{x,y\in \Lambda _{L},\mathrm{s}\in 
\mathrm{S}}h\left( x-y\right) a_{x,\mathrm{s}}^{\ast }a_{y,\mathrm{s}%
}+\sum\limits_{x,y\in \Lambda _{L},\mathrm{s},\mathrm{t}\in \mathrm{S}%
}\gamma ^{d}f\left( \gamma \left( x-y\right) \right) a_{y,\mathrm{t}}^{\ast
}a_{y,\mathrm{t}}a_{x,\mathrm{s}}^{\ast }a_{x,\mathrm{s}}\ ,\qquad L\in 
\mathbb{N}_{0}\ .  \label{SR}
\end{equation}%
The function $h$ represents the hopping term\footnote{%
For instance, take $h\left( x-y\right) =-\langle \mathfrak{e}_{x},\Delta _{%
\mathrm{d}}\mathfrak{e}_{y}\rangle _{\ell ^{2}\left( \mathfrak{L}\right) }\ $%
for $x,y\in \mathfrak{L}$, where $\Delta _{\mathrm{d}}\in \mathcal{B}(\ell
^{2}(\mathfrak{L}))$ is the usual $d$--dimensional discrete Laplacian and
with $\left\{ \mathfrak{e}_{x}\right\} _{x\in \ell ^{2}\left( \mathfrak{L}%
\right) }$ being the canonical orthonormal basis of $\ell ^{2}(\mathfrak{L})$%
.}, or kinetic part, of the lattice fermion system and is assumed to decay
fast at large distances, while $f$ is a pair potential characterizing the
interparticle interaction, whose range is tuned by the parameter $\gamma \in
(0,1)$. As is usual in theoretical physics, $f$ is assumed to be fast
decaying\footnote{%
Take for instance a compactly supported function $f$.}, reflection-symmetric%
\footnote{%
I.e., $f\left( x\right) =f\left( -x\right) $. Usually, $f\left( x\right)
=v\left( \left\vert x\right\vert \right) $ for some function $v:\mathbb{R}%
_{0}^{+}\rightarrow \mathbb{R}$.}, and positive definite (i.e., the Fourier\
transform $\hat{f}$ of $f$ is a positive function on $\mathbb{R}^{d}$). This
choice for $f$ is reminiscent of a superstability condition, which is
essential in the bosonic case \cite[Section 2.2 and Appendix G]%
{BruZagrebnov8}.

For any inverse temperature $\beta \in \mathbb{R}^{+}$, the infinite volume
pressure 
\begin{equation*}
P^{\mathrm{SR}}\left( \gamma \right) \doteq \underset{L\rightarrow \infty }{%
\lim }\frac{1}{\beta |\Lambda _{L}|}\ln \mathrm{Trace}(\mathrm{e}^{-\beta
H_{L}^{\mathrm{SR}}}),\qquad \gamma \in \left( 0,1\right) \ ,
\end{equation*}%
is well-defined and given by a variational problem for translation-invariant
states. See, e.g., \cite[Theorem 2.12]{BruPedra2}. Equilibrium states can
thus be naturally defined as being the solutions (minimizers) to this
variational problem. However, a mathematically rigorous computation of the
pressure and equilibrium states to show possible phase transitions is 
\textbf{elusive}, beyond perturbative arguments, even after decades of
mathematical studies. \medskip

\noindent \textbf{Mean-Field model.} On the other hand, instead of the above
model, one may study a mean-field version of it, defined by the box
Hamiltonians%
\begin{equation}
H_{L}^{\mathrm{MF}}\doteq \sum\limits_{x,y\in \Lambda _{L},\mathrm{s}\in 
\mathrm{S}}h\left( x-y\right) a_{x,\mathrm{s}}^{\ast }a_{y,\mathrm{s}}+\frac{%
\eta }{\left\vert \Lambda _{L}\right\vert }\sum_{x,y\in \Lambda _{L},\mathrm{%
s},\mathrm{t}\in \mathrm{S}}a_{y,\mathrm{t}}^{\ast }a_{y,\mathrm{t}}a_{x,%
\mathrm{s}}^{\ast }a_{x,\mathrm{s}}\ ,\qquad L\in \mathbb{N}_{0}\ ,
\label{MF0}
\end{equation}%
for some positive parameter $\eta \in \mathbb{R}_{0}^{+}$. For any $\beta
\in \mathbb{R}^{+}$, the infinite volume pressure 
\begin{equation*}
P^{\mathrm{MF}}\left( \eta \right) \doteq \underset{L\rightarrow \infty }{%
\lim }\frac{1}{\beta |\Lambda _{L}|}\ln \mathrm{Trace}(\mathrm{e}^{-\beta
H_{L}^{\mathrm{MF}}})\ ,\qquad \eta \in \mathbb{R}_{0}^{+}\ ,
\end{equation*}%
is well-defined and given by a variational problem for translation-invariant
states, thanks to \cite[Theorem 2.12]{BruPedra2}. A general notion of
(generalized) equilibrium states can again be given via this variational
problem. What is more, they can be explicitly computed through an
alternative version of the approximating Hamiltonian method \cite%
{Bogjunior,AHM-non-poly1,AHM-non-poly2,approx-hamil-method0,approx-hamil-method,approx-hamil-method2}
(see \cite[Sections 2.10 and 10.2]{BruPedra2} for more details), named \emph{%
thermodynamic game} (Section \ref{Section thermo game}), introduced in \cite[%
Section 2.7]{BruPedra2}. In this case, \cite[Theorem 2.36]{BruPedra2} shows
that 
\begin{equation}
P^{\mathrm{MF}}\left( \eta \right) =\inf_{c\in \mathbb{C}}\left\{ \eta
\left\vert c\right\vert ^{2}+P\left( c,\eta \right) \right\} \ ,\qquad \eta
\in \mathbb{R}_{0}^{+}\ ,  \label{variational studied}
\end{equation}%
where $P:\mathbb{C}\times \mathbb{R}_{0}^{+}\rightarrow \mathbb{R}$ is the
infinite volume pressure defined by 
\begin{equation}
P\left( c,\eta \right) \doteq \underset{L\rightarrow \infty }{\lim }\frac{1}{%
\beta |\Lambda _{L}|}\ln \mathrm{Trace}(\mathrm{e}^{-\beta H_{L}\left(
c\right) })\ ,\qquad c\in \mathbb{C},\ \eta \in \mathbb{R}_{0}^{+}\ ,
\label{functionsdsldslkdj}
\end{equation}%
and 
\begin{equation}
H_{L}\left( c\right) \doteq \sum\limits_{x,y\in \Lambda _{L},\mathrm{s}\in 
\mathrm{S}}h\left( x-y\right) a_{x,\mathrm{s}}^{\ast }a_{y,\mathrm{s}}+2\eta 
\func{Re}\{c\}\sum_{x\in \Lambda _{L},\mathrm{s}\in \mathrm{S}}a_{x,\mathrm{s%
}}^{\ast }a_{x,\mathrm{s}}\ .  \label{var pb}
\end{equation}%
This Hamiltonian is quadratic in terms of creation and annihilations
operators and can thus be explicitly diagonalized, as is well-known. It
means that the function (\ref{functionsdsldslkdj}) can be computed and the
variational problem (\ref{variational studied}) studied by analytic or
rigorous numerical methods. Last but not least, \cite[Theorems 2.21 and 2.39]%
{BruPedra2} also show that (generalized) equilibrium states can be obtained
by using self-consistency conditions, which refer, in a sense, to
Euler-Lagrange equations for the variational problem (\ref{variational
studied}). See Section \ref{Section effective theories} for a concise
explanation of this point.\medskip 

\noindent \textbf{Kac Limit.} The Kac, or long-range, limit refers here to
the limit $\gamma \rightarrow 0^{+}$ of short-range models that are already
in the thermodynamic limit. For small parameters $\gamma \ll 1$, the
short-range model defined in finite volume by (\ref{SR}) has an
interparticle interaction with very large range ($\mathcal{O}(\gamma ^{-1})$%
), but the interaction strength is small as $\gamma ^{d}$, in such a way
that the first Born approximation\footnote{%
I.e., $\int_{\mathbb{R}^{d}}\gamma ^{d}f\left( \gamma x\right) \mathrm{d}%
x=\int_{\mathbb{R}^{d}}f\left( x\right) \mathrm{d}x\doteq \hat{f}(0)$.} to
the scattering length of the interparticle potential remains constant, as is
usual. One therefore expects to have some effective long-range model in the
limit $\gamma \rightarrow 0^{+}$. This is explicitly what we prove here by
showing, among other things, that 
\begin{equation*}
\lim_{\gamma \rightarrow 0^{+}}P^{\mathrm{SR}}\left( \gamma \right) =P^{%
\mathrm{MF}}(\hat{f}(0))\ ,
\end{equation*}%
where $\hat{f}$ is the (positive) Fourier\ transform of the two-body
interaction potential $f$. See Equation (\ref{dddfdfdf}). Explicit
computations of the rate of convergence can also be done by using our
estimates, see Proposition \ref{propuni copy(2)} as well as the proof of
Theorem \ref{propuni copy(3)}. More importantly, the equilibrium states of
the short-range model can also be approximated by generalized equilibrium
states of the corresponding mean-field model, as shown by Theorem \ref%
{propuni copy(3)}. Recall that a precise study of phase transitions for
short-range models is notoriously difficult, but we show here that it can be
done\footnote{%
The existence of a phase transition, like a first-order one, in the limiting
mean-field model does not necessarily imply a phase transition in the
corresponding short-range model for small, but nonzero $\gamma \ll 1$.
However, this convergence can imply important properties on correlation
functions of the short-range model for small $\gamma $. This is similar to
what occurs in the thermodynamic limit: Recall that the equilibrium states
are always unique at finite volume, whereas first order phase transitions do
take place at infinite volume. See discussions in \cite[Section 2.6]%
{BruPedra2}.} in the Kac limit via a mean-field model for which efficient
mathematical methods (for instance, what we call the thermodynamic game) can
be used. \medskip

The results referring to the above examples are already highly non-trivial
and use in an essential way various fundamental outcomes of \cite{BruPedra2}%
. Nevertheless, our present results go \textbf{far beyond} this example, by
showing, for instance, that the kinetic part (represented by the function $h$%
) can be replaced with a very general short-range fermionic Hamiltonian. See
Theorem \ref{propuni copy(3)}, which is in turn generalized by Theorem \ref%
{propuni copy(5)} to include the case of possibly infinitely many long-range
repulsions. The main limitation in this example is the fact that the
function $f$ has to be a positive definite function, which leads to a purely
repulsive mean-field model (Section \ref{Section purely attrac}), in the
long-range (Kac) limit. This situation excludes some important models of
physics like BCS-type\ models (e.g., the reduced BCS Hamiltonian or the
strong coupling BCS Hamiltonian), which have purely attractive mean-field
(interparticle) interactions, see discussions after Theorem \ref{propuni
copy(6)0}.

In the present paper, we provide results that do not depend upon the
positive definiteness of the pair potential $f$. In particular, general
purely attractive interparticle interactions, like the BCS model (see (\ref%
{BCS interaction1}) and (\ref{dddfdfdfdddfdfdf})), are included. See
Theorems \ref{propuni copy(6)0} and \ref{propuni copy(6)}. The general case,
i.e., the case of long-range interaction forces which are neither purely
attractive nor purely repulsive, refers to Proposition \ref{propuni copy(10)}%
, Theorems \ref{propuni copy(11)}, \ref{propuni copy(12)bis} and \ref%
{propuni copy(12)} as well as Corollaries \ref{propuni copy(13)} and \ref%
{propuni copy(14)}. See also Section \ref{General Case}. As expected, any
Kac limit leads to mean-field pressures and equilibrium states. However, the
limit mean-field model is \textbf{not necessarily} what one traditionally
guesses. In fact, it strongly depends upon the hierarchy of ranges between
attractive and repulsive interparticle forces. For instance, if the range of
repulsive forces is much larger than the range of the attractive ones, then
in the Kac limit for these forces one may get a mean-field model that is 
\textbf{non-conventional}\footnote{%
Such models were introduced for the first time in \cite{BruPedra2}.}. See,
e.g., Theorems \ref{propuni copy(12)bis} and Sections \ref{section
non-conventnioal pressure}, \ref{section non-conventnioal pressure2}.
However, we also show in Theorems \ref{propuni copy(5)} and \ref{propuni
copy(6)} that, in the case of purely attractive or purely repulsive
long-range forces, the formally expected mean-field model is in fact the
correct effective long-range model for the Kac limit. See also Theorem \ref%
{propuni copy(15)}.

Our paper follows a rather old sequence of studies on the Kac limit,
basically starting from 1959 with Kac's work on classical one-dimensional
spin systems. The first important result \cite{Penrose-Lebowitz0} in this
period was provided by Penrose and Lebowitz in 1966, who proved the
convergence of the free energy of a classical system towards the one of the
van der Waals theory. Shortly after, the results of this seminal paper were
extended to quantum systems (Boltzmann, Bose, or Fermi statistics) by Lieb
in \cite{Lieb-Kac}. In 1971, Penrose and Lebowitz went considerably further
than \cite{Penrose-Lebowitz0} with the paper \cite{Penrose-Lebowitz}. See
also \cite{Hemmen-Lebowitz} for a review of all these results of classical
statistical mechanics. These outcomes form the mainstays of the subsequent
results on the Kac limit and we recommend the book \cite{Presutti},
published in 2009, for a more recent review on the subject in classical
statistical mechanics, including the so-called Lebowitz-Penrose theorem and
a more exhaustive list of references.

Studies on the Kac limit are still performed nowadays in classical
statistical mechanics, see, e.g., \cite{tonnelli,tonelli2,Franz}. By
contrast, to our knowledge, not much was done on the subject for quantum
systems after the first main important results \cite{Lieb-Kac} in 1966
(which refer to quantum particles in the continuum, but may certainly be
extended to lattice systems). The second important results \cite%
{Smedt-Zagrebnov} concern the Kac limit\ (i.e., the van der Waals limit) of
a Bose gas in presence of two-body interactions in the grand-canonical
ensemble and in the continuum, in the spirit of \cite{Lieb-Kac}. \cite%
{Smedt-Zagrebnov} is liminal because, by using a scaled external field, the
authors show for the first time that the Kac limit can lead to another
mean-field-type model than the expected mean-field gas at high densities.
See discussions at the end of \cite{Smedt-Zagrebnov} for more details. More
than three decades later, we recover this general observation by combining
instead repulsive and attractive interactions, without any
scaled-external-field perturbation method as used in \cite{Smedt-Zagrebnov}.
Since the 2000's we are only aware of a few papers in relation to the Kac
limit for quantum systems: In 2003 \cite{Martin0,Martin1}, the Kac limit is
used within a diagrammatic approach and in 2005 \cite{Martin} the same
authors study corrections to the Kac limit for the Bose system used in \cite%
{Smedt-Zagrebnov}, but without scaled external fields. In 2019, the Kac
limit is also used again for the Bose gas in the Hartree-Fock approximation
to analyze (again) the Bose--Einstein condensation \cite{Alastuey}. This
shows that the research activity on the subject is still nowadays quite
restricted.

The main innovation of our present study is the fact that we are able to
show convergence in the Kac limit not only for pressure-like quantities (for
instance, the thermodynamic limit of the logarithm of canonical or
grand-canononical partition functions), as in previous works, but also for
equilibrium states, i.e., for \textbf{all} correlation functions. Our
results\ on states were made possible by the variational approach of \cite%
{BruPedra2} for equilibrium states of mean-field models. Additionally, also
in contrast with previous results on Kac limits, our method allows for
coexistence of both attractive and repulsive long-range forces. This
important extension is related to the game theoretical characterization of
equilibrium states of mean-field models (cf. thermodynamic game) also
introduced in \cite{BruPedra2}. Our study thus paves the way for studying
phase transitions\footnote{%
Mean-field repulsions have generally a geometrical effect by possibly
breaking the face structure of the set of (generalized) equilibrium states
(see \cite[Lemma 9.8]{BruPedra2}). When this appears, we have\ phase
transitions with long-range order. See \cite[Section 2.9]{BruPedra2}.}, or
at least important fingerprints of them like strong correlations at long
distances, for models having interactions whose ranges are finite, but very
large. It also sheds a new light on mean-field models by connecting them
with short-range ones, in a mathematically precise manner (like in \cite%
{Smedt-Zagrebnov}). Even on the level of the partition functions (here the
grand-canonical pressure), our results go considerably further than previous
ones, by allowing for instance a continuum of mean-field components (cf.
Definitions \ref{definition Kac interaction} and \ref{definition Kac
interaction copy(1)}), as well as very general short-range Hamiltonians for
the \textquotedblleft free\textquotedblright\ part of the model.

To conclude, similar to the seminal\footnote{%
See all studies done during the last two decades in relation with the
Gross-Pitaevskii theory and mean-field systems for indistinguishable
particles (bosons).} paper \cite{LY1998} reviving in 1998 an old problem on
dilute Bose gases studied by Dyson in 1957, our paper sheds new light on old
questions about long-range limits of short-range interactions. We think that
such studies will be important for future theoretical developments in
many-body theory, since long-range interactions are expected to imply
effective, classical background fields, in the spirit of the Higgs mechanism
of quantum field theory. This is shown in \cite%
{BruPedra-MFII,BruPedra-MFIII,Bru-pedra-MF-IV} for mean-field models.\medskip

The paper is organized as follows: Section \ref{Section FERMI0} presents the
general mathematical framework. Observe that we focus on lattice fermion
systems. See Remark \ref{Quantum spin systems}. In Section \ref{Section
Banach space interaction}, we give a brief account on the theory of fermion
systems on the lattice with short-range interactions, which is used as a
springboard to introduce the theory of mean-field models in Section \ref%
{Long-Range Models}. Note that the mathematical setting used in the current
paper is -- up to minor modifications -- the one of \cite%
{BruPedra-MFII,BruPedra2}, including the notation. We thus provide it in a
concise way. Section \ref{Short-Range to Long-Range Models} establishes
precise mathematical relations between short-range and mean-field models via
Kac limits, in which concerns both the pressure and equilibrium states. This
section includes our main results, which are Theorems \ref{propuni copy(11)}%
, \ref{propuni copy(12)bis}, \ref{propuni copy(12)} and \ref{propuni
copy(15)}, Propositions \ref{propuni copy(10)} and \ref{propuni copy(17)},
as well as Corollaries \ref{propuni copy(13)} and \ref{propuni copy(14)}.
Since the results are general and abstract, for convenience, in Section \ref%
{Illustration} we illustrate them on models of the form (\ref{SR}), to which
an attractive long-range interaction term is incorporated (so that a
competition between attractive and repulsive long-range forces takes place).
In fact, this section was designed to be read almost independently of the
other ones, allowing the (possibly non-expert) reader to catch more easily
the main results of the paper. Section \ref{Appendix} is an appendix that
gathers some useful technical results.

\begin{remark}[Quantum spin systems]
\label{Quantum spin systems}\mbox{ }\newline
Our study focuses on lattice fermion systems, which are, from a technical
point of view, slightly more difficult than quantum spin systems, because of
a non-commutativity issue at different lattice sites. However, all the
results presented here hold true for quantum spin systems, via obvious
modifications.
\end{remark}

\begin{remark}[Periodic quantum lattice systems]
\label{Periodic quantum lattice systems}\mbox{ }\newline
Our study focuses on lattice fermion systems that are translation-invariant
(in space). However, all the results presented here hold true for
(space-)periodic lattice fermion systems, by appropriately redefining the
spin set\footnote{%
In fact, one can see the lattice points in a (space) period as a single
point in an equivalent lattice on which particles have an enlarged spin set.}%
. A similar argument holds true for quantum spin systems.
\end{remark}

\section{Algebraic Formulation of Lattice Fermion Systems\label{Section
FERMI0}}

\subsection{CAR Algebra for Lattice Fermions\label{Algebra of Lattices}}

\subsubsection{Background Lattice}

Fix once and for all the dimension $d\in \mathbb{N}$ of the (cubic) lattice.
Let $\mathfrak{L}\doteq \mathbb{Z}^{d}$ and $\mathcal{P}_{\mathrm{f}%
}\subseteq 2^{\mathfrak{L}}$ be the set of all non-empty finite subsets of
the lattice $\mathfrak{L}$. In order to define the thermodynamic limit, we
use the cubic boxes%
\begin{equation}
\Lambda _{L}\doteq \{(x_{1},\ldots ,x_{d})\in \mathfrak{L}:|x_{1}|,\ldots
,|x_{d}|\leq L\}\in \mathcal{P}_{\mathrm{f}}\ ,\qquad L\in \mathbb{N}_{0},
\label{eq:def lambda n}
\end{equation}%
as a so-called van Hove sequence.

\subsubsection{The CAR $C^{\ast }$-Algebra}

For any nonempty subset $\Lambda \subseteq \mathfrak{L}$, $\mathcal{U}%
_{\Lambda }$ denotes the universal unital $C^{\ast }$-algebra generated by
elements $\{a_{x,\mathrm{s}}\}_{x\in \Lambda ,\mathrm{s}\in \mathrm{S}}$
satisfying the canonical anti-commutation relations (CAR): 
\begin{equation}
\left\{ 
\begin{array}{ccc}
a_{x,\mathrm{s}}a_{y,\mathrm{t}}+a_{y,\mathrm{t}}a_{x,\mathrm{s}} & = & 0 \\ 
a_{x,\mathrm{s}}^{\ast }a_{y,\mathrm{t}}+a_{y,\mathrm{t}}a_{x,\mathrm{s}%
}^{\ast } & = & \delta _{x,y}\delta _{\mathrm{s},\mathrm{t}}\mathfrak{1}%
\end{array}%
\right. ,\text{\qquad }x,y\in \Lambda ,\text{ }\mathrm{s},\mathrm{t}\in 
\mathrm{S},  \label{CARbis}
\end{equation}%
where $\mathfrak{1}$ stands for the unit of the algebra, $\delta _{\cdot
,\cdot }$ is the Kronecker delta and $\mathrm{S}$\ is some finite set
(representing an orthonormal basis of spin modes), which is fixed once and
for all. If $\Lambda =\emptyset $ then $\mathcal{U}_{\Lambda }\doteq \mathbb{%
C}$. We use the notation 
\begin{equation}
\left\vert A\right\vert ^{2}\doteq A^{\ast }A,\qquad A\in \mathcal{U}%
_{\Lambda },\ \Lambda \subseteq \mathfrak{L},  \label{carre de A}
\end{equation}%
to shorten mathematical expressions, in particular in the context of
mean-field models.

By identifying the units and the generators $\{a_{x,\mathrm{s}}\}_{x\in
\Lambda \cap \Lambda ^{\prime },\mathrm{s}\in \mathrm{S}}$ in any two $%
C^{\ast }$-algebras $\mathcal{U}_{\Lambda }$ and $\mathcal{U}_{\Lambda
^{\prime }}$, $\{\mathcal{U}_{\Lambda }\}_{\Lambda \in 2^{\mathfrak{L}}}$
canonically forms a net of unital $C^{\ast }$-algebras with respect to
inclusion: For all subsets $\Lambda ,\Lambda ^{\prime }\subseteq \mathfrak{L}
$ so that $\Lambda \subseteq \Lambda ^{\prime }$, one has $\mathcal{U}%
_{\Lambda }\subseteq \mathcal{U}_{\Lambda ^{\prime }}$. For $\Lambda =%
\mathfrak{L}$ we use the notation $\mathcal{U}\equiv \mathcal{U}_{\mathfrak{L%
}}$. Observe additionally that the subspace 
\begin{equation}
\mathcal{U}_{0}\doteq \bigcup_{\Lambda \in \mathcal{P}_{\mathrm{f}}}\mathcal{%
U}_{\Lambda }\subseteq \mathcal{U}\equiv \mathcal{U}_{\mathfrak{L}}
\label{simple}
\end{equation}%
is a dense $\ast $-algebra of the CAR $C^{\ast }$-algebra $\mathcal{U}$ of
the infinite lattice. In particular, $\mathcal{U}$\ is separable, because $%
\mathcal{U}_{\Lambda }$ has finite dimension for all (finite subsets) $%
\Lambda \in \mathcal{P}_{\mathrm{f}}$ and the collection $\mathcal{P}_{%
\mathrm{f}}$\ of sets is countable. Elements of $\mathcal{U}_{0}$ are called
local elements of $\mathcal{U}$. The (real) Banach subspace of all
self-adjoint elements of $\mathcal{U}$ is denoted by $\mathcal{U}^{\mathbb{R}%
}\varsubsetneq \mathcal{U}$.

The local causality of quantum field theory is broken in CAR algebras and
physical quantities are therefore defined from the subalgebra of even
elements, which are defined as follows: Given a fixed parameter $\theta \in 
\mathbb{R}/(2\pi \mathbb{Z)}$, the condition 
\begin{equation}
\mathrm{g}_{\theta }(a_{x,\mathrm{s}})=\mathrm{e}^{-i\theta }a_{x,\mathrm{s}%
}\ ,\qquad x\in \mathbb{Z}^{d},\ \mathrm{s}\in \mathrm{S},
\label{automorphism gauge invariance}
\end{equation}%
defines a unique $\ast $-automorphism $\mathrm{g}_{\theta }$ of the $C^{\ast
}$-algebra $\mathcal{U}$. Note that, for any $\Lambda \subseteq \mathfrak{L}$%
, $\mathrm{g}_{\theta }(\mathcal{U}_{\Lambda })\subseteq \mathcal{U}%
_{\Lambda }$ and thus $\mathrm{g}_{\theta }$ canonically defines a $\ast $%
-automorphism of the subalgebra $\mathcal{U}_{\Lambda }$. A special role is
played by $\mathrm{g}_{\pi }$. Elements $A,B\in \mathcal{U}_{\Lambda }$, $%
\Lambda \subseteq \mathfrak{L}$, satisfying $\mathrm{g}_{\pi }(A)=A$ and $%
\mathrm{g}_{\pi }(B)=-B$ are respectively called even and odd. Every element
of the algebra can be decomposed into a sum of even and odd terms. (Elements 
$A\in \mathcal{U}_{\Lambda }$ satisfying $\mathrm{g}_{\theta }(A)=A$ for any 
$\theta \in \mathbb{R}/(2\pi \mathbb{Z)}$ are called gauge invariant.) The
space of even elements of $\mathcal{U}$ is denoted by 
\begin{equation}
\mathcal{U}^{+}\doteq \{A\in \mathcal{U}:A=\mathrm{g}_{\pi }(A)\}\subseteq 
\mathcal{U}\text{ }.  \label{definition of even operators}
\end{equation}%
It is a unital $C^{\ast }$-subalgebra of the $C^{\ast }$-algebra $\mathcal{U}
$. In physics, $\mathcal{U}^{+}$ is seen as more fundamental than $\mathcal{U%
}$, because of the local causality in quantum field theory, which holds in
the first $C^{\ast }$-algebra, but not in the second one. See, e.g.,
discussions in \cite[Section 2.3]{BruPedra-MFII}.

\subsection{States of Lattice Fermion Systems}

\subsubsection{Even States}

States on the $C^{\ast }$-algebra $\mathcal{U}$ are, by definition, linear
functionals $\rho :\mathcal{U}\rightarrow \mathbb{C}$ which are positive,
i.e., for all elements $A\in \mathcal{U}$, $\rho (\left\vert A\right\vert
^{2})\geq 0$, and normalized, i.e., $\rho (\mathfrak{1})=1$. Equivalently,
the linear functional $\rho $ is a state iff $\rho (\mathfrak{1})=1$ and $%
\Vert \rho \Vert _{\mathcal{U}^{\ast }}=1$. The set of all states on $%
\mathcal{U}$ is denoted by 
\begin{equation}
E\doteq \bigcap\limits_{A\in \mathcal{U}}\{\rho \in \mathcal{U}^{\ast }:\rho
(\mathfrak{1})=1,\;\rho (|A|^{2})\geq 0\}\ .  \label{set of states}
\end{equation}%
This convex set is metrizable and compact with respect to the weak$^{\ast }$
topology. Mutatis mutandis, for every $\Lambda \subseteq \mathfrak{L}$, we
define the set $E_{\Lambda }$ of all states on the sub-algebra $\mathcal{U}%
_{\Lambda }\subseteq \mathcal{U}$. For any $\Lambda \subseteq \mathfrak{L}$,
the symbol $\rho _{\Lambda }$ denotes the restriction of any $\rho \in E$ to
the sub-algebra $\mathcal{U}_{\Lambda }$. This restriction is clearly a
state on $\mathcal{U}_{\Lambda }$.

Even states on $\mathcal{U}_{\Lambda }$, $\Lambda \subseteq \mathfrak{L}$,
are, by definition, the states $\rho \in E_{\Lambda }$ satisfying $\rho
\circ \mathrm{g}_{\pi }=\rho $. In other words, the even states\ on $%
\mathcal{U}_{\Lambda }$ are exactly those vanishing on all odd elements of $%
\mathcal{U}_{\Lambda }$. The set of even states on $\mathcal{U}$ can be
canonically identified with the set of states on the $C^{\ast }$-subalgebra $%
\mathcal{U}^{+}$ of even elements, by \cite[Proof of Proposition 2.1]%
{BruPedra-MFII}. As a consequence, physically relevant states on $\mathcal{U}
$ are even.

\subsubsection{Translation-Invariant States\label{Sect Periodic-State Space}}

Lattice translations refer to the group homomorphism $x\mapsto \alpha _{x}$
from $(\mathbb{Z}^{d},+)$ to the group of $\ast $-automorphisms of the CAR $%
C^{\ast }$-algebra $\mathcal{U}$ of the (infinite) lattice $\mathfrak{L}$,
which is uniquely defined by the condition%
\begin{equation}
\alpha _{x}(a_{y,\mathrm{s}})=a_{y+x,\mathrm{s}}\ ,\quad y\in \mathfrak{L},\;%
\mathrm{s}\in \mathrm{S}\text{ }.  \label{transl}
\end{equation}%
Via this group homomorphism we define the translation invariance of states
and interactions of lattice fermion systems.

The state $\rho \in E$ is said to be translation-invariant iff it satisfies $%
\rho \circ \alpha _{x}=\rho $ for all $x\in \mathbb{Z}^{d}$. The space of
translation-invariant states on $\mathcal{U}$ is the convex set%
\begin{equation}
E_{1}\doteq \bigcap\limits_{x\in \mathbb{Z}^{d},\text{ }A\in \mathcal{U}%
}\{\rho \in \mathcal{U}^{\ast }:\rho (\mathfrak{1})=1,\;\rho (|A|^{2})\geq
0,\;\rho =\rho \circ \alpha _{x}\}\ ,  \label{periodic invariant states}
\end{equation}%
which is again metrizable and compact with respect to the weak$^{\ast }$
topology. Any translation-invariant state is even. See, for instance, \cite[%
Lemma 1.8]{BruPedra2}. Thanks to the Krein-Milman theorem \cite[Theorem 3.23]%
{Rudin}, $E_{1}$ is the weak$^{\ast }$-closure of the convex hull of the
(non-empty) set $\mathcal{E}\left( E_{1}\right) $ of its extreme points,
which turns out to be a weak$^{\ast }$-dense ($G_{\delta }$) subset \cite[%
Corollary 4.6]{BruPedra2}: 
\begin{equation}
E_{1}=\overline{\mathrm{co}}\left( \mathcal{E}\left( E_{1}\right) \right) =%
\overline{\mathcal{E}\left( E_{1}\right) }\ ,  \label{cov heull l perio}
\end{equation}%
where $\overline{\mathrm{co}}(K)$ denotes the weak$^{\ast }$-closed convex
hull of a set $K$. This fact is well-known and is also true for quantum spin
systems on lattices \cite[Example 4.3.26 and discussions p. 464]%
{BrattelliRobinsonI}.

Since $E_{1}$ is metrizable (because of the separability of $\mathcal{U}$),
the Choquet theorem applies: By \cite[Theorem 1.9]{BruPedra2}, for any $\rho
\in E_{1}$, there is a unique probability measure\footnote{%
For $E$ is a metrizable compact space, any finite Borel measure is regular
and tight. Thus, here, probability measures are just the same as normalized
Borel measures.} $\mu _{\rho }$ on $E_{1}$ such that $\mu _{\rho }\left( 
\mathcal{E}\left( E_{1}\right) \right) =1$ and%
\begin{equation}
\rho \left( A\right) =\int_{E_{1}}\hat{\rho}\left( A\right) \mu _{\rho
}\left( \mathrm{d}\hat{\rho}\right) \text{ },\qquad A\in \mathcal{U}\text{ }.
\label{choquet theorem}
\end{equation}%
In particular, $E_{1}$ is a Choquet simplex. In fact, up to an affine
homeomorphism, $E_{1}$ is the so-called Poulsen simplex \cite[Theorem 1.12]%
{BruPedra2}. (Note in passing that $\mu _{\rho }\ $is an orthogonal measure,
thanks to \cite[Theorem 5.1]{BruPedra-MFIII}.)

The unique decomposition of a translation-invariant state $\rho \in E_{1}$
in terms of extreme translation-invariant states $\hat{\rho}\in \mathcal{E}%
(E_{1})$ is also called the \emph{ergodic} decomposition of $\rho $ because
of the following fact: Define the space-averages of any element $A\in 
\mathcal{U}$ by%
\begin{equation}
A_{L}\doteq \frac{1}{|\Lambda _{L}|}\sum\limits_{x\in \Lambda _{L}}\alpha
_{x}\left( A\right) \ ,\mathrm{\qquad }L\in \mathbb{N}_{0}\text{ }.
\label{Limit of Space-Averages}
\end{equation}%
Then, by definition, a translation-invariant state $\hat{\rho}\in E_{1}$ is
said to be \emph{ergodic} if 
\begin{equation}
\lim\limits_{L\rightarrow \infty }\hat{\rho}(\left\vert A_{L}\right\vert
^{2})=|\hat{\rho}(A)|^{2}\ ,\mathrm{\qquad }A\in \mathcal{U}\text{ }.
\label{Ergodicity}
\end{equation}%
(Recall Equation (\ref{carre de A}).) By \cite[Theorem 1.16]{BruPedra2}, any
extreme translation-invariant state is ergodic and vice-versa. In other
words, the set of extreme translation-invariant states is equal to%
\begin{equation}
\mathcal{E}(E_{1})=\left\{ \hat{\rho}\in E_{1}:\hat{\rho}\text{ is ergodic}%
\right\} =\bigcap\limits_{A\in \mathcal{U}}\left\{ \hat{\rho}\in
E_{1}:\lim\limits_{L\rightarrow \infty }\hat{\rho}(\left\vert
A_{L}\right\vert ^{2})=|\hat{\rho}(A)|^{2}\right\} \ .  \label{Ergodicity2}
\end{equation}

\section{Short-Range Models\label{Section Banach space interaction}}

\subsection{The Banach Space of Interactions\label{Section Banach space
interaction copy(1)}}

In the algebraic setting, a (complex) \emph{interaction} is, by definition,
any mapping $\Phi :\mathcal{P}_{\mathrm{f}}\rightarrow \mathcal{U}^{+}$ from
the set $\mathcal{P}_{\mathrm{f}}\subseteq 2^{\mathfrak{L}}$ of all finite
subsets of $\mathfrak{L}$ to the $C^{\ast }$-subalgebra $\mathcal{U}^{+}$ (%
\ref{definition of even operators}) of even elements such that $\Phi
_{\Lambda }\in \mathcal{U}_{\Lambda }$ for all $\Lambda \in \mathcal{P}_{%
\mathrm{f}}$. The set $\mathcal{V}$ of all interactions can be naturally
endowed with the structure of a complex vector space (via the usual
point-wise vector space operations), as well as with the antilinear
involution 
\begin{equation}
\Phi \mapsto \Phi ^{\ast }\doteq (\Phi _{\Lambda }^{\ast })_{\Lambda \in 
\mathcal{P}_{\mathrm{f}}}\ .  \label{involution}
\end{equation}%
An interaction $\Phi $ is said to be self-adjoint iff $\Phi =\Phi ^{\ast }$.
The set $\mathcal{V}^{\mathbb{R}}$ of all self-adjoint interactions forms a
real subspace of the space of all (complex) interactions.

By definition, the interaction $\Phi \in \mathcal{V}$ is
translation-invariant\ if%
\begin{equation}
\Phi _{\Lambda +x}=\alpha _{x}\left( \Phi _{\Lambda }\right) \ ,\qquad x\in 
\mathbb{Z}^{d},\ \Lambda \in \mathcal{P}_{\mathrm{f}}\text{ },
\label{ti interaction}
\end{equation}%
where $\{\alpha _{x}\}_{x\in \mathbb{Z}^{d}}$ is the set of (translation) $%
\ast $-automorphisms of $\mathcal{U}$ defined by (\ref{transl}), while%
\begin{equation}
\Lambda +x\doteq \left\{ y+x\in \mathfrak{L}:y\in \Lambda \right\} \ ,\qquad
x\in \mathbb{Z}^{d},\ \Lambda \in \mathcal{P}_{\mathrm{f}}\text{ }.
\label{translation box}
\end{equation}%
The space of translation-invariant interactions is denoted by 
\begin{equation*}
\mathcal{V}_{1}\doteq \bigcap\limits_{x\in \mathbb{Z}^{d},\ \Lambda \in 
\mathcal{P}_{\mathrm{f}}}\left\{ \Phi \in \mathcal{V}:\Phi _{\Lambda
+x}=\alpha _{x}\left( \Phi _{\Lambda }\right) \right\} \varsubsetneq 
\mathcal{V}\ .
\end{equation*}%
Using the norm 
\begin{equation}
\left\Vert \Phi \right\Vert _{\mathcal{W}_{1}}\doteq \sum\limits_{\Lambda
\in \mathcal{P}_{\mathrm{f}},\;\Lambda \supseteq \{0\}}\left\vert \Lambda
\right\vert ^{-1}\left\Vert \Phi _{\Lambda }\right\Vert _{\mathcal{U}}\
,\qquad \Phi \in \mathcal{V}_{1}\text{ },  \label{iteration0}
\end{equation}%
we obtain a separable Banach space 
\begin{equation}
\mathcal{W}_{1}\doteq \left\{ \Phi \in \mathcal{V}_{1}:\left\Vert \Phi
\right\Vert _{\mathcal{W}_{1}}<\infty \right\}  \label{W1}
\end{equation}%
of translation-invariant interactions. The (real) Banach subspace of
interactions that are simultaneously self-adjoint and translation-invariant,
is denoted by 
\begin{equation}
\mathcal{W}_{1}^{\mathbb{R}}\doteq \mathcal{V}^{\mathbb{R}}\cap \mathcal{W}%
_{1}\ ,  \label{W1Re}
\end{equation}%
similar to $\mathcal{U}^{\mathbb{R}}\varsubsetneq \mathcal{U}$ and $\mathcal{%
V}^{\mathbb{R}}\varsubsetneq \mathcal{V}$.

Additionally, for any $\Lambda \in \mathcal{P}_{\mathrm{f}}$ with $\Lambda
\ni 0$, we define the closed subspace\footnote{$\mathcal{W}_{\Lambda }$ is a
closed subspace of $\mathcal{W}_{1}$ because of the continuity and linearity
of the mappings $\Phi \mapsto \Phi _{\mathcal{Z}}$ for all $\mathcal{Z}\in 
\mathcal{P}_{f}$.}%
\begin{equation}
\mathcal{W}_{\Lambda }\doteq \left\{ \Phi \in \mathcal{W}_{1}:\Phi _{%
\mathcal{Z}}=0\text{ whenever }\mathcal{Z}\nsubseteq \Lambda \text{, }%
\mathcal{Z}\ni 0\right\}  \label{eq:enpersitebis0}
\end{equation}%
of interactions that are simultaneously finite-range and
translation-invariant. Similar to (\ref{simple}),%
\begin{equation}
\mathcal{W}_{0}\doteq \bigcup_{L\in \mathbb{N}_{0}}\mathcal{W}_{\Lambda
_{L}}\subseteq \mathcal{W}_{1}  \label{W0}
\end{equation}%
is a dense subspace of $\mathcal{W}_{1}$. A translation-invariant
interaction $\Phi \in \mathcal{W}_{1}$ is said to be \emph{finite-range} if
it lies in $\mathcal{W}_{0}$. As usual, $\mathcal{W}_{0}^{\mathbb{R}}\doteq 
\mathcal{V}^{\mathbb{R}}\cap \mathcal{W}_{0}$.

To conclude, note that the norm defined by (\ref{iteration0}) is rather
weak. In fact, the Banach space $\mathcal{W}_{1}$ is quite large and
includes interactions having a long-range character in which concerns
dynamics, in the sense that usual Lieb-Robinson bounds may not hold true for
them. Models associated with $\mathcal{W}_{1}$ can however be properly named
(quasi-)short-range, in contrast with the mean-field models of Sections \ref%
{Long-Range Models}, which refer to an idealization of interactions that are
weak, but act across extremely long distances.

\subsection{Energy Density Functionals on Translation-Invariant States\label%
{energy density1}}

Local energy elements associated with a given complex interaction $\Phi \in 
\mathcal{V}$ correspond to the following sequence within the $C^{\ast }$%
-subalgebra $\mathcal{U}^{+}$ (\ref{definition of even operators}) of even
elements:%
\begin{equation}
U_{L}^{\Phi }\doteq \sum\limits_{\Lambda \subseteq \Lambda _{L}}\Phi
_{\Lambda }\in \mathcal{U}_{\Lambda _{L}}\cap \mathcal{U}^{+}\ ,\qquad L\in 
\mathbb{N}_{0}\text{ },  \label{equation fininte vol dynam0}
\end{equation}%
where we recall that $\Lambda _{L}$, $L\in \mathbb{N}_{0}$, are the cubic
boxes (\ref{eq:def lambda n}) used to define the thermodynamic limit. The
energy density of a state $\rho \in E$ with respect to a given interaction $%
\Phi \in \mathcal{V}$ is defined by 
\begin{equation*}
e_{\Phi }\left( \rho \right) \doteq \underset{L\rightarrow \infty }{\lim
\sup }\frac{\mathrm{Re}\{\rho \left( U_{L}^{\Phi }\right) \}}{\left\vert
\Lambda _{L}\right\vert }+i\ \underset{L\rightarrow \infty }{\lim \sup }%
\frac{\mathrm{Im}\{\rho \left( U_{L}^{\Phi }\right) \}}{\left\vert \Lambda
_{L}\right\vert }\in \lbrack -\infty ,\infty ]+i[-\infty ,\infty ]\ .
\end{equation*}%
If $\Phi \in \mathcal{V}^{\mathbb{R}}$ is self-adjoint then $(U_{L}^{\Phi
})_{L\in \mathbb{N}_{0}}$ is a sequence (of local Hamiltonians) in $\mathcal{%
U}^{\mathbb{R}}$ and thus, $e_{\Phi }(\rho )$ belongs to $[-\infty ,\infty ]$
for all states $\rho \in E$.

By \cite[Proposition 3.2]{BruPedra-MFII}, for any translation-invariant
state $\rho \in E_{1}$ (\ref{periodic invariant states}) and each
translation-invariant interaction$\ \Phi \in \mathcal{W}_{1}$, 
\begin{equation}
e_{\Phi }\left( \rho \right) =\lim\limits_{L\rightarrow \infty }\frac{\rho
\left( U_{L}^{\Phi }\right) }{\left\vert \Lambda _{L}\right\vert }=\rho
\left( \mathfrak{e}_{\Phi }\right) \ ,  \label{ssssssssss}
\end{equation}%
where $\mathfrak{e}_{(\cdot )}:\mathcal{W}_{1}\rightarrow \mathcal{U}^{+}$
is the continuous mapping from the Banach space $\mathcal{W}_{1}$ to the $%
C^{\ast }$-algebra $\mathcal{U}^{+}\subseteq \mathcal{U}$, defined by 
\begin{equation}
\mathfrak{e}_{\Phi }\doteq \sum\limits_{\mathcal{Z}\in \mathcal{P}_{\mathrm{f%
}},\;\mathcal{Z}\ni 0}\frac{\Phi _{\mathcal{Z}}}{\left\vert \mathcal{Z}%
\right\vert }\in \mathcal{U}^{+}\ ,\text{\qquad }\Phi \in \mathcal{W}_{1}%
\text{ }.  \label{eq:enpersite}
\end{equation}%
In particular, for any fixed $\Phi \in \mathcal{W}_{1}$, the mapping $\rho
\mapsto e_{\Phi }(\rho )$ from $E_{1}$ to $\mathbb{C}$ is a weak$^{\ast }$%
-continuous affine functional. Last but not least, by straightforward
estimates using Equation (\ref{iteration0}), note that, for all
translation-invariant interactions $\Phi ,\Psi \in \mathcal{W}_{1}$, 
\begin{equation}
\left\Vert U_{L}^{\Phi }-U_{L}^{\Psi }\right\Vert _{\mathcal{U}}=\left\Vert
U_{L}^{\Phi -\Psi }\right\Vert _{\mathcal{U}}\leq \left\vert \Lambda
_{L}\right\vert \left\Vert \Phi -\Psi \right\Vert _{\mathcal{W}_{1}}\
,\qquad L\in \mathbb{N}_{0}\text{ },  \label{norm Uphi}
\end{equation}%
and, for all translation-invariant states $\rho \in E_{1}$, 
\begin{equation}
\left\vert e_{\Phi }\left( \rho \right) -e_{\Psi }\left( \rho \right)
\right\vert \leq \left\Vert \Phi -\Psi \right\Vert _{\mathcal{W}_{1}}\
,\qquad \Phi ,\Psi \in \mathcal{W}_{1}\ .  \label{inequality a la con}
\end{equation}%
In particular, given any fixed translation-invariant state $\rho \in E_{1}$,
the linear mapping $\Phi \mapsto e_{\Phi }\left( \rho \right) $ from $%
\mathcal{W}_{1}$ to $\mathbb{C}$ is Lipschitz continuous.

\subsection{Entropy Density Functional on Translation-Invariant States \label%
{energy density2}}

The entropy density functional $s:E_{1}\rightarrow \mathbb{R}_{0}^{+}$ maps
any translation-invariant state to its von Neumann entropy per unit volume
in the thermodynamic limit, that is,%
\begin{equation}
s(\rho )\doteq -\lim\limits_{L\rightarrow \infty }\left\{ \frac{1}{|\Lambda
_{L}|}\mathrm{Trace}\,\left( \mathrm{d}_{\rho _{\Lambda _{L}}}\ln \mathrm{d}%
_{\rho _{\Lambda _{L}}}\right) \right\} \ ,\qquad \rho \in E_{1}\text{ },
\label{entropy density}
\end{equation}%
where we recall that $\rho _{\Lambda _{L}}$ is the restriction of the
translation-invariant state $\rho \in E_{1}$ to the finite-dimensional CAR $%
C^{\ast }$-algebra $\mathcal{U}_{\Lambda _{L}}$ of the cubic box $\Lambda
_{L}$ defined by (\ref{eq:def lambda n}). Here, $\mathrm{d}_{\rho _{\Lambda
_{L}}}\in \mathcal{U}_{\Lambda _{L}}$ is the (uniquely defined) density
matrix representing the state $\rho _{\Lambda _{L}}$ via a trace\footnote{%
For $\Lambda \in \mathcal{P}_{f}$, the trace on the finite-dimensional $%
C^{\ast }$-algebra $\mathcal{U}_{\Lambda }$ refers to the usual trace on the
fermionic Fock space representation.}: 
\begin{equation*}
\rho _{\Lambda _{L}}(\cdot )=\mathrm{Trace}\,\left( \;\cdot \;\mathrm{d}%
_{\rho _{\Lambda _{L}}}\right) \ .
\end{equation*}%
By \cite[Lemma 4.15]{BruPedra2}, the functional $s$ is well-defined on the
set $E_{1}$ of translation-invariant states. See also \cite[Section 10.2]%
{Araki-Moriya}. By \cite[Lemma 1.29]{BruPedra2}, the entropy density
functional $s$ is a weak$^{\ast }$-upper semicontinuous affine functional.

\subsection{Equilibrium States as Minimizers of the Free Energy Density\label%
{Translation Invariant Equilibrium States}}

Equilibrium states of lattice fermion systems are always defined from a
fixed self-adjoint interaction, which determines the energy density of
states as well as the microscopic dynamics. Here, we define equilibrium
states as minimizers of the free energy density functional, in direct
relation with the notion of (grand-canonical) pressure: For a fixed $\beta
\in \mathbb{R}^{+}$, the infinite volume pressure $\mathrm{P}$ is the
real-valued function on the real Banach subspace $\mathcal{W}_{1}^{\mathbb{R}%
}$ (\ref{W1Re}) of interactions that are self-adjoint and
translation-invariant, defined by 
\begin{equation}
\Phi \mapsto \mathrm{P}_{\Phi }\doteq \underset{L\rightarrow \infty }{\lim }%
\frac{1}{\beta |\Lambda _{L}|}\ln \mathrm{Trace}(\mathrm{e}^{-\beta
U_{L}^{\Phi }})\ .  \label{pressure short range}
\end{equation}%
Recall that the parameter $\beta \in \mathbb{R}^{+}$ is the inverse
temperature of the system. It is fixed once and for all and, therefore, it
is often omitted in our discussions or notation. By \cite[Theorem 2.12]%
{BruPedra2}, the above pressure is well-defined and, for any $\Phi \in 
\mathcal{W}_{1}^{\mathbb{R}}$,%
\begin{equation}
\mathrm{P}_{\Phi }=-\inf f_{\Phi }(E_{1})<\infty \text{ },
\label{pressure free energy}
\end{equation}%
where the mapping $f_{\Phi }:E_{1}\rightarrow \mathbb{R}$ is the free energy
density functional defined on the set $E_{1}$ of translation-invariant
states by%
\begin{equation}
f_{\Phi }\doteq e_{\Phi }-\beta ^{-1}s\ .  \label{map free energy}
\end{equation}%
Recall that $e_{\Phi }:E_{1}\rightarrow \mathbb{R}$ is the energy density
functional defined in Section \ref{energy density1} for any $\Phi \in 
\mathcal{W}_{1}$, while $s:E_{1}\rightarrow \mathbb{R}_{0}^{+}$ is the
entropy density functional presented in Section \ref{energy density2}.

As explained in Sections \ref{energy density1}--\ref{energy density2}, the
functionals $e_{\Phi }$, $\Phi \in \mathcal{W}_{1}^{\mathbb{R}}$, and $%
-\beta ^{-1}s$, $\beta \in \mathbb{R}^{+}$, are weak$^{\ast }$-lower
semicontinuous and affine (with $e_{\Phi }$ being actually weak$^{\ast }$%
-continuous). In particular, the functional $f_{\Phi }$ (\ref{map free
energy}) is weak$^{\ast }$-lower semicontinuous and affine. Therefore, for
any $\Phi \in \mathcal{W}_{1}^{\mathbb{R}}$, this functional has minimizers
in the weak$^{\ast }$-compact set $E_{1}$ of translation-invariant states.
Similarly to what is done for translation-invariant quantum spin systems
(see, e.g., \cite{BrattelliRobinson,Sewell}), for any $\Phi \in \mathcal{W}%
_{1}^{\mathbb{R}}$, the set $\mathit{M}_{\Phi }$ of translation-invariant
equilibrium states\ of fermions on the lattice is, by definition, the
(non-empty) set 
\begin{equation}
\mathit{M}_{\Phi }\doteq \left\{ \omega \in E_{1}:f_{\Phi }\left( \omega
\right) =\inf \,f_{\Phi }(E_{1})=-\mathrm{P}_{\Phi }\right\}
\label{minimizer short range}
\end{equation}%
of all minimizers of the free energy density functional $f_{\Phi }$ over the
set $E_{1}$. By affineness and weak$^{\ast }$-lower semicontinuity of $%
f_{\Phi }$, $\mathit{M}_{\Phi }$ is a (non-empty) weak$^{\ast }$-closed face%
\footnote{%
Recall that a face $F$ of a convex set $K$ is a subset of $K$ with the
property that, if $\rho =\lambda _{1}\rho _{1}+\cdots +\lambda _{n}\rho
_{n}\in F$ with $\rho _{1},\ldots ,\rho _{n}\in K$, $\lambda _{1},\ldots
,\lambda _{n}\in (0,1)$ and $\lambda _{1}+\cdots +\lambda _{n}=1$, then $%
\rho _{1},\ldots ,\rho _{n}\in F$.} of $E_{1}$ for any $\Phi \in \mathcal{W}%
_{1}^{\mathbb{R}}$.

Recall that this is not the only reasonable way of defining equilibrium
states. For fixed interactions, they can also be defined as tangent
functionals to the corresponding pressure functional or via other conditions
like the local stability, the Gibbs condition or the Kubo-Martin-Schwinger
(KMS) condition. All these definitions are generally not completely
equivalent to each other. For more details, we recommend the paper \cite%
{Araki-Moriya}.

\section{Mean-Field Models\label{Long-Range Models}}

\subsection{The Banach Space of Mean-Field Models}

Let 
\begin{equation}
\mathbb{S}\doteq \left\{ \Phi \in \mathcal{W}_{1}:\left\Vert \Phi
\right\Vert _{\mathcal{U}}=1\right\}  \label{unit sphere}
\end{equation}%
be the unit sphere of the Banach space $\mathcal{W}_{1}$ (\ref{W1}) of
translation-invariant short-range interactions. Let $\mathcal{S}_{1}$ denote
the space of signed Borel measures of bounded variation on $\mathbb{S}$,
which is a real Banach space whose norm is the total variation of measures 
\begin{equation*}
\left\Vert \mathfrak{a}\right\Vert _{\mathcal{S}_{1}}\doteq \left\vert 
\mathfrak{a}\right\vert \left( \mathbb{S}\right) \text{ },\qquad \mathfrak{a}%
\in \mathcal{S}_{1}\text{ }.
\end{equation*}%
The space of translation-invariant mean-field, or long-range, models is the
separable (real) Banach space%
\begin{equation}
\mathcal{M}_{1}\doteq \left\{ \mathfrak{m}\in \mathcal{W}_{1}^{\mathbb{R}%
}\times \mathcal{S}_{1}:\left\Vert \mathfrak{m}\right\Vert _{\mathcal{M}%
_{1}}<\infty \right\} \ ,  \label{def long range1}
\end{equation}%
whose norm is 
\begin{equation}
\left\Vert \mathfrak{m}\right\Vert _{\mathcal{M}_{1}}\doteq \left\Vert \Phi
\right\Vert _{\mathcal{W}_{1}}+\left\Vert \mathfrak{a}\right\Vert _{\mathcal{%
S}_{1}}\,,\qquad \mathfrak{m}\doteq \left( \Phi ,\mathfrak{a}\right) \in 
\mathcal{M}_{1}\text{ }.  \label{def long range2}
\end{equation}%
The spaces $\mathcal{W}_{1}^{\mathbb{R}}$ and $\mathcal{S}_{1}$ are seen as
subspaces of $\mathcal{M}_{1}$, i.e., 
\begin{equation}
\mathcal{W}_{1}^{\mathbb{R}}\subseteq \mathcal{M}_{1}\qquad \text{and}\qquad 
\mathcal{S}_{1}\subseteq \mathcal{M}_{1}\ ,  \label{identification model}
\end{equation}%
with the canonical identifications $\Phi \equiv \left( \Phi ,0\right) \in 
\mathcal{M}_{1}$ for $\Phi \in \mathcal{W}_{1}^{\mathbb{R}}$ and $\mathfrak{a%
}\equiv \left( 0,\mathfrak{a}\right) \in \mathcal{M}_{1}$ for $\mathfrak{a}%
\in \mathcal{S}_{1}$.

The local Hamiltonians associated with any mean-field model $\mathfrak{m}%
\doteq \left( \Phi ,\mathfrak{a}\right) \in \mathcal{M}_{1}$ are the
(well-defined) self-adjoint elements%
\begin{equation}
U_{L}^{\mathfrak{m}}\doteq U_{L}^{\Phi }+\frac{1}{\left\vert \Lambda
_{L}\right\vert }\int_{\mathbb{S}}\left\vert U_{L}^{\Psi }\right\vert ^{2}%
\mathfrak{a}\left( \mathrm{d}\Psi \right) \,,\qquad L\in \mathbb{N}_{0}\text{
},  \label{equation long range energy}
\end{equation}%
where we recall that $|A|^{2}\doteq A^{\ast }A$ for any $A\in \mathcal{U}$,
see (\ref{carre de A}). Note that $U_{L}^{\left( \Phi ,0\right)
}=U_{L}^{\Phi }$ for any self-adjoint interaction $\Phi \in \mathcal{W}_{1}^{%
\mathbb{R}}$ (cf. (\ref{identification model})) and straightforward
estimates yield the bound 
\begin{equation}
\left\Vert U_{L}^{\mathfrak{m}}\right\Vert _{\mathcal{U}}\leq \left\vert
\Lambda _{L}\right\vert \left\Vert \mathfrak{m}\right\Vert _{\mathcal{M}%
_{1}}\ ,\qquad L\in \mathbb{N}_{0}\text{ },\ \mathfrak{m}\in \mathcal{M}_{1}%
\text{ },  \label{energy bound long range}
\end{equation}%
by Equations (\ref{norm Uphi}) and (\ref{def long range2}).

\begin{remark}[Equivalent definitions of translation-invariant mean-field
models]
\label{remark temp copy(1)}\mbox{ }\newline
\cite{BruPedra2} and \cite{BruPedra-MFII,BruPedra-MFIII} have different
definitions of mean-field models. We use here the formalism introduced in 
\cite{BruPedra-MFII,BruPedra-MFIII}. Compare (\ref{def long range1}) with 
\cite[Definition 2.1]{BruPedra2}. In fact, \cite[Section 8]{BruPedra-MFII}
shows that the results of \cite{BruPedra2} apply equally to all
translation-invariant mean-field models $\mathfrak{m}\in \mathcal{M}_{1}$.
\end{remark}

\subsection{Purely Repulsive and Purely Attractive Mean-Field Models\label%
{Section purely attrac}}

By the Hahn decomposition theorem, any signed measure $\mathfrak{a}$ of
bounded variation on the unit sphere $\mathbb{S}$ of the Banach space $%
\mathcal{W}_{1}$ has a unique decomposition%
\begin{equation}
\mathfrak{a=}\underset{\text{mean-field repulsion}}{\underbrace{\mathfrak{a}%
_{+}}}-\underset{\text{mean-field attraction}}{\underbrace{\mathfrak{a}_{-}}}
\label{Hahn decomposition}
\end{equation}%
$\mathfrak{a}_{\pm }$ being two positive finite measures vanishing on
disjoint Borel sets, respectively denoted by $\mathbb{S}_{\mp }\subseteq 
\mathbb{S}$. Recall that such a decomposition is called the Jordan
decomposition of the measure of bounded variation $\mathfrak{a}$ and $|%
\mathfrak{a}|=\mathfrak{a}_{+}+\mathfrak{a}_{-}$. Mean-field attractions are
represented by the measure $\mathfrak{a}_{-}$, whereas $\mathfrak{a}_{+}$
refers to mean-field repulsions. A mean-field model $\mathfrak{m}\doteq
\left( \Phi ,\mathfrak{a}\right) \in \mathcal{M}_{1}$ is said to be \emph{%
purely attractive} iff $\mathfrak{a}_{+}=0$, while it is \emph{purely
repulsive} iff $\mathfrak{a}_{-}=0$.

Distinguishing between these two special types of models is important
because the effects of mean-field attractions and repulsions on the
structure of the corresponding sets of (generalized) equilibrium states can
be very different: By \cite[Theorem 2.25]{BruPedra2}, mean-field attractions
have no particular effect on the structure of the set of (generalized,
translation-invariant) equilibrium states, which is still a (non-empty) weak$%
^{\ast }$-closed face of the set $E_{1}$ of translation-invariant states,
like for interactions in $\mathcal{W}_{1}^{\mathbb{R}}$. By contrast,
mean-field repulsions have generally a geometrical effect by possibly
breaking the face structure of the set of (generalized) equilibrium states
(see \cite[Lemma 9.8]{BruPedra2}).

\subsection{The Space-Averaging Functional on Translation-Invariant States 
\label{Section space averaging}}

In addition to the energy density and entropy density functionals, defined
respectively in Sections \ref{energy density1}--\ref{energy density2}, we
need the so-called space-averaging functional in order to study the
thermodynamic properties of long-range models. This new density functional
is defined on the set $E_{1}$ of translation-invariant states as follows:
For any $A\in \mathcal{U}$, the mapping $\Delta _{A}:E_{1}\rightarrow 
\mathbb{R}$ is (well-)defined by%
\begin{equation}
\rho \mapsto \Delta _{A}\left( \rho \right) \doteq \lim\limits_{L\rightarrow
\infty }\rho \left( \left\vert A_{L}\right\vert ^{2}\right) \in \left[ |\rho
(A)|^{2},\Vert A\Vert _{\mathcal{U}}^{2}\right] \text{ },
\label{space averaging}
\end{equation}%
where $|A_{L}|^{2}\doteq A_{L}^{\ast }A_{L}$ (see (\ref{carre de A})) and $%
A_{L}$ is defined by (\ref{Limit of Space-Averages}) for any $L\in \mathbb{N}%
_{0}$. Compare this last definition with Equation (\ref{Ergodicity}). See
also \cite[Section 1.3]{BruPedra2}. By \cite[Theorem 1.18]{BruPedra2}, the
functional $\Delta _{A}$ is affine and weak$^{\ast }$-upper semicontinuous.
Thanks again to \cite[Theorem 1.18]{BruPedra2}, note additionally that, at
any fixed translation-invariant state $\rho \in E_{1}$,%
\begin{equation*}
|\Delta _{A}\left( \rho \right) -\Delta _{B}\left( \rho \right) |\leq (\Vert
A\Vert _{\mathcal{U}}+\Vert B\Vert _{\mathcal{U}})\Vert A-B\Vert _{\mathcal{U%
}}\ ,\qquad A,B\in \mathcal{U}\text{ }.
\end{equation*}%
In particular, at fixed $\rho \in E_{1}$, the mapping $A\mapsto \Delta
_{A}\left( \rho \right) $ from $\mathcal{U}$ to $\mathbb{R}$ is locally
Lipschitz continuous.

For any $A\in \mathcal{U}$, \cite[Theorem 1.19]{BruPedra2} also proves,
among other things, that $\Delta _{A}$ can be decomposed in terms of an
integral over the set $\mathcal{E}_{1}$: For all $\rho \in E_{1}$ with
ergodic (or extreme) decomposition given by the probability measure $\mu
_{\rho }$ of Equation (\ref{choquet theorem}), 
\begin{equation*}
\Delta _{A}\left( \rho \right) =\int_{\mathcal{E}_{1}}\left\vert \hat{\rho}%
\left( A\right) \right\vert ^{2}\mathrm{d}\mu _{\rho }\left( \hat{\rho}%
\right) \ .
\end{equation*}%
Its $\Gamma $-regularization $\Gamma \left( \Delta _{A}\right) $ on $E_{1}$\
is the weak$^{\ast }$-continuous convex mapping $\rho \mapsto \left\vert
\rho \left( A\right) \right\vert ^{2}$. Recall that the so-called $\Gamma $%
-regularization of a functional $h$ on $E_{1}$, denoted here by $\Gamma (h)$%
, is defined by 
\begin{equation}
\Gamma \left( h\right) \left( \rho \right) \doteq \sup \left\{ m(\rho ):m\in 
\mathrm{A}\left( \mathcal{U}^{\ast }\right) \;\text{and }m|_{E_{1}}\leq
h|_{E_{1}}\right\} \ ,\qquad \rho \in E_{1}\ ,  \label{gamma regulirisation}
\end{equation}%
with $\mathrm{A}\left( \mathcal{U}^{\ast }\right) $ being the set of all
affine and weak$^{\ast }$-continuous functionals on the dual space $\mathcal{%
U}^{\ast }$ of the $C^{\ast }$-algebra $\mathcal{U}$. The $\Gamma $%
-regularization $\Gamma \left( h\right) $ of a functional $h$ on $E_{1}$
equals its twofold Legendre-Fenchel transform, also called the biconjugate%
\emph{\ }(functional) of $h$. Indeed, $\Gamma \left( h\right) $ is the
largest lower weak$^{\ast }$-semicontinuous and convex minorant of $h$. Note
that the $\Gamma $--regularization $\Gamma \left( h_{1}+h_{2}\right) $ of
the sum of two functionals $h_{1}$ and $h_{2}$ is generally not equal to the
sum $\Gamma \left( h_{1}\right) +\Gamma \left( h_{2}\right) $. For more
details on $\Gamma $-regularizations, we recommend \cite[Section 10.5]%
{BruPedra2} and, in particular, \cite[Corollary 10.30]{BruPedra2}.

For any signed Borel measure $\mathfrak{a}\in \mathcal{S}_{1}$ of bounded
variation on $\mathbb{S}$, we define the space-averaging functional $\Delta
_{\mathfrak{a}}:E_{1}\rightarrow \mathbb{R}$ on translation-invariant states
by 
\begin{equation}
\Delta _{\mathfrak{a}}\left( \rho \right) \doteq \int_{\mathbb{S}}\Delta _{%
\mathfrak{e}_{\Psi }}\left( \rho \right) \mathfrak{a}\left( \mathrm{d}\Psi
\right) \text{ },\qquad \rho \in E_{1}\ .
\label{Free-energy density long range0}
\end{equation}%
Recall that the continuous mapping $\mathfrak{e}_{(\cdot )}:\mathcal{W}%
_{1}\rightarrow \mathcal{U}$ is\ defined by Equation (\ref{eq:enpersite}).
By \cite[Theorem 1.18]{BruPedra2}, $\Delta _{\mathfrak{a}}$ is a
well-defined, affine and weak$^{\ast }$-upper semicontinuous functional on
the set $E_{1}$ of translation-invariant states. By \cite[Theorem 1.19]%
{BruPedra2}, for any positive measure $\mathfrak{a}\in \mathcal{S}_{1}$,
i.e., $\mathfrak{a=\mathfrak{a}_{+}-a}_{-}=\mathfrak{\mathfrak{a}_{+}}$, its 
$\Gamma $-regularization equals the functional%
\begin{equation}
\Gamma \left( \Delta _{\mathfrak{a}}\right) \left( \rho \right) =\int_{%
\mathbb{S}}\left\vert \rho \left( \mathfrak{e}_{\Psi }\right) \right\vert
^{2}\mathfrak{a}_{+}\left( \mathrm{d}\Psi \right) \text{ },\qquad \rho \in
E_{1}\ .  \label{gammareguliration of delata}
\end{equation}

\subsection{Infinite Volume Pressures\label{Section Pressures Longrange}}

\subsubsection{Conventional Pressures}

As is usual in quantum statistical mechanics, at any given inverse
temperature $\beta \in \mathbb{R}^{+}$, the conventional infinite volume
pressure for translation-invariant mean-field models is, by definition, the
real-valued function $\mathrm{P}^{\sharp }$ on the Banach space $\mathcal{M}%
_{1}$ (\ref{def long range1}) of translation-invariant mean-field models,
defined by 
\begin{equation}
\mathfrak{m}\mapsto \mathrm{P}_{\mathfrak{m}}^{\sharp }\doteq \underset{%
L\rightarrow \infty }{\lim }\frac{1}{\beta |\Lambda _{L}|}\ln \mathrm{Trace}(%
\mathrm{e}^{-\beta U_{L}^{\mathfrak{m}}})\ .  \label{pressure long range}
\end{equation}%
By \cite[Theorem 2.12]{BruPedra2}, this mapping is well-defined and, for any 
$\mathfrak{m}=(\Phi ,\mathfrak{\mathfrak{a}})\in \mathcal{M}_{1}$, 
\begin{equation}
\mathrm{P}_{\mathfrak{m}}^{\sharp }=-\inf f_{\mathfrak{m}}^{\sharp }\left(
E_{1}\right) \in \mathbb{R}\text{ },  \label{BCS main theorem 1eq}
\end{equation}%
where $f_{\mathfrak{m}}^{\sharp }:E_{1}\rightarrow \mathbb{R}$ is the free
energy density functional defined by%
\begin{equation}
f_{\mathfrak{m}}^{\sharp }\doteq \Delta _{\mathfrak{a}}+f_{\Phi }=\Delta _{%
\mathfrak{a}}+e_{\Phi }-\beta ^{-1}s\ .
\label{Free-energy density long range}
\end{equation}%
See Equation (\ref{map free energy}), defining the free energy density
functional $f_{\Phi }$ for any self-adjoint translation-invariant and
short-range interaction $\Phi \in \mathcal{W}_{1}^{\mathbb{R}}$. Observe
that Equation (\ref{BCS main theorem 1eq}) is an extension of (\ref{pressure
free energy}) -- which refers to\ the space $\mathcal{W}_{1}^{\mathbb{R}}$
of (quasi-)short-range models only -- to the space $\mathcal{M}_{1}\supseteq 
\mathcal{W}_{1}^{\mathbb{R}}$ of mean-field models.

\subsubsection{Non-conventional Pressures\label{section non-conventnioal
pressure}}

In \cite[Section 2.5]{BruPedra2}, we introduce a new free energy density
functional, as well as the corresponding pressure associated with mean-field
models via a variational problem. More precisely, we refer to \cite[%
Equations (2.16) and (2.18)]{BruPedra2}. They were used to study the
structure of generalized equilibrium states for mean-field models.

The new free energy density functional on $E_{1}$ is defined, for any fixed $%
\mathfrak{m}=(\Phi ,\mathfrak{\mathfrak{a}})\in \mathcal{M}_{1}$, by%
\begin{equation}
f_{\mathfrak{m}}^{\flat }\left( \rho \right) \doteq \int_{\mathbb{S}%
}\left\vert \rho \left( \mathfrak{e}_{\Psi }\right) \right\vert ^{2}%
\mathfrak{a}_{+}\left( \mathrm{d}\Psi \right) -\Delta _{\mathfrak{a}%
_{-}}\left( \rho \right) +e_{\Phi }(\rho )-\beta ^{-1}s(\rho )\ ,\qquad \rho
\in E_{1}\ ,  \label{convex functional g_m}
\end{equation}%
where $\mathfrak{a}_{\pm }\in \mathcal{S}_{1}$ are the two positive finite
measures resulting from the Jordan decomposition of the signed measure $%
\mathfrak{a}\in \mathcal{S}_{1}$. Here, $\Delta _{\mathfrak{a}_{-}}$ is the
space-averaging functional defined by (\ref{Free-energy density long range0}%
) for $\mathfrak{a}_{-}\in \mathcal{S}_{1}$. By Equation (\ref%
{gammareguliration of delata}), note that 
\begin{equation*}
\Delta _{\mathfrak{a}_{+}}\left( \rho \right) \geq \Gamma \left( \Delta _{%
\mathfrak{a}_{+}}\right) \left( \rho \right) =\int_{\mathbb{S}}\left\vert
\rho \left( \mathfrak{e}_{\Psi }\right) \right\vert ^{2}\mathfrak{a}%
_{+}\left( \mathrm{d}\Psi \right) \ ,\qquad \rho \in E_{1}\ ,
\end{equation*}%
while $\Delta _{\mathfrak{a}}=\Delta _{\mathfrak{a}_{+}}-\Delta _{\mathfrak{a%
}_{-}}$, by Equation (\ref{Hahn decomposition}) and linearity of the mapping 
$\mathfrak{a}\mapsto \Delta _{\mathfrak{a}}$. It follows from (\ref%
{Free-energy density long range}) and (\ref{convex functional g_m}) that 
\begin{equation}
f_{\mathfrak{m}}^{\flat }\leq f_{\mathfrak{m}}^{\sharp }\ ,\qquad \mathfrak{m%
}\in \mathcal{M}_{1}\ ,  \label{bound free bemol diese}
\end{equation}%
inspiring the notation $\flat $ (\textquotedblleft lower in
pitch\textquotedblright ) and $\sharp $ (\textquotedblleft higher in
pitch\textquotedblright ). In fact, we replace the functional $\Delta _{%
\mathfrak{a}_{+}}$ in $f_{\mathfrak{m}}^{\sharp }$ by its $\Gamma $%
-regularization (\ref{gammareguliration of delata}) to define the new free
energy density functional $f_{\mathfrak{m}}^{\flat }$.

Similar to (\ref{BCS main theorem 1eq}), the non-conventional pressure
associated with a mean-field model $\mathfrak{m}\in \mathcal{M}_{1}$ is then
defined to be the real-valued function 
\begin{equation}
\mathfrak{m}\mapsto \mathrm{P}_{\mathfrak{m}}^{\flat }\doteq -\inf f_{%
\mathfrak{m}}^{\flat }\left( E_{1}\right) \in \mathbb{R}
\label{Pressure bemol}
\end{equation}%
on the Banach space $\mathcal{M}_{1}$ (\ref{def long range1}) of
translation-invariant mean-field models. Note from (\ref{pressure free
energy}), (\ref{BCS main theorem 1eq}) and (\ref{Pressure bemol}) that $%
\mathrm{P}_{\Phi }^{\sharp }=\mathrm{P}_{\Phi }^{\flat }=\mathrm{P}_{\Phi }$
for any $\Phi \in \mathcal{W}_{1}\subseteq \mathcal{M}_{1}$. Moreover, for
any purely repulsive or purely attractive mean-field model $\mathfrak{m}\in 
\mathcal{M}_{1}$, as defined in Section \ref{Section purely attrac}, $%
\mathrm{P}_{\mathfrak{m}}^{\sharp }=\mathrm{P}_{\mathfrak{m}}^{\flat }\doteq 
\mathrm{P}_{\mathfrak{m}}$, thanks to \cite[Theorem 2.25]{BruPedra2}.
However, for mean-field models with non-trivial attractive and repulsive
mean-field interactions, one only has 
\begin{equation}
\mathrm{P}_{\mathfrak{m}}^{\sharp }\leq \mathrm{P}_{\mathfrak{m}}^{\flat }\
,\qquad \mathfrak{m}\in \mathcal{M}_{1}\ ,  \label{Pressure bemol2}
\end{equation}%
by virtue of Equation (\ref{bound free bemol diese}). In other words, the
non-conventional pressure is only an upper bound of the conventional one, in
general. In fact, in \cite[Section 2.7]{BruPedra2} we give examples of
models $\mathfrak{m}\in \mathcal{M}_{1}$ for which $\mathrm{P}_{\mathfrak{m}%
}^{\sharp }<\mathrm{P}_{\mathfrak{m}}^{\flat }$.

For general mean-field models, the non-conventional pressure is not equal to
the thermodynamic limit of lattice fermion systems at finite-volume, in
contrast with the conventional one (cf. (\ref{pressure long range})--(\ref%
{BCS main theorem 1eq})). However, in the present paper, we show that the
non-conventional pressure introduced in \cite{BruPedra2} is not only an
interesting mathematical object, but also \emph{physically relevant} in the
context of the Kac limit studied in Section \ref{Short-Range to Long-Range
Models}.

\subsubsection{Thermodynamic Games\label{Section thermo game}}

As stated above, the conventional and non-conventional pressures are given
via infima over translation-invariant states, see Equations (\ref{BCS main
theorem 1eq}) and (\ref{Pressure bemol}). They are pivotal in \cite%
{BruPedra2} for the study of infinite volume equilibrium states.
Nonetheless, it is a priori not clear how useful these variational formulae
are to study phase transitions. To tackle this question, it is convenient to
consider the so-called Bogoliubov approximations of mean-field models, which
refer to the approximating Hamiltonian method used in the past to compute
the conventional pressure associated with particular mean-field models, as
explained in \cite[Section 2.10]{BruPedra2}. In \cite{BruPedra2}, we
generalize this method to all models of the Banach space $\mathcal{M}_{1}$
and all corresponding equilibrium states. We use the view point of game
theory by interpreting the mean-field attractions $\mathfrak{a}_{-}$ and
repulsions $\mathfrak{a}_{+}$ of any model $\mathfrak{m}=(\Phi ,\mathfrak{%
\mathfrak{a}}=\mathfrak{a}_{+}-\mathfrak{a}_{-})\in \mathcal{M}_{1}$ as
attractive and repulsive players, respectively. This leads to a two-person
zero-sum game named (by us) the \emph{thermodynamic game}, which is defined
as follows:

For any mean-field model $\mathfrak{m}=(\Phi ,\mathfrak{\mathfrak{a}})\in 
\mathcal{M}_{1}$ and every function $c=(c_{\Psi })_{\Psi \in \mathbb{S}}\in
L^{2}(\mathbb{S};\mathbb{C};|\mathfrak{\mathfrak{a}}|)$, we define a
so-called approximating (self-adjoint, short-range) interaction by 
\begin{equation}
\Phi _{\mathfrak{m}}(c)\doteq \Phi +2\int_{\mathbb{S}}\mathrm{Re}\left\{ 
\overline{c_{\Psi }}\Psi \right\} \mathfrak{a}\left( \mathrm{d}\Psi \right)
\in \mathcal{W}_{1}^{\mathbb{R}}\ .  \label{approx interaction}
\end{equation}%
The integral in the last definition, which refers to a self-adjoint
interaction, i.e., an element of the space $\mathcal{W}_{1}^{\mathbb{R}}$,
has to be understood as follows: 
\begin{equation}
\left( \int_{\mathbb{S}}\mathrm{Re}\left\{ \overline{c_{\Psi }}\Psi \right\} 
\mathfrak{a}\left( \mathrm{d}\Psi \right) \right) _{\Lambda }\doteq \int_{%
\mathbb{S}}\mathrm{Re}\left\{ \overline{c_{\Psi }}\Psi _{_{\Lambda
}}\right\} \mathfrak{a}\left( \mathrm{d}\Psi \right) \ ,\qquad \Lambda \in 
\mathcal{P}_{\mathrm{f}}\text{ }.  \label{def integral approx interaction}
\end{equation}%
Note that the integral in the definiens is well-defined because, for each $%
\Lambda \in \mathcal{P}_{\mathrm{f}}$, the integrand is an absolutely
integrable (measurable) function taking values in a finite-dimensional
normed space, which is $\mathcal{U}_{\Lambda }$. By Equation (\ref{equation
fininte vol dynam0}), the energy observables associated with $\Phi _{%
\mathfrak{m}}(c)$ equal%
\begin{equation}
U_{L}^{\Phi _{\mathfrak{m}}(c)}=U_{L}^{\Phi }+\int_{\mathbb{S}}2\mathrm{Re}%
\left\{ \overline{c_{\Psi }}U_{L}^{\Psi }\right\} \mathfrak{a}\left( \mathrm{%
d}\Psi \right) \ ,\qquad L\in \mathbb{N}_{0}\ .  \label{approx hamil}
\end{equation}%
One then deduces from Equations (\ref{pressure free energy})--(\ref{map free
energy}) that%
\begin{equation}
\mathrm{P}_{\Phi _{\mathfrak{m}}(c)}=-\inf f_{\Phi _{\mathfrak{m}}(c)}\left(
E_{1}\right) \text{ },\qquad c\in L^{2}(\mathbb{S};\mathbb{C};|\mathfrak{%
\mathfrak{a}}|)\text{ },  \label{variational problem approx}
\end{equation}%
where, for any translation-invariant state $\rho \in E_{1}$,%
\begin{equation*}
f_{\Phi _{\mathfrak{m}}(c)}(\rho )=\int_{\mathbb{S}}2\mathrm{Re}\left\{ 
\overline{c_{\Psi }}e_{\Psi }(\rho )\right\} \mathfrak{a}\left( \mathrm{d}%
\Psi \right) +e_{\Phi }(\rho )-\beta ^{-1}s(\rho )\ .
\end{equation*}%
As compared to the pressure $\mathrm{P}_{\mathfrak{m}}^{\sharp }$ for
translation-invariant mean-field models $\mathfrak{m}\in \mathcal{M}_{1}$, $%
\mathrm{P}_{\Phi _{\mathfrak{m}}(c)}$ is, in principle, easier to analyze,
because it comes from a purely short-range interaction $\Phi _{\mathfrak{m}%
}(c)\in \mathcal{W}_{1}^{\mathbb{R}}$. In many cases that are important for
condensed matter, $\mathrm{P}_{\Phi _{\mathfrak{m}}(c)}$ is even explicitly
known or given by converging perturbative expansions around some simple
(unperturbed) object.

Recall the Jordan decomposition of the measure $\mathfrak{a}$ stated in
Equation (\ref{Hahn decomposition}): $\mathfrak{a=a}_{+}-\mathfrak{a}_{-}$
with $\mathfrak{a}_{\pm }$ being two positive finite measures vanishing on
disjoint Borel subsets $\mathbb{S}_{\mp }\subseteq \mathbb{S}$,
respectively, referring to the Hahn decomposition theorem. Then, we define
two Hilbert spaces corresponding respectively to the repulsive and
attractive components, $\mathfrak{a}_{+}$ and $\mathfrak{a}_{-}$, of any
mean-field model $\mathfrak{m}\in \mathcal{M}_{1}$: 
\begin{equation}
L_{\pm }^{2}(\mathbb{S};\mathbb{C})\doteq L^{2}(\mathbb{S};\mathbb{C};%
\mathfrak{a}_{\pm })\ .  \label{definition of positive-negative L2 space}
\end{equation}%
Note that we canonically have the identification%
\begin{equation*}
L^{2}(\mathbb{S};\mathbb{C};|\mathfrak{a}|)=L_{+}^{2}(\mathbb{S};\mathbb{C}%
)\oplus L_{-}^{2}(\mathbb{S};\mathbb{C})\ .
\end{equation*}%
The approximating free energy density functional%
\begin{equation*}
\mathfrak{f}_{\mathfrak{m}}:L_{-}^{2}(\mathbb{S};\mathbb{C})\times L_{+}^{2}(%
\mathbb{S};\mathbb{C})\rightarrow \mathbb{R}
\end{equation*}%
is then defined by 
\begin{equation}
\mathfrak{f}_{\mathfrak{m}}\left( c_{-},c_{+}\right) \doteq -\left\Vert
c_{+}\right\Vert _{2}^{2}+\left\Vert c_{-}\right\Vert _{2}^{2}-\mathrm{P}%
_{\Phi _{\mathfrak{m}}\left( c_{-}+c_{+}\right) }\ ,\qquad c_{\pm }\in
L_{\pm }^{2}(\mathbb{S};\mathbb{C})\text{ }.
\label{approximating free energy density functional}
\end{equation}%
The \emph{thermodynamic game}\ is the two-person zero-sum game defined from $%
\mathfrak{f}_{\mathfrak{m}}$, with its conservative values being equal to $-%
\mathrm{P}_{\mathfrak{m}}^{\sharp }$ and $-\mathrm{P}_{\mathfrak{m}}^{\flat
} $: By \cite[Theorem 2.36]{BruPedra2}, the conventional and
non-conventional pressures associated with any mean-field model $\mathfrak{m}%
\in \mathcal{M}_{1}$ are equal to 
\begin{equation}
\mathrm{P}_{\mathfrak{m}}^{\sharp }=-\inf_{c_{-}\in L_{-}^{2}(\mathbb{S};%
\mathbb{C})}\sup_{c_{+}\in L_{+}^{2}(\mathbb{S};\mathbb{C})}\mathfrak{f}_{%
\mathfrak{m}}\left( c_{-},c_{+}\right) \qquad \text{and}\qquad \mathrm{P}_{%
\mathfrak{m}}^{\flat }=-\sup_{c_{+}\in L_{+}^{2}(\mathbb{S};\mathbb{C}%
)}\inf_{c_{-}\in L_{-}^{2}(\mathbb{S};\mathbb{C})}\mathfrak{f}_{\mathfrak{m}%
}\left( c_{-},c_{+}\right) \ .  \label{pressures}
\end{equation}%
Compare these equalities with Equations (\ref{BCS main theorem 1eq})-(\ref%
{convex functional g_m}) and (\ref{Pressure bemol}).

Note that the $\sup $ and the $\inf $ in (\ref{pressures}) do not commute in
general. See \cite[p. 42]{BruPedra2}. A sufficient condition for $\sup $ and 
$\inf $ to commute is given through Sion's minimax theorem \cite{HIDETOSHI
KOMIYA} as follows:

\begin{lemma}[Sufficient condition for $\mathrm{P}_{\mathfrak{m}}^{\sharp }=%
\mathrm{P}_{\mathfrak{m}}^{\flat }$]
\label{theorem structure of omega copy(1)}\mbox{ }\newline
Let $\mathfrak{m}\in \mathcal{M}_{1}$ be any mean-field model. If, for any
fixed $c_{+}\in L_{+}^{2}(\mathbb{S};\mathbb{C})$, the function $\mathfrak{f}%
_{\mathfrak{m}}\left( \cdot ,c_{+}\right) $ on $L_{-}^{2}(\mathbb{S};\mathbb{%
C})$ is quasi-convex, i.e., for all $r\in \mathbb{R}$, the level set 
\begin{equation*}
\left\{ c_{-}\in L_{-}^{2}(\mathbb{S};\mathbb{C}):\mathfrak{f}_{\mathfrak{m}%
}\left( c_{-},c_{+}\right) \leq r\right\}
\end{equation*}%
is convex, then $\mathrm{P}_{\mathfrak{m}}^{\sharp }=\mathrm{P}_{\mathfrak{m}%
}^{\flat }$.
\end{lemma}

\begin{proof}
Take $\mathfrak{m}\in \mathcal{M}_{1}$. By \cite[Lemma 8.1]{BruPedra2}, for
any $c_{+}\in L_{+}^{2}(\mathbb{S};\mathbb{C})$, the function $\mathfrak{f}_{%
\mathfrak{m}}\left( \cdot ,c_{+}\right) $ on $L_{-}^{2}(\mathbb{S};\mathbb{C}%
)$ is weakly lower semicontinuous, while, for any $c_{-}\in L_{-}^{2}(%
\mathbb{S};\mathbb{C})$, the function $\mathfrak{f}_{\mathfrak{m}}\left(
c_{-},\cdot \right) $ on $L_{+}^{2}(\mathbb{S};\mathbb{C})$ is weakly upper
semicontinuous and concave. By \cite[Lemmata 8.3-8.4]{BruPedra2}, there is a
closed ball $\mathcal{B}_{R}\left( 0\right) \subseteq L_{-}^{2}(\mathbb{S};%
\mathbb{C})$ of radius $R<\infty $ such that%
\begin{eqnarray*}
\inf_{c_{-}\in L_{-}^{2}(\mathbb{S};\mathbb{C})}\sup_{c_{+}\in L_{+}^{2}(%
\mathbb{S};\mathbb{C})}\mathfrak{f}_{\mathfrak{m}}\left( c_{-},c_{+}\right)
&=&\inf_{c_{-}\in \mathcal{B}_{R}\left( 0\right) }\sup_{c_{+}\in L_{+}^{2}(%
\mathbb{S};\mathbb{C})}\mathfrak{f}_{\mathfrak{m}}\left( c_{-},c_{+}\right)
\ , \\
\sup_{c_{+}\in L_{+}^{2}(\mathbb{S};\mathbb{C})}\inf_{c_{-}\in L_{-}^{2}(%
\mathbb{S};\mathbb{C})}\mathfrak{f}_{\mathfrak{m}}\left( c_{-},c_{+}\right)
&=&\sup_{c_{+}\in L_{+}^{2}(\mathbb{S};\mathbb{C})}\inf_{c_{-}\in \mathcal{B}%
_{R}\left( 0\right) }\mathfrak{f}_{\mathfrak{m}}\left( c_{-},c_{+}\right) \ .
\end{eqnarray*}%
Note that any closed ball of $L_{-}^{2}(\mathbb{S};\mathbb{C})$ is convex
and weakly compact, by the Banach-Alaoglu theorem and the reflexivity of
Hilbert spaces\footnote{%
The weak and the weak$^{\ast }$ topologies are the same in this case.}.
Therefore, if we additionally assume the quasi-convexity of the function $%
\mathfrak{f}_{\mathfrak{m}}\left( \cdot ,c_{+}\right) $ at fixed $c_{+}\in
L_{+}^{2}(\mathbb{S};\mathbb{C})$, then the lemma follows from (\ref%
{pressures}) and Sion's minimax theorem \cite{HIDETOSHI KOMIYA}.
\end{proof}

\subsection{Generalized Equilibrium States\label{Translation Invariant
Equilibrium States copy(1)}}

\subsubsection{Conventional Equilibrium States\label{Translation Invariant
Equilibrium States copy(2)}}

We give here the extension of the notion of equilibrium states of Section %
\ref{Translation Invariant Equilibrium States} to mean-field models, by
using the variational principle associated with the conventional pressure,
as is usual in quantum statistical mechanics at equilibrium. An important
issue appears in this situation, because of the lack of weak$^{\ast }$-lower
semicontinuity of the free energy density functional in presence of
mean-field repulsions (Section \ref{Section purely attrac}).

In fact, similar to (\ref{minimizer short range}), for any
translation-invariant mean-field model $\mathfrak{m}\in \mathcal{M}_{1}$,
one might define the set of equilibrium states by%
\begin{equation}
\mathit{M}_{\mathfrak{m}}\doteq \left\{ \omega \in E_{1}:f_{\mathfrak{m}%
}^{\sharp }\left( \omega \right) =\inf \,f_{\mathfrak{m}}^{\sharp }(E_{1})=-%
\mathrm{P}_{\mathfrak{m}}^{\sharp }\right\} \ .  \label{minimizer long range}
\end{equation}%
Note however that the free energy density functional $f_{\mathfrak{m}%
}^{\sharp }$ is in general not weak$^{\ast }$-lower semicontinuous on $E_{1}$
and it is thus a priori not clear whether $\mathit{M}_{\mathfrak{m}}$ is
empty or not. In fact, by Equation (\ref{Hahn decomposition}), for any
translation-invariant mean-field model $\mathfrak{m}=(\Phi ,\mathfrak{%
\mathfrak{a}})\in \mathcal{M}_{1}$, 
\begin{equation}
f_{\mathfrak{m}}^{\sharp }=\underset{\text{weak}^{\ast }\text{-upper
semicont.}}{\underbrace{\Delta _{\mathfrak{a}_{+}}}}+\underset{\text{weak}%
^{\ast }\text{-lower semicont.}}{\underbrace{\left( -\Delta _{\mathfrak{a}%
_{-}}+e_{\Phi }-\beta ^{-1}s\right) }}\ .  \label{sdsdsdssd}
\end{equation}%
See Sections \ref{energy density1}--\ref{energy density2} and \ref{Section
space averaging}. Therefore, instead of considering $\mathit{M}_{\mathfrak{m}%
}$, we define 
\begin{equation}
\mathit{\Omega }_{\mathfrak{m}}^{\sharp }\doteq \left\{ \omega \in
E_{1}:\exists \{\rho _{n}\}_{n=1}^{\infty }\subseteq E_{1}\mathrm{\ }\text{%
weak}^{\ast }\text{ converging to}\ \omega \text{ such\ that\ }\underset{%
n\rightarrow \infty }{\lim }f_{\mathfrak{m}}^{\sharp }(\rho _{n})=\inf \,f_{%
\mathfrak{m}}^{\sharp }(E_{1})\right\}  \label{definition equilibirum state}
\end{equation}%
as being the (conventional) set of \emph{generalized} equilibrium states of
any fixed translation-invariant mean-field model $\mathfrak{m}\in \mathcal{M}%
_{1}$ (at inverse temperature $\beta \in \mathbb{R}^{+}$). Observe for
instance that, under periodic boundary conditions, the weak$^{\ast }$
accumulation points of (finite-volume) Gibbs states associated with any
mean-field model $\mathfrak{m}\in \mathcal{M}_{1}$ and $\beta \in \mathbb{R}%
^{+}$\ always belong to $\mathit{\Omega }_{\mathfrak{m}}^{\sharp }$, but not
necessarily to $\mathit{M}_{\mathfrak{m}}$, by \cite[Theorem 3.13]{BruPedra2}%
.

Obviously, by weak$^{\ast }$-compactness of $E_{1}$, the set $\mathit{\Omega 
}_{\mathfrak{m}}^{\sharp }$ is non-empty and $\mathit{\Omega }_{\mathfrak{m}%
}^{\sharp }\supseteq \mathit{M}_{\mathfrak{m}}$. This definition can be
expressed in terms of the graph of $f_{\mathfrak{m}}^{\sharp }$: 
\begin{equation*}
\mathit{\Omega }_{\mathfrak{m}}^{\sharp }\times \{\inf \,f_{\mathfrak{m}%
}^{\sharp }(E_{1})\}=\left( E_{1}\times \{\inf \,f_{\mathfrak{m}}^{\sharp
}(E_{1})\}\right) \cap \overline{\mathrm{Graph}(f_{\mathfrak{m}}^{\sharp })}%
\text{ },
\end{equation*}%
where the closure of the graph of $f_{\mathfrak{m}}^{\sharp }$ refers to the
product topology of the weak$^{{\ast }}$ topology on $E_{1}$ and the usual
topology on $\mathbb{R}$. It follows that $\mathit{\Omega }_{\mathfrak{m}%
}^{\sharp }$ is weak$^{{\ast }}$-closed and convex, by affineness of $f_{%
\mathfrak{m}}^{\sharp }$. Thus, $\mathit{\Omega }_{\mathfrak{m}}^{\sharp }$\
is a (non-empty) weak$^{\ast }$-compact convex subset of $E_{1}$. See \cite[%
Lemma 2.16]{BruPedra2}. If $\mathfrak{a}_{+}=0$ then $\mathit{\Omega }_{%
\mathfrak{m}}^{\sharp }=\mathit{M}_{\mathfrak{m}}$ is a (non-empty)\ weak$%
^{\ast }$-closed face of the Poulsen simplex $E_{1}$. By contrast, as
already mentioned above, a mean-field repulsion $\mathfrak{a}_{+}$ has
generally a \emph{geometrical} effect on the set $\mathit{\Omega }_{%
\mathfrak{m}}^{\sharp }$, by possibly breaking its face structure in $E_{1}$%
. This effect can lead to long-range order of generalized equilibrium
states. See \cite[Section 2.9]{BruPedra2}.

\subsubsection{Non-conventional Equilibrium States\label{section
non-conventnioal pressure2}}

In Section \ref{section non-conventnioal pressure} we introduce a
non-conventional pressure by means of a new free energy density functional.
See Equations (\ref{convex functional g_m}) and (\ref{Pressure bemol}). We
show in Section \ref{Short-Range to Long-Range Models} that these
mathematical objects are physically relevant in the context of the Kac
limit. Therefore, at inverse temperature $\beta \in \mathbb{R}^{+}$ and for
any mean-field model $\mathfrak{m}\in \mathcal{M}_{1}$, similar to the set $%
\mathit{\Omega }_{\mathfrak{m}}^{\sharp }$ of generalized equilibrium states
defined by (\ref{definition equilibirum state}), we define a set $\mathit{%
\Omega }_{\mathfrak{m}}^{\flat }$ of non-conventional equilibrium states by 
\begin{equation}
\mathit{\Omega }_{\mathfrak{m}}^{\flat }\doteq \left\{ \omega \in E_{1}:f_{%
\mathfrak{m}}^{\flat }\left( \omega \right) =\inf \,f_{\mathfrak{m}}^{\flat
}(E_{1})=-\mathrm{P}_{\mathfrak{m}}^{\flat }\right\}
\label{definition equilibirum state nonconventional}
\end{equation}%
with $f_{\mathfrak{m}}^{\flat }$ being the free energy density functional (%
\ref{convex functional g_m}). In contrast with the affine functional $f_{%
\mathfrak{m}}^{\sharp }$ (see (\ref{sdsdsdssd})), note that $f_{\mathfrak{m}%
}^{\flat }$ is weak$^{\ast }$-lower semicontinuous but only convex (and not
affine). To prove these properties, use Equation (\ref{convex functional g_m}%
) and Sections \ref{energy density1}--\ref{energy density2} and \ref{Section
space averaging} together with the obvious inequality 
\begin{equation}
\left\Vert \mathfrak{e}_{\Phi }\right\Vert _{\mathcal{U}}\leq \left\Vert
\Phi \right\Vert _{\mathcal{W}_{1}}\ ,\qquad \Phi \in \mathcal{W}_{1}\ ,
\label{ssdsdsdsdsddsdsds}
\end{equation}%
and Lebesgue's dominated convergence theorem to deduce the weak$^{\ast }$%
-continuity of the convex functional 
\begin{equation*}
\rho \mapsto \int_{\mathbb{S}}\left\vert \rho \left( \mathfrak{e}_{\Psi
}\right) \right\vert ^{2}\mathfrak{a}_{+}\left( \mathrm{d}\Psi \right)
\end{equation*}%
on $E_{1}$. Hence, $\mathit{\Omega }_{\mathfrak{m}}^{\flat }$\ is a nonempty
weak$^{\ast }$-compact convex subset of $E_{1}$.

If $\mathfrak{m}=(\Phi ,\mathfrak{\mathfrak{a}}=\mathfrak{a}_{+}-\mathfrak{a}%
_{-})\in \mathcal{M}_{1}$ with $\mathfrak{a}_{+}=0$ then $\mathit{\Omega }_{%
\mathfrak{m}}^{\sharp }=\mathit{\Omega }_{\mathfrak{m}}^{\flat }=\mathit{M}_{%
\mathfrak{m}}$ is a (nonempty)\ weak$^{\ast }$-closed face of the Poulsen
simplex $E_{1}$. In fact, for any purely repulsive or attractive mean-field
model $\mathfrak{m}\in \mathcal{M}_{1}$, as defined in Section \ref{Section
purely attrac}, the equality $\mathit{\Omega }_{\mathfrak{m}}^{\sharp }=%
\mathit{\Omega }_{\mathfrak{m}}^{\flat }$ always holds true, thanks to \cite[%
Theorem 2.25]{BruPedra2}. However, for mean-field models $\mathfrak{m}\in 
\mathcal{M}_{1}$ with non-trivial attractive and repulsive mean-field
interactions, the sets $\mathit{\Omega }_{\mathfrak{m}}^{\sharp }$ and $%
\mathit{\Omega }_{\mathfrak{m}}^{\flat }$ of conventional and
non-conventional equilibrium states are \textbf{not} equal to each other, in
general.

\subsubsection{Self-Consistency of Generalized Equilibrium States \label%
{Section effective theories}}

In Section \ref{Section thermo game} we introduce the thermodynamic game,
which provides an efficient mathematical method to study phase transitions
induced by mean-field interactions. It refers to a two-person zero-sum game
defined from the approximating free energy density functional (\ref%
{approximating free energy density functional}). By Equation (\ref{pressures}%
), the conventional and non-conventional pressures associated with any
mean-field model are, up to a minus sign, the conservative values of this
game. The thermodynamic game also provides a complete characterization of
the sets (\ref{definition equilibirum state}) and (\ref{definition
equilibirum state nonconventional}) of (translation-invariant) conventional
and non-conventional equilibrium states, as follows:

The $\sup $ and the $\inf $ in the variational problems (\ref{pressures})
are attained, i.e., they are respectively a $\max $ and a $\min $. For any
mean-field model $\mathfrak{m}\in \mathcal{M}_{1}$, the sets 
\begin{eqnarray}
\mathcal{C}_{\mathfrak{m}}^{\sharp } &\doteq &\left\{ d_{-}\in L_{-}^{2}(%
\mathbb{S};\mathbb{C}):\max_{c_{+}\in L_{+}^{2}(\mathbb{S};\mathbb{C})}%
\mathfrak{f}_{\mathfrak{m}}\left( d_{-},c_{+}\right) =-\mathrm{P}_{\mathfrak{%
m}}^{\sharp }\right\}  \label{eq conserve strategy} \\
\mathcal{C}_{\mathfrak{m}}^{\flat } &\doteq &\left\{ d_{+}\in L_{+}^{2}(%
\mathbb{S};\mathbb{C}):\min_{c_{-}\in L_{-}^{2}(\mathbb{S};\mathbb{C})}%
\mathfrak{f}_{\mathfrak{m}}\left( c_{-},d_{+}\right) =-\mathrm{P}_{\mathfrak{%
m}}^{\flat }\right\}  \label{eq conserve strategy2}
\end{eqnarray}%
of conservative strategies of the repulsive and attractive players,
respectively, are non-empty. By \cite[Lemma 8.4]{BruPedra2}, $\mathcal{C}_{%
\mathfrak{m}}^{\sharp }$ is norm-bounded and weakly compact, while the set $%
\mathcal{C}_{\mathfrak{m}}^{\flat }$ has exactly one element $d_{+}$ when $%
\mathfrak{m}=(\Phi ,\mathfrak{\mathfrak{a}}=\mathfrak{a}_{+}-\mathfrak{a}%
_{-})\in \mathcal{M}_{1}$ with $\mathfrak{a}_{+}\neq 0$. In the particular
case of purely repulsive mean-field models, i.e., when $\mathfrak{a}_{-}=0$, 
$\mathcal{C}_{\mathfrak{m}}^{\sharp }=\{0\}=L_{-}^{2}(\mathbb{S};\mathbb{C})$%
, as $\mathfrak{f}_{\mathfrak{m}}$ is independent of $c_{-}$. Similarly, if $%
\mathfrak{a}_{+}=0$ then $\mathcal{C}_{\mathfrak{m}}^{\flat
}=\{0\}=L_{+}^{2}(\mathbb{S};\mathbb{C})$.

In \cite[Lemma 8.3 ($\sharp $)]{BruPedra2} it is proven that, when $%
\mathfrak{a}_{+}\neq 0$, for all functions $c_{-}\in L_{-}^{2}(\mathbb{S};%
\mathbb{C})$, the set 
\begin{equation}
\left\{ d_{+}\in L_{+}^{2}(\mathbb{S};\mathbb{C}):\max_{c_{+}\in L_{+}^{2}(%
\mathbb{S};\mathbb{C})}\mathfrak{f}_{\mathfrak{m}}\left( c_{-},c_{+}\right) =%
\mathfrak{f}_{\mathfrak{m}}\left( c_{-},d_{+}\right) \right\}
\label{eq conserve strategybis}
\end{equation}%
has exactly one element, which is denoted by $\mathrm{r}_{+}(c_{-})$. By 
\cite[Lemma 8.8]{BruPedra2}, if $\mathfrak{a}_{+}\neq 0$ then the mapping%
\begin{equation}
\mathrm{r}_{+}:c_{-}\mapsto \mathrm{r}_{+}\left( c_{-}\right)
\label{thermodyn decision rule}
\end{equation}%
defines a continuous functional from $L_{-}^{2}(\mathbb{S};\mathbb{C})$ to $%
L_{+}^{2}(\mathbb{S};\mathbb{C})$, where $L_{-}^{2}(\mathbb{S};\mathbb{C})$
and $L_{+}^{2}(\mathbb{S};\mathbb{C})$ are endowed with the weak and norm
topologies, respectively. This mapping is called (by us) the thermodynamic
decision rule\ of the mean-field model $\mathfrak{m}\in \mathcal{M}_{1}$. In
the particular case of purely attractive mean-field models, i.e., when $%
\mathfrak{a}_{+}=0$, $\mathfrak{f}_{\mathfrak{m}}$ is independent of $c_{+}$
and one trivially has $\mathrm{r}_{+}=0$, since $L_{+}^{2}(\mathbb{S};%
\mathbb{C})=\{0\}$ in this case.

For any mean-field model $\mathfrak{m}=(\Phi ,\mathfrak{\mathfrak{a}}=%
\mathfrak{a}_{+}-\mathfrak{a}_{-})\in \mathcal{M}_{1}$, it is convenient to
introduce a family of approximating, purely attractive mean-field models by%
\begin{equation}
\mathfrak{m}\left( c_{+}\right) \doteq (\Phi +\Phi _{\mathfrak{m}}\left(
c_{+}\right) ,-\mathfrak{a}_{-})\in \mathcal{M}_{1}\ ,\qquad c_{+}\in
L_{+}^{2}(\mathbb{S};\mathbb{C})\ .  \label{m approche 2}
\end{equation}%
Then, for every pair of functions $c_{\pm }\in L_{\pm }^{2}(\mathbb{S};%
\mathbb{C})$, we define the (possibly empty) sets 
\begin{equation}
\mathit{\Omega }_{\mathfrak{m}}\left( c_{-},c_{+}\right) \doteq \left\{
\omega \in \mathit{M}_{\Phi _{\mathfrak{m}}(c_{+}+c_{-})}:e_{(\cdot )}\left(
\omega \right) =c_{+}+c_{-}\quad \left\vert \mathfrak{a}\right\vert \text{%
--a.e.}\right\} \subseteq E_{1}  \label{subset of a face}
\end{equation}%
and 
\begin{equation}
\mathit{\Omega }_{\mathfrak{m}}\left( c_{+}\right) \doteq \left\{ \omega \in 
\mathit{\Omega }_{\mathfrak{m}\left( c_{+}\right) }^{\sharp }:e_{(\cdot
)}\left( \omega \right) =c_{+}\quad \mathfrak{a}_{+}\text{--a.e.}\right\}
\subseteq E_{1}\ ,  \label{subset of a face2}
\end{equation}%
where, for any fixed translation-invariant state $\rho \in E_{1}$, the
continuous and bounded mapping $e_{(\cdot )}\left( \rho \right) :\mathbb{%
S\rightarrow C}$ is defined from (\ref{ssssssssss})--(\ref{eq:enpersite}) by%
\begin{equation}
e_{\Psi }\left( \rho \right) \doteq \rho \left( \mathfrak{e}_{\Psi }\right)
\ ,\qquad \Psi \in \mathbb{S}\ ,  \label{ssssssssssssssssssss}
\end{equation}%
while $\mathit{M}_{\Phi _{\mathfrak{m}}(c)}$ and $\mathit{\Omega }_{%
\mathfrak{m}\left( c_{+}\right) }^{\sharp }$ are the sets respectively
defined by (\ref{minimizer short range}) and (\ref{definition equilibirum
state}). Note that $\mathit{M}_{\Phi _{\mathfrak{m}}(c)}$ and $\mathit{%
\Omega }_{\mathfrak{m}\left( c_{+}\right) }^{\sharp }$ are (non-empty) weak$%
^{\ast }$-closed faces of $E_{1}$, since $\mathfrak{m}\left( c_{+}\right) $
is a purely attractive mean-field model. Then, we obtain a (static)
self-consistency condition\ for (generalized) equilibrium states, which
refers, in a sense, to Euler-Lagrange equations for the variational problem
defining the thermodynamic game. More precisely, we have the following
statements:

\begin{theorem}[Self-consistency of equilibrium states]
\label{theorem structure of omega}\mbox{ }\newline
Let $\mathfrak{m}\in \mathcal{M}_{1}$ be any translation-invariant
mean-field model. \newline
\emph{(}$\sharp $\emph{-i)} 
\begin{equation*}
\mathit{\Omega }_{\mathfrak{m}}^{\sharp }=\overline{\mathrm{co}}\left( 
\underset{d_{-}\in \mathcal{C}_{\mathfrak{m}}^{\sharp }}{\cup }\mathit{%
\Omega }_{\mathfrak{m}}\left( d_{-},\mathrm{r}_{+}(d_{-})\right) \right) \ .
\end{equation*}%
\emph{(}$\sharp $\emph{-ii)} The set $\mathcal{E}(\mathit{\Omega }_{%
\mathfrak{m}}^{\sharp })$ of extreme points of the weak$^{\ast }$-compact
convex set $\mathit{\Omega }_{\mathfrak{m}}^{\sharp }$ is included in the
union of the sets%
\begin{equation*}
\mathcal{E}\left( \mathit{\Omega }_{\mathfrak{m}}\left( d_{-},\mathrm{r}%
_{+}(d_{-})\right) \right) \text{ },\qquad d_{-}\in \mathcal{C}_{\mathfrak{m}%
}^{\sharp }\text{ },
\end{equation*}%
of all extreme points of $\mathit{\Omega }_{\mathfrak{m}}\left( d_{-},%
\mathrm{r}_{+}(d_{-})\right) $, $d_{-}\in \mathcal{C}_{\mathfrak{m}}^{\sharp
}$, which are non-empty, convex, mutually disjoint, weak$^{\ast }$-closed
subsets of $E_{1}$.\newline
\emph{(}$\flat $\emph{)} 
\begin{equation*}
\mathcal{C}_{\mathfrak{m}}^{\flat }=\left\{ d_{+}\right\} \qquad \text{and}%
\qquad \mathit{\Omega }_{\mathfrak{m}}^{\flat }=\mathit{\Omega }_{\mathfrak{m%
}}\left( d_{+}\right) \ .
\end{equation*}
\end{theorem}

\begin{proof}
Assertion ($\sharp $-i) results from \cite[Theorem 2.21 (i)]{BruPedra2} and 
\cite[Theorem 2.39 (i)]{BruPedra2}, while ($\sharp $-ii) corresponds to \cite%
[Theorem 2.39 (ii)]{BruPedra2}. It remains to prove Assertion ($\flat $). As
already mentioned, the fact that $\mathcal{C}_{\mathfrak{m}}^{\flat
}=\{d_{+}\}$ refers to \cite[Lemma 8.4]{BruPedra2}. However, the identity $%
\mathit{\Omega }_{\mathfrak{m}}^{\flat }=\mathit{\Omega }_{\mathfrak{m}%
}\left( d_{+}\right) $ was not considered in \cite{BruPedra2}, but its proof
is similar to the one of \cite[Lemma 9.2]{BruPedra2}: Fix $\mathfrak{m}\in 
\mathcal{M}_{1}$. The set $\mathit{\Omega }_{\mathfrak{m}}\left(
d_{+}\right) \cap \mathit{\Omega }_{\mathfrak{m}}^{\flat }$ is non-empty, by 
\cite[Lemma 8.5]{BruPedra2}. So, take some $\omega \in \mathit{\Omega }_{%
\mathfrak{m}}\left( d_{+}\right) \cap \mathit{\Omega }_{\mathfrak{m}}^{\flat
}$. In particular, the equilibrium state $\omega $ satisfies the
self-consistency condition 
\begin{equation}
e_{(\cdot )}\left( \omega \right) \doteq \omega \left( \mathfrak{e}_{(\cdot
)}\right) =d_{+}\quad \mathfrak{a}_{+}\text{--a.e.}
\label{gap equation repulsif}
\end{equation}%
We now observe that, for any $\rho \in E_{1}$, 
\begin{equation}
2\int_{\mathbb{S}}\mathrm{Re}\left\{ \overline{d_{+,\Psi }}\rho \left( 
\mathfrak{e}_{\Psi }\right) \right\} \mathfrak{a}_{+}\left( \mathrm{d}\Psi
\right) =\int_{\mathbb{S}}\left\vert \rho \left( \mathfrak{e}_{\Psi }\right)
\right\vert ^{2}\mathfrak{a}_{+}\left( \mathrm{d}\Psi \right) +\left\Vert
d_{+}\right\Vert _{2}^{2}-\int_{\mathbb{S}}\left\vert \rho \left( \mathfrak{e%
}_{\Psi }\right) -d_{+,\Psi }\right\vert ^{2}\mathfrak{a}_{+}\left( \mathrm{d%
}\Psi \right)  \label{equality super bisbis}
\end{equation}%
and since $\omega \in \mathit{\Omega }_{\mathfrak{m}}\left( d_{+}\right)
\cap \mathit{\Omega }_{\mathfrak{m}}^{\flat }$ satisfies (\ref{gap equation
repulsif}), we conclude that 
\begin{eqnarray}
&&\inf\limits_{\rho \in E_{1}}\left\{ 2\int_{\mathbb{S}}\mathrm{Re}\left\{ 
\overline{d_{+,\Psi }}\rho \left( \mathfrak{e}_{\Psi }\right) \right\} 
\mathfrak{a}_{+}\left( \mathrm{d}\Psi \right) -\Delta _{\mathfrak{a}%
_{-}}\left( \rho \right) +e_{\Phi }(\rho )-\beta ^{-1}s(\rho )\right\}
\label{equality super bis} \\
&=&\int_{\mathbb{S}}\left\vert \omega \left( \mathfrak{e}_{\Psi }\right)
\right\vert ^{2}\mathfrak{a}_{+}\left( \mathrm{d}\Psi \right) -\Delta _{%
\mathfrak{a}_{-}}\left( \omega \right) +e_{\Phi }(\omega )-\beta
^{-1}s(\omega )+\left\Vert d_{+}\right\Vert _{2}^{2}  \notag \\
&=&f_{\mathfrak{m}}^{\flat }(\omega )+\left\Vert d_{+}\right\Vert _{2}^{2} 
\notag \\
&=&\inf f_{\mathfrak{m}}^{\flat }(E_{1})+\left\Vert d_{+}\right\Vert
_{2}^{2}\ .  \label{equality super bisequality super bis}
\end{eqnarray}%
Going backwards from (\ref{equality super bisequality super bis}) to (\ref%
{equality super bis}) and using then (\ref{equality super bisbis}), for any
non-conventional equilibrium state $\omega \in \mathit{\Omega }_{\mathfrak{m}%
}^{\flat }$, we obtain the inequality 
\begin{equation*}
f_{\mathfrak{m}}^{\flat }(\omega )+\left\Vert d_{+}\right\Vert _{2}^{2}\leq
f_{\mathfrak{m}}^{\flat }(\omega )-\int_{\mathbb{S}}\left\vert \omega \left( 
\mathfrak{e}_{\Psi }\right) -d_{+,\Psi }\right\vert ^{2}\mathfrak{a}%
_{+}\left( \mathrm{d}\Psi \right) +\left\Vert d_{+}\right\Vert _{2}^{2}\ ,
\end{equation*}%
i.e., 
\begin{equation*}
\int_{\mathbb{S}}\left\vert \omega \left( \mathfrak{e}_{\Psi }\right)
-d_{+,\Psi }\right\vert ^{2}\mathfrak{a}_{+}\left( \mathrm{d}\Psi \right)
\leq 0\ .
\end{equation*}%
As a consequence, any non-conventional equilibrium state $\omega \in \mathit{%
\Omega }_{\mathfrak{m}}^{\flat }$ satisfies the self-consistency condition (%
\ref{gap equation repulsif}) with $d_{+}\in \mathcal{C}_{\mathfrak{m}%
}^{\flat }$. Combining this with (\ref{equality super bisbis})--(\ref%
{equality super bisequality super bis}), it follows that $\mathit{\Omega }_{%
\mathfrak{m}}^{\flat }\subseteq \mathit{\Omega }_{\mathfrak{m}}\left(
d_{+}\right) $. Conversely, take any $\omega \in \mathit{\Omega }_{\mathfrak{%
m}}\left( d_{+}\right) $ with $d_{a,+}\in \mathcal{C}_{\mathfrak{m}}^{\flat
} $. Such a state $\omega \in \mathit{\Omega }_{\mathfrak{m}}\left(
d_{+}\right) $ is a solution to the variational problem (\ref{equality super
bis}) and we easily deduce that $\omega \in \mathit{\Omega }_{\mathfrak{m}%
}^{\flat }$, again from (\ref{equality super bis})--(\ref{equality super
bisequality super bis}).
\end{proof}

Theorem \ref{theorem structure of omega} implies in particular that, for any
extreme state $\hat{\omega}\in \mathcal{E}(\mathit{\Omega }_{\mathfrak{m}%
}^{\sharp })$ of $\mathit{\Omega }_{\mathfrak{m}}^{\sharp }$, there is a
unique $d_{-}\in \mathcal{C}_{\mathfrak{m}}^{\sharp }$ such that 
\begin{equation}
d\doteq d_{-}+\mathrm{r}_{+}(d_{-})=e_{(\cdot )}(\hat{\omega})\ .
\label{gap equations}
\end{equation}%
In the physics literature on superconductors, the above equality refers to
the so-called gap equations. Conversely, for any $d_{-}\in \mathcal{C}_{%
\mathfrak{m}}^{\sharp }$, there is some generalized equilibrium state $%
\omega $ satisfying the condition above, but $\omega $ is not necessarily an
extreme point of $\mathit{\Omega }_{\mathfrak{m}}^{\sharp }$.

To conclude, note that Theorem \ref{theorem structure of omega} yields the
equality $\mathit{\Omega }_{\mathfrak{m}}^{\sharp }=\mathit{\Omega }_{%
\mathfrak{m}}^{\flat }$ for any purely repulsive or purely attractive
mean-field model $\mathfrak{m}\in \mathcal{M}_{1}$, as defined in Section %
\ref{Section purely attrac}. However, for mean-field models $\mathfrak{m}\in 
\mathcal{M}_{1}$ with both non-trivial attractive and repulsive mean-field
interactions, there is no reason to have this equality, in general.

\section{From Short-Range to Mean-Field Models\label{Short-Range to
Long-Range Models}}

\subsection{Banach Spaces of Reflection-Symmetric Functions\label{sect
banach refl inv functions}}

Recall that the dimension of the (cubic) lattice $\mathfrak{L}\doteq \mathbb{%
Z}^{d}$ is denoted by $d\in \mathbb{N}$ and is always fixed. In \cite[%
Equation (2.8)]{Lieb-Kac}, the function $f:\mathbb{R}^{d}\rightarrow \mathbb{%
R}$ used to introduce the Kac limit as in (\ref{SR}) is reflection-symmetric
and has to decay like $\left\vert x\right\vert ^{-(d+\varepsilon )}$ in the
limit $\left\vert x\right\vert \rightarrow \infty $, for some strictly
positive parameter $\varepsilon \in \mathbb{R}^{+}$, being bounded and
continuous at $x=0$, cf. \cite[Conditions $(\varphi 1)$ and $(\varphi 2)$]%
{Lieb-Kac}. In the sequel, we use a similar, albeit stronger, condition. In
fact, we do not only impose the decay of $f$ itself, but also of its
derivatives up to order $2d$, at least.

Given the parameters $\varepsilon \in \mathbb{R}^{+}$ and $\kappa \in
\left\{ 2d,2d+1,\ldots ,\infty \right\} $, we consider the real space%
\begin{eqnarray}
\mathfrak{D}_{\varepsilon ,\kappa } &\doteq &%
%TCIMACRO{\TeXButton{\Big\{}{\Big\{}}%
%BeginExpansion
\Big\{%
%EndExpansion
f\in C^{\kappa }\left( \mathbb{R}^{d},\mathbb{R}\right) :\forall \ell \in 
\mathbb{N}_{0}^{d},\ \left\vert \ell \right\vert \leq \kappa ,\quad
\lim_{\left\vert x\right\vert \rightarrow \infty }\left\vert x\right\vert
^{d+\varepsilon +|\ell |}\partial _{x}^{\ell }f\left( x\right) =0,\text{ } 
\notag \\
&&\qquad \qquad \qquad \qquad \qquad \qquad \qquad \text{and}\qquad \forall
x\in \mathbb{R}^{d},\quad f\left( -x\right) =f\left( x\right) 
%TCIMACRO{\TeXButton{\Big\}}{\Big\}} }%
%BeginExpansion
\Big\}
%EndExpansion
\label{D fracepsilon}
\end{eqnarray}%
of continuously $\kappa $-differentiable, reflection-symmetric, real-valued
functions on $\mathbb{R}^{d}$ that decay faster than $\left\vert
x\right\vert ^{-(d+\varepsilon )}$ as $\left\vert x\right\vert \rightarrow
\infty $. We use above the standard multi-index notation\footnote{$\partial
_{x_{l}}^{0}\doteq \mathbf{1}$ for $l=1,\ldots ,d$.}: 
\begin{equation}
\ell !\doteq \ell _{1}!\ldots \ell _{d}!\ ,\qquad \left\vert \ell
\right\vert \doteq \ell _{1}+\cdots +\ell _{d},\qquad \partial _{x}^{\ell
}\doteq \partial _{x_{1}}^{\ell _{1}}\cdots \partial _{x_{d}}^{\ell _{d}}\ ,
\label{multi-index notation}
\end{equation}%
for any $\ell =\left( \ell _{1},\ldots ,\ell _{d}\right) \in \mathbb{N}%
_{0}^{d}$. Using the norm 
\begin{equation}
\left\Vert f\right\Vert _{\mathfrak{D}_{\varepsilon ,\kappa }}\doteq
\sum_{\ell \in \mathbb{N}_{0}^{d}:\left\vert \ell \right\vert \leq \kappa }%
\frac{1}{\ell !}\sup_{x\in \mathbb{R}^{d}}\left\vert \left( 1+\left\vert
x\right\vert \right) ^{d+\varepsilon +|\ell |}\partial _{x}^{\ell }f\left(
x\right) \right\vert \ ,  \label{norm definition function kac}
\end{equation}%
the space $\mathfrak{D}_{\varepsilon ,\kappa }$ becomes a separable real
Banach space: The fact that the norm $\left\Vert \cdot \right\Vert _{%
\mathfrak{D}_{\varepsilon ,\kappa }}$ is complete follows from the
closedness of differential operators. In order to prove the separability of
the Banach space $\mathfrak{D}_{\varepsilon ,\kappa }$, take a positive,
compactly supported and smooth real-valued function $g$ on $\mathbb{R}^{d}$
with $\left\Vert g\right\Vert _{1}=1$, and define a second function $h$ of
this type by 
\begin{equation*}
h\left( x_{1},\ldots ,x_{d}\right) \doteq \int_{\left[ -1/2,1/2\right]
^{d}}g\left( x_{1}-y_{1},\ldots ,x_{d}-y_{d}\right) \mathrm{d}^{d}y\ ,\qquad
x_{1},\ldots ,x_{d}\in \mathbb{R}\ .
\end{equation*}%
For any $n\in \mathbb{N}$, $z\in \mathbb{Z}^{d}$ and all multi-indices $\ell
\in \mathbb{N}_{0}^{d}$ , we then define the function 
\begin{equation*}
h_{n,z,\ell }\left( x\right) \doteq \left( z+x\right) ^{\ell }h\left(
2^{n}\left( z+x\right) \right) +\left( z-x\right) ^{\ell }h\left(
2^{n}\left( z-x\right) \right) \ ,\qquad x\in \mathbb{R}^{d}\ .
\end{equation*}%
The set of all rational linear combinations of the subset $\{h_{n,z,\ell
}\}_{n\in \mathbb{N},z\in \mathbb{Z}^{d},\ell \in \mathbb{N}_{0}^{d}}$ is
dense in the Banach space $\mathfrak{D}_{\varepsilon ,\kappa }$. Here, for
any $x=(x_{1},\ldots ,x_{d})\in \mathbb{Z}^{d}$ and $\ell \in \mathbb{N}%
_{0}^{d}$, as is usual, $x^{\ell }$ stands for the product $x_{1}^{\ell
_{1}}\cdots x_{d}^{\ell _{d}}$.

Unless it is not explicitly mentioned, from now on, we omit in the notation
the parameter $\kappa $, which is taken \emph{by default} as $\kappa =2d$.
I.e., $\mathfrak{D}_{\varepsilon }\equiv \mathfrak{D}_{\varepsilon ,2d}$ if
the parameter $\kappa $\ is not explicitly specified. In fact, a parameter $%
\kappa >2d$ is only relevant in Section \ref{General Case}. For the moment,
note only that $\kappa =2d$ corresponds to the biggest space in the scale $(%
\mathfrak{D}_{\varepsilon ,\kappa })_{\kappa =2d,2d+1,\ldots }$ of vector
spaces: For any $\kappa _{1},\kappa _{2}=2d,2d+1,\ldots $ with $\kappa
_{2}\geq \kappa _{1}$, $\mathfrak{D}_{\varepsilon ,\kappa _{2}}$ is a vector
subspace of $\mathfrak{D}_{\varepsilon ,\kappa _{1}}$ and 
\begin{equation*}
\left\Vert f\right\Vert _{\mathfrak{D}_{\varepsilon ,\kappa _{1}}}\leq
\left\Vert f\right\Vert _{\mathfrak{D}_{\varepsilon ,\kappa _{2}}}\ ,\qquad
f\in \mathfrak{D}_{\varepsilon ,\kappa _{1}}\ ,
\end{equation*}%
by Equation (\ref{norm definition function kac}).

A closed convex cone in $\mathfrak{D}_{\varepsilon }\equiv \mathfrak{D}%
_{\varepsilon ,2d}$ for any $\varepsilon \in \mathbb{R}^{+}$ is given by the
subset $\mathfrak{D}_{\varepsilon ,+}\subseteq \mathfrak{D}_{\varepsilon }$
of all positive definite functions, i.e., of all those functions whose
Fourier transforms are everywhere non-negative. Note that any function $f\in 
\mathfrak{D}_{\varepsilon }$ has a well-defined Fourier transform, which is
a function $\mathbb{R}^{d}\rightarrow \mathbb{R}$ denoted by 
\begin{equation}
F\left( f\right) \left( k\right) \equiv \hat{f}\left( k\right) \doteq \int_{%
\mathbb{R}^{d}}f\left( x\right) \mathrm{e}^{-ik\cdot x}\mathrm{d}^{d}x\
,\qquad k\in \mathbb{R}^{d}\ .  \label{fourier transform}
\end{equation}%
The notation $F(f)$ for the Fourier transform of $f$ is only used to make
some expressions simpler, $\hat{f}$ being the default notation. Recall that
the Fourier transform of any reflection-symmetric real-valued function is
again a reflection-symmetric real-valued function.

In order to study approximations of mean-field attractions via Kac
interactions, we also need the following property of functions: We say that
a function $g:\mathbb{R}^{d}\rightarrow \mathbb{R}$ is \emph{scaling-monotone%
} if%
\begin{equation}
g\left( \gamma ^{-1}k\right) \leq g\left( k\right) \ ,\qquad k\in \mathbb{R}%
^{d},\ \gamma \in (0,1)\ .  \label{secaling monotone}
\end{equation}%
In fact, if $g$ is differentiable, then this condition is equivalent to 
\begin{equation*}
\nabla g\left( k\right) \cdot k\leq 0\ ,\qquad k\in \mathbb{R}^{d}\ .
\end{equation*}%
Using this definition, for any $\varepsilon \in \mathbb{R}^{+}$, we
introduce the closed convex cones 
\begin{equation*}
\mathfrak{C}_{\varepsilon ,+}\doteq \left\{ f\in \mathfrak{D}_{\varepsilon
,+}:\hat{f}\text{ is scaling-monotone}\right\} \ ,\qquad \varepsilon \in 
\mathbb{R}^{+}\ .
\end{equation*}%
Examples of functions in the above cone are given as follows:\ 

\begin{itemize}
\item The Yukawa-type potential $f_{1}$ defined by its Fourier transform 
\begin{equation*}
\hat{f}_{1}\left( k\right) \doteq \frac{c_{0}}{\left\vert k\right\vert
^{2}+c_{1}}\mathrm{e}^{-c_{2}\left\vert k\right\vert ^{2}}\ ,\qquad k\in 
\mathbb{R}^{d}\ ,
\end{equation*}%
for constants $c_{0},c_{1},c_{2}\in \mathbb{R}^{+}$, belongs to $\mathfrak{C}%
_{\varepsilon ,+}$ for all $\varepsilon \in \mathbb{R}^{+}$. Note that\ $%
c_{2}=0$ corresponds to the usual Yukawa potential, whereas the case $%
c_{2}\in \mathbb{R}^{+}$ refers to a regularization at short distances.

\item Take any finite positive Borel measure $\mu $ on $(\mathbb{R}^{+})^{d}$
such that, for some $\eta \in \mathbb{R}^{+}$, 
\begin{equation*}
\mu (\left\{ \left( s_{1},\ldots ,s_{d}\right) :\exists j\in \left\{
1,\ldots ,d\right\} ,\quad s_{j}<\eta \right\} )=0
\end{equation*}%
and define the function $f_{2}$ by%
\begin{equation*}
f_{2}\left( x\right) \doteq \int_{(\mathbb{R}^{+})^{d}}\mathrm{e}%
^{-(s_{1}x_{1}^{2}+\cdots +s_{d}x_{d}^{2})}\mu \left( \mathrm{d}s\right) \
,\qquad x=\left( x_{1},\ldots ,x_{d}\right) \in \mathbb{R}^{d}\ .
\end{equation*}%
Then, $f_{2}\in \mathfrak{C}_{\varepsilon ,+}$ for all strictly positive
parameters $\varepsilon \in \mathbb{R}^{+}$.
\end{itemize}

In the sequel, we use the notation 
\begin{equation}
\mathfrak{D}_{0}\doteq \bigcup_{\varepsilon \in \mathbb{R}^{+}}\mathfrak{D}%
_{\varepsilon }\ ,\qquad \mathfrak{D}_{0,+}\doteq \bigcup_{\varepsilon \in 
\mathbb{R}^{+}}\mathfrak{D}_{\varepsilon ,+}\qquad \text{and}\qquad 
\mathfrak{C}_{0,+}\doteq \bigcup_{\varepsilon \in \mathbb{R}^{+}}\mathfrak{C}%
_{\varepsilon ,+}\ .  \label{D frac0}
\end{equation}

\subsection{Kac Interactions}

\subsubsection{Simple-Field Case}

The link between interactions (in the sense of Section \ref{Section Banach
space interaction copy(1)}) and mean-field models is explicitly established
in the sequel by using the so-called Kac limit. Like, for instance, in \cite%
{Lieb-Kac}, a Kac model is a short-range model that depends upon parameters
that fix the range of some of its interaction terms. The Kac limit then
refers to taking these ranges to infinity, \emph{after} the thermodynamic
limit. By contrast, in the usual mean-field setting described in Section \ref%
{Long-Range Models}, both the range of interaction terms and the size of the
system are taken simultaneously to infinity, in the thermodynamic limit.

The study presented here uses arguments that are of abstract nature and,
hence, not model-dependent. We define a general notion of \emph{Kac
interactions} as follows:

\begin{defn}[Kac interactions -- simple-field case]
\label{definition Kac interaction}\mbox{ }\newline
For any parameter $\gamma \in (0,1)$, we define\footnote{%
In the definition of Kac interactions and functions, we have a factor of $2$
as compared to Lieb's and Penrose's papers: Their sum is basically over
pairs of enumerated particles $i,j$ such that $i<j$, while our definition
applied to their special case would refer to a full sum over $i,j$.} the Kac
function $\mathcal{K}_{\gamma }$ to be the mapping from $\mathcal{W}%
_{1}\times \mathfrak{D}_{0}$ to $\mathcal{W}_{1}^{\mathbb{R}}$ defined, for
any $\Phi \in \mathcal{W}_{1}$ and $f\in \mathfrak{D}_{0}$, by 
\begin{equation*}
\mathcal{K}_{\gamma }\left( \Phi ,f\right) _{\Lambda }\doteq \sum_{\mathcal{Z%
}_{1},\mathcal{Z}_{2}\in \mathcal{P}_{\mathrm{f}}:\mathcal{Z}_{1}\cup 
\mathcal{Z}_{2}=\Lambda }\frac{\left\vert \mathcal{Z}_{1}\cup \mathcal{Z}%
_{2}\right\vert }{\left\vert \mathcal{Z}_{1}\right\vert +\left\vert \mathcal{%
Z}_{2}\right\vert }\Phi _{\mathcal{Z}_{1}}^{\ast }\Phi _{\mathcal{Z}%
_{2}}\sum_{x\in \mathcal{Z}_{1},y\in \mathcal{Z}_{2}}\frac{\gamma
^{d}f\left( \gamma \left( x-y\right) \right) }{\left\vert \mathcal{Z}%
_{1}\right\vert \left\vert \mathcal{Z}_{2}\right\vert },\qquad \Lambda \in 
\mathcal{P}_{\mathrm{f}}\ .
\end{equation*}%
This interaction is named the Kac interaction associated with $\Phi $ and $f$%
.
\end{defn}

\noindent The fact that $\mathcal{K}_{\gamma }$ maps\ elements of $\mathcal{W%
}_{1}\times \mathfrak{D}_{0}$ to the real Banach space $\mathcal{W}_{1}^{%
\mathbb{R}}$ is proven in Lemma \ref{lemma Kac norm} (i), which additionally
shows that it is a locally Lipschitz continuous function. Definition \ref%
{definition Kac interaction} can be extended to all interactions $\Phi \in 
\mathcal{V\supseteq W}_{1}$, real-valued functions $f$ on $\mathbb{R}^{d}$
and $\gamma \in \mathbb{R}$. This generalization is not considered here
since we focus on the limit $\gamma \rightarrow 0^{+}$ for Kac interactions
in the Banach space $\mathcal{W}_{1}$.

Definition \ref{definition Kac interaction} includes, mutatis mutandis, much
more general models than \cite{Lieb-Kac}, where, translated to the lattice
fermion case, only the following particular example is considered: Using the
finite-range interaction $\mathbf{N}\in \mathcal{W}_{0}^{\mathbb{R}%
}\subseteq \mathcal{W}_{1}^{\mathbb{R}}$ defined by 
\begin{equation}
\mathbf{N}_{\Lambda }\doteq \left\{ 
\begin{array}{ll}
\sum_{\mathrm{s}\in \mathrm{S}}a_{x,\mathrm{s}}^{\ast }a_{x,\mathrm{s}} & 
\text{if }\Lambda =\left\{ x\right\} \text{ for }x\in \mathfrak{L} \\ 
0 & \text{otherwise}%
\end{array}%
\right. ,\qquad \Lambda \in \mathcal{P}_{\mathrm{f}}\ ,
\label{N interaction0}
\end{equation}%
for any $f\in \mathfrak{D}_{0}$ and $\gamma \in (0,1)$, the associated Kac
interaction is defined\footnote{%
Here, $\delta _{x,y}$ denote the Kronecker delta.} by 
\begin{equation}
\mathcal{K}_{\gamma }\left( \mathbf{N},f\right) _{\Lambda }\doteq 2\sum_{%
\mathrm{s},\mathrm{t}\in \mathrm{S}}\left( 1-\frac{3}{4}\delta _{x,y}\right)
\gamma ^{d}f\left( \gamma \left( x-y\right) \right) a_{y,\mathrm{t}}^{\ast
}a_{y,\mathrm{t}}a_{x,\mathrm{s}}^{\ast }a_{x,\mathrm{s}}
\label{N interaction1}
\end{equation}%
whenever $\Lambda =\left\{ x,y\right\} $ for $x,y\in \mathfrak{L}$, and $%
\mathcal{K}_{\gamma }\left( \mathbf{N},f\right) _{\Lambda }\doteq 0$
otherwise. Therefore, for any $\Phi \in \mathcal{V}^{\mathbb{R}}$, the
finite-volume Hamiltonian (\ref{equation fininte vol dynam0}) associated
with the interaction $\Phi +\mathcal{K}_{\gamma }\left( \mathbf{N},f\right) $
is equal in this case to 
\begin{equation}
U_{L}^{\Phi +\mathcal{K}_{\gamma }\left( \mathbf{N},f\right) }\doteq
U_{L}^{\Phi }-\frac{\gamma ^{d}f\left( 0\right) }{2}\sum_{x\in \Lambda _{L},%
\mathrm{s}\in \mathrm{S}}a_{x,\mathrm{s}}^{\ast }a_{x,\mathrm{s}%
}+\sum_{x,y\in \Lambda _{L},\mathrm{s},\mathrm{t}\in \mathrm{S}}\gamma
^{d}f\left( \gamma \left( x-y\right) \right) a_{y,\mathrm{t}}^{\ast }a_{y,%
\mathrm{t}}a_{x,\mathrm{s}}^{\ast }a_{x,\mathrm{s}}\ ,\quad L\in \mathbb{N}%
_{0}\text{ }.  \label{N interaction2}
\end{equation}%
In other words, the function $f\in \mathfrak{D}_{0}$ is interpreted as a
pair potential whose range\footnote{%
Take for instance a compactly supported function $f\in \mathfrak{D}_{0}$.
See, more generally, Equation (\ref{norm definition function kac}).} is
tuned by the parameter $\gamma \in (0,1)$.

Another interesting example is given by BCS-type models of
superconductivity. They refer to the following Kac interactions: Let the
spin set $\mathrm{S}=\{\uparrow ,\downarrow \}$. Consider the finite-range
interaction $\mathbf{C}\in \mathcal{W}_{0}^{\mathbb{R}}\subseteq \mathcal{W}%
_{1}^{\mathbb{R}}$ defined by 
\begin{equation}
\mathbf{C}_{\Lambda }\doteq \left\{ 
\begin{array}{ll}
a_{x,\uparrow }a_{x,\downarrow } & \text{if }\Lambda =\left\{ x\right\} 
\text{ for }x\in \mathfrak{L} \\ 
0 & \text{otherwise}%
\end{array}%
\right. ,\qquad \Lambda \in \mathcal{P}_{\mathrm{f}}\ .
\label{BCS interaction}
\end{equation}%
For any $f\in \mathfrak{D}_{0}$ and $\gamma \in (0,1)$, the associated Kac
interaction is defined by 
\begin{equation*}
\mathcal{K}_{\gamma }\left( \mathbf{C},-f\right) _{\Lambda }\doteq \gamma
^{d}f\left( \gamma \left( x-y\right) \right) \left( \frac{3}{4}\delta
_{x,y}-1\right) \left( a_{y,\downarrow }^{\ast }a_{y,\uparrow }^{\ast
}a_{x,\uparrow }a_{x,\downarrow }+a_{x,\downarrow }^{\ast }a_{x,\uparrow
}^{\ast }a_{y,\uparrow }a_{y,\downarrow }\right)
\end{equation*}%
whenever $\Lambda =\left\{ x,y\right\} $ for $x,y\in \mathfrak{L}$, and $%
\mathcal{K}_{\gamma }\left( \mathbf{C},-f\right) _{\Lambda }\doteq 0$
otherwise. Therefore, for any $\Phi \in \mathcal{V}^{\mathbb{R}}$, the
finite-volume Hamiltonian (\ref{equation fininte vol dynam0}) associated
with the interaction $\Phi +\mathcal{K}_{\gamma }\left( \mathbf{C},-f\right) 
$ is equal in this case to 
\begin{equation}
U_{L}^{\Phi +\mathcal{K}_{\gamma }\left( \mathbf{C},-f\right) }\doteq
U_{L}^{\Phi }+\frac{\gamma ^{d}f\left( 0\right) }{2}\sum_{x\in \Lambda
_{L}}a_{x,\downarrow }^{\ast }a_{x,\uparrow }^{\ast }a_{x,\uparrow
}a_{x,\downarrow }-\sum_{x,y\in \Lambda _{L}}\gamma ^{d}f\left( \gamma
\left( x-y\right) \right) a_{y,\downarrow }^{\ast }a_{y,\uparrow }^{\ast
}a_{x,\uparrow }a_{x,\downarrow }\ ,\quad L\in \mathbb{N}_{0}\text{ }.
\label{BCS interaction1}
\end{equation}%
The function $f\in \mathfrak{D}_{0}$ encodes the hopping strength of Cooper
pairs. This model thus implements a BCS interaction whose range is tuned by
the parameter $\gamma \in (0,1)$.

\begin{remark}
\label{remark idiote}\mbox{ }\newline
The second term of the right-hand side of Equation (\ref{N interaction2})
can be absorbed in the short-range component $\Phi $. Additionally, it tends
to zero like $\gamma ^{d}$ in the sense of $\mathcal{W}_{1}$, as $\gamma
\rightarrow 0^{+}$, and it is therefore irrelevant for the pressure in the
Kac limit. The same is true for the second term of the right-hand side of
Equation (\ref{BCS interaction1}).
\end{remark}

\subsubsection{Multiple-Field Case\label{Multiple-Field Case}}

In order to approximate models with possibly infinitely many long-range
components, we generalize Definition \ref{definition Kac interaction}. To
this end, recall that $\mathbb{S}$ is the unit sphere of the Banach space $%
\mathcal{W}_{1}$ (\ref{W1}) of translation-invariant short-range
interactions, see Equation (\ref{unit sphere}). $\mathcal{S}_{1}$ is the
(real) Banach space of signed Borel measures of bounded variation on $%
\mathbb{S}$. Similar to this construction, for any $\varepsilon \in \mathbb{R%
}^{+}$, we introduce the function sets%
\begin{equation*}
\mathbb{D}_{\varepsilon }\doteq \left\{ f\in \mathfrak{D}_{\varepsilon
}:\left\Vert f\right\Vert _{\mathfrak{D}_{\varepsilon }}=1\right\} \ ,\qquad 
\mathbb{D}_{\varepsilon ,+}\doteq \mathbb{D}_{\varepsilon }\cap \mathfrak{D}%
_{\varepsilon ,+}\qquad \text{and}\qquad \mathbb{C}_{\varepsilon ,+}\doteq 
\mathbb{D}_{\varepsilon }\cap \mathfrak{C}_{\varepsilon ,+}
\end{equation*}%
and consider the (real) Banach space of Borel measures of bounded variation
on $\mathbb{S}\times \mathbb{D}_{\varepsilon }$, $\mathbb{S}\times \mathbb{D}%
_{\varepsilon ,+}$ and $\mathbb{S}\times \mathbb{C}_{\varepsilon ,+}$
denoted by $\mathcal{D}_{\varepsilon }$, $\mathcal{D}_{\varepsilon
,+}\subseteq \mathcal{D}_{\varepsilon }$ and $\mathcal{C}_{\varepsilon
,+}\subseteq \mathcal{D}_{\varepsilon }$, respectively. Here, we identify
each element of $\mathcal{D}_{\varepsilon ,+}$ and $\mathcal{C}_{\varepsilon
,+}$ with an element of $\mathcal{D}_{\varepsilon }$ whose support lies in $%
\mathbb{S}\times \mathbb{D}_{\varepsilon ,+}$ and $\mathbb{S}\times \mathbb{C%
}_{\varepsilon ,+}$, respectively. Similar to (\ref{D frac0}), let 
\begin{equation}
\mathcal{D}_{0}\doteq \bigcup_{\varepsilon \in \mathbb{R}^{+}}\mathcal{D}%
_{\varepsilon }\ ,\qquad \mathcal{D}_{0,+}\doteq \bigcup_{\varepsilon \in 
\mathbb{R}^{+}}\mathcal{D}_{\varepsilon ,+}\qquad \text{and}\qquad \mathcal{C%
}_{0,+}\doteq \bigcup_{\varepsilon \in \mathbb{R}^{+}}\mathcal{C}%
_{\varepsilon ,+}\ .  \label{D0}
\end{equation}

Recall that $\mathfrak{D}_{\varepsilon }\equiv \mathfrak{D}_{\varepsilon
,\kappa }$ still depends upon the parameter $\kappa =2d,2d+1,\ldots $, which
is by default set to $2d$. See Section \ref{sect banach refl inv functions}.
Hence, all the objects $\mathbb{D}_{\varepsilon }$, $\mathbb{D}_{\varepsilon
,+}$, $\mathbb{C}_{\varepsilon ,+}$, $\mathcal{D}_{\varepsilon }$, $\mathcal{%
D}_{\varepsilon ,+}$, $\mathcal{C}_{\varepsilon ,+}$, $\mathcal{D}_{0}$, $%
\mathcal{D}_{0,+}$ and $\mathcal{C}_{0,+}$ depend upon $\kappa $. As
explained in Section \ref{sect banach refl inv functions}, this parameter is
omitted in the notation, unless we have to refer to it explicitly. For any $%
\varepsilon \in \mathbb{R}^{+}$ and $\kappa _{1},\kappa _{2}=2d,2d+1,\ldots $
with $\kappa _{2}\geq \kappa _{1}$, we canonically identify $\mathcal{D}%
_{\varepsilon ,\kappa _{2}}$with a subspace of $\mathcal{D}_{\varepsilon
,\kappa _{1}}$: Let $\Xi $ denote the continuous mapping 
\begin{equation*}
(\Psi ,f)\mapsto (\Psi ,\left\Vert f\right\Vert _{\mathfrak{D}_{\varepsilon
,\kappa _{1}}}^{-1}f)
\end{equation*}%
from $\mathbb{S}\times \mathbb{D}_{\varepsilon ,\kappa _{2}}$ to $\mathbb{S}%
\times \mathbb{D}_{\varepsilon ,\kappa _{1}}$ and let the continuous,
bounded, positive valued function $\mathfrak{d}:\mathbb{S}\times \mathbb{D}%
_{\varepsilon ,\kappa _{2}}\rightarrow \mathbb{R}_{0}^{+}$ be defined by $%
\mathfrak{d}(\Psi ,f)\doteq \left\Vert f\right\Vert _{\mathfrak{D}%
_{\varepsilon ,\kappa _{1}}}$. Then any $\mathfrak{b}\in \mathcal{D}%
_{\varepsilon ,\kappa _{2}}$ is identified with the pushforward 
\begin{equation*}
\Xi _{\ast }(\mathfrak{db})\in \mathcal{D}_{\varepsilon ,\kappa _{1}}
\end{equation*}%
of $\mathfrak{db}$ under the (Borel measurable) mapping $\Xi $, where $%
\mathfrak{db}\in \mathcal{D}_{\varepsilon ,\kappa _{2}}$ is the Borel
measure of finite variation defined by%
\begin{equation*}
\mathfrak{db}(\mathcal{A})\doteq \int_{\mathcal{A}}\mathfrak{d}(\Psi ,f)%
\mathfrak{b}(\mathrm{d}(\Psi ,f))
\end{equation*}%
for any Borel set $\mathcal{A}\subseteq \mathbb{S}\times \mathbb{D}%
_{\varepsilon ,\kappa _{2}}$. Canonical vector space inclusions $\mathcal{D}%
_{0,\kappa _{2}}\subseteq \mathcal{D}_{0,\kappa _{1}}$, $\mathcal{D}%
_{0,\kappa _{2},+}\subseteq \mathcal{D}_{0,\kappa _{1},+}$, and $\mathcal{C}%
_{0,\kappa _{2},+}\subseteq \mathcal{C}_{0,\kappa _{1},+}$ are defined in a
similar manner.

Recall that the Kac function $\mathcal{K}_{\gamma }$ is the mapping from $%
\mathcal{W}_{1}\times \mathfrak{D}_{0}$ to $\mathcal{W}_{1}^{\mathbb{R}}$ of
Definition \ref{definition Kac interaction} for any fixed $\gamma \in (0,1)$%
. By Lemma \ref{lemma Kac norm} (i), it is locally Lipschitz continuous.
Therefore, we can extend the definition of Kac interactions by replacing
simple interactions with Borel measures in $\mathcal{D}_{0}$:

\begin{defn}[Kac interactions -- multiple-field case]
\label{definition Kac interaction copy(1)}\mbox{ }\newline
For any $\varepsilon \in \mathbb{R}^{+}$, $\mathfrak{b}\in \mathcal{D}%
_{\varepsilon }$ and $\gamma \in (0,1)$, we define the corresponding Kac
interaction to be%
\begin{equation*}
\Phi ^{\mathfrak{b},\gamma }\doteq \int_{\mathbb{S}\times \mathbb{D}%
_{\varepsilon }}\mathcal{K}_{\gamma }\left( \Psi ,f\right) \mathfrak{b}%
\left( \mathrm{d}\left( \Psi ,f\right) \right) \in \mathcal{W}_{1}^{\mathbb{R%
}}\ .
\end{equation*}
\end{defn}

\noindent Thanks to Lemma \ref{lemma Kac norm} (i), the above integral is a
Bochner integral because the measure $\mathfrak{b}$ has bounded variation
and $\mathcal{W}_{1}$ and $\mathfrak{D}_{\varepsilon }$ are both separable
Banach spaces. See, e.g., \cite[Chapter III, Theorems 1.1 and 1.2]{pettis}.

Note that the mappings $\mathfrak{b\mapsto }\Phi ^{\mathfrak{b},\gamma }$, $%
\gamma \in (0,1)$, from $\mathcal{D}_{\varepsilon }$\ to $\mathcal{W}_{1}^{%
\mathbb{R}}$, are (real) linear. With the canonically identification of $%
\mathcal{D}_{\varepsilon ,\kappa _{2}}$ with a subspace of $\mathcal{D}%
_{\varepsilon ,\kappa _{1}}$ for $\kappa _{2}\geq \kappa _{1}$, as explained
above, observe also that for all $\mathfrak{b}\in \mathcal{D}_{\varepsilon
,\kappa _{2}}$ and $\gamma \in (0,1)$, one has%
\begin{equation*}
\int_{\mathbb{S}\times \mathbb{D}_{\varepsilon ,\kappa _{1}}}\mathcal{K}%
_{\gamma }\left( \Psi ,f\right) \mathfrak{b}\left( \mathrm{d}\left( \Psi
,f\right) \right) =\int_{\mathbb{S}\times \mathbb{D}_{\varepsilon ,\kappa
_{2}}}\mathcal{K}_{\gamma }\left( \Psi ,f\right) \mathfrak{b}\left( \mathrm{d%
}\left( \Psi ,f\right) \right) \text{ },
\end{equation*}%
i.e., in the above definition of $\Phi ^{\mathfrak{b},\gamma }\in \mathcal{W}%
_{1}^{\mathbb{R}}$ it does not matter on which space $\mathbb{S}\times 
\mathbb{D}_{\varepsilon ,\kappa }$, $\kappa \in 2d,2d+1,\ldots $, one sees $%
\mathfrak{b}$ as a measure.

Recall also that the subset $\mathcal{W}_{0}$ of finite-range interactions
defined by (\ref{W0}) is dense in the Banach space $\mathcal{W}_{1}$. We
have a similar property for Kac interactions, in the following sense:\ 

\begin{lemma}[Approximation by finite-range interactions]
\label{definition Kac interaction copy(2)}\mbox{ }\newline
For any $\varepsilon \in \mathbb{R}^{+}$, $\mathfrak{b}\in \mathcal{D}%
_{\varepsilon }$ and $\eta \in \mathbb{R}^{+}$, there is a finite sequence $%
\left( \Psi _{1},f_{1}\right) ,\ldots ,\left( \Psi _{n},f_{n}\right) $ in $%
\mathbb{S}\times \mathbb{D}_{\varepsilon }$ such that, for any $\gamma \in
\left( 0,1\right) $, 
\begin{equation*}
\left\{ \Psi _{1},\ldots ,\Psi _{n}\right\} \subseteq \mathcal{W}_{0}\qquad 
\text{and}\qquad \left\Vert \Phi ^{\mathfrak{b},\gamma }-\Phi ^{\mathfrak{b}%
_{\eta },\gamma }\right\Vert _{\mathcal{W}_{1}}\leq \eta \ ,
\end{equation*}%
where $\mathfrak{b}_{\eta }\in \mathcal{D}_{\varepsilon }$ is the finite sum
of Dirac measures $\delta _{\left( \Psi ,f\right) }$ on $\mathbb{S}\times 
\mathbb{D}_{\varepsilon }$: 
\begin{equation*}
\mathfrak{b}_{\eta }\doteq \sum_{j=1}^{n}\delta _{\left( \Psi
_{j},f_{j}\right) }\ .
\end{equation*}%
If $\mathfrak{b}\in \mathcal{C}_{\varepsilon ,+}$ then the sequence can be
chosen such that $f_{1},\ldots ,f_{n}\in \mathfrak{C}_{\varepsilon ,+}$.
\end{lemma}

\begin{proof}
Fix $\varepsilon \in \mathbb{R}^{+}$, $\mathfrak{b}\in \mathcal{D}%
_{\varepsilon }$ and $\eta \in \mathbb{R}^{+}$. Again, $\mathcal{W}_{1}$ and 
$\mathfrak{D}_{\varepsilon }$ are both separable Banach spaces. Then, for
any $\vartheta \in \mathbb{R}^{+}$, there is a step function $\mathbf{1}%
_{\vartheta }$ from $\mathbb{S}\times \mathbb{D}_{\varepsilon }$ to itself
such that 
\begin{equation*}
\int_{\mathbb{S}\times \mathbb{D}_{\varepsilon }}\left\Vert \mathbf{1}%
_{\vartheta }\left( \Psi ,f\right) -\left( \Psi ,f\right) \right\Vert _{%
\mathcal{W}_{1}\times \mathfrak{D}_{\varepsilon }}\mathfrak{b}\left( \mathrm{%
d}\left( \Psi ,f\right) \right) \leq \vartheta \ ,
\end{equation*}%
thanks to \cite[Chapter III, Theorems 1.1 and 1.2]{pettis}. Because of the
density of $\mathcal{W}_{0}\subseteq \mathcal{W}_{1}$, we can assume that 
\begin{equation*}
\mathbf{1}_{\vartheta }\left( \mathbb{S}\times \mathbb{D}_{\varepsilon
}\right) \subseteq \mathcal{W}_{0}\times \mathbb{D}_{\varepsilon }\ .
\end{equation*}%
We then define $\mathfrak{b}_{\eta }$ to be the pushforward of $\mathfrak{b}$
through $\mathbf{1}_{\vartheta }$. By Lemma \ref{lemma Kac norm} (i), 
\begin{equation*}
\left\Vert \Phi ^{\mathfrak{b},\gamma }-\Phi ^{\mathfrak{b}_{\eta },\gamma
}\right\Vert _{\mathcal{W}_{1}}\leq \eta \ ,\qquad \gamma \in \left(
0,1\right) \ ,
\end{equation*}%
for sufficiently small $\vartheta \in \mathbb{R}^{+}$. The proof for the
special case $\mathfrak{b}\in \mathcal{C}_{\varepsilon ,+}$ only needs an
obvious adaptation of the above arguments.
\end{proof}

\subsection{Kac Limit of Energy Densities\label{Kac Limit of Energy
Densities}}

\subsubsection{Simple-Field Case}

Recall that the energy density functional on translation-invariant states is
the mapping $e_{\Psi }:E_{1}\rightarrow \mathbb{C}$, defined by (\ref%
{ssssssssss}) for each translation-invariant interaction $\Psi \in \mathcal{W%
}_{1}$, that is, 
\begin{equation}
e_{\Psi }\left( \rho \right) =\rho \left( \mathfrak{e}_{\Psi }\right) \qquad 
\text{with}\qquad \mathfrak{e}_{\Psi }\doteq \sum\limits_{\mathcal{Z}\in 
\mathcal{P}_{\mathrm{f}},\;\mathcal{Z}\ni 0}\frac{\Psi _{\mathcal{Z}}}{%
\left\vert \mathcal{Z}\right\vert }\in \mathcal{U}  \label{equation utilise}
\end{equation}%
for any $\Psi \in \mathcal{W}_{1}$ and $\rho \in E_{1}$. Note that any
translation-invariant interaction$\ \Phi \in \mathcal{W}_{1}$ yields a
self-adjoint and translation-invariant Kac interaction $\mathcal{K}_{\gamma
}\left( \Phi ,f\right) \in \mathcal{W}_{1}^{\mathbb{R}}$ for each $f\in 
\mathfrak{D}_{0}$ and $\gamma \in (0,1)$, see Definition \ref{definition Kac
interaction}. In particular, Equation (\ref{equation utilise}) with $\Psi =%
\mathcal{K}_{\gamma }\left( \Phi ,f\right) $ gives the energy density of the
Kac interaction in the simple-field case. In the Kac limit $\gamma
\rightarrow 0^{+}$, we obtain a multiple of the space-averaging functional $%
\Delta _{\mathfrak{e}_{\Phi }}:E_{1}\rightarrow \mathbb{R}$ defined on
translation-invariant states by (\ref{space averaging}) for $A=\mathfrak{e}%
_{\Phi }$:\ 

\begin{theorem}[Energy density in the Kac limit]
\label{propuni copy(1)}\mbox{ }\newline
For any interaction $\Phi \in \mathcal{W}_{1}$ and function$\ f\in \mathfrak{%
D}_{0}$, the family $(e_{\mathcal{K}_{\gamma }\left( \Phi ,f\right)
})_{\gamma \in (0,1)}$ of energy density functionals converges pointwise to
the functional $\hat{f}(0)\Delta _{\mathfrak{e}_{\Phi }}$ in the Kac limit,
i.e., 
\begin{equation*}
\lim_{\gamma \rightarrow 0^{+}}e_{\mathcal{K}_{\gamma }\left( \Phi ,f\right)
}\left( \rho \right) =\hat{f}\left( 0\right) \Delta _{\mathfrak{e}_{\Phi
}}\left( \rho \right) \ ,\qquad \rho \in E_{1}\ ,
\end{equation*}%
where $\hat{f}$ denotes the Fourier\ transform of $f$.
\end{theorem}

\begin{proof}
Fix $\Phi \in \mathcal{W}_{1}$ and $f\in \mathfrak{D}_{0}$. The proof is
done in several steps:\medskip

\noindent \underline{Step 1:} The element $\mathfrak{e}_{\Psi }$ given by
Equation (\ref{equation utilise}) for the Kac interaction $\Psi =\mathcal{K}%
_{\gamma }\left( \Phi ,f\right) \in \mathcal{W}_{1}^{\mathbb{R}}$ of
Definition \ref{definition Kac interaction} satisfies the equality%
\begin{eqnarray}
\mathfrak{e}_{\mathcal{K}_{\gamma }\left( \Phi ,f\right) }+\mathfrak{R}%
_{\Phi }^{f,\gamma } &=&\sum_{\mathcal{Z}_{1},\mathcal{Z}_{2}\in \mathcal{P}%
_{\mathrm{f}}:\mathcal{Z}_{1}\cap \mathcal{Z}_{2}\supseteq \left\{ 0\right\}
}\left( \frac{\left\vert \mathcal{Z}_{1}\right\vert }{\left\vert \mathcal{Z}%
_{1}\right\vert +\left\vert \mathcal{Z}_{2}\right\vert }\frac{\Phi _{%
\mathcal{Z}_{1}}^{\ast }}{\left\vert \mathcal{Z}_{1}\right\vert }\sum_{z\in 
\mathfrak{L}}\frac{\alpha _{z}\left( \Phi _{\mathcal{Z}_{2}}\right) }{%
\left\vert \mathcal{Z}_{2}\right\vert }\sum_{x\in \mathcal{Z}_{1},y\in 
\mathcal{Z}_{2}}\frac{\gamma ^{d}f\left( \gamma \left( x-y\right) -\gamma
z\right) }{\left\vert \mathcal{Z}_{1}\right\vert \left\vert \mathcal{Z}%
_{2}\right\vert }\right.  \notag \\
&&+\left. \frac{\left\vert \mathcal{Z}_{2}\right\vert }{\left\vert \mathcal{Z%
}_{1}\right\vert +\left\vert \mathcal{Z}_{2}\right\vert }\sum_{z\in 
\mathfrak{L}}\frac{\alpha _{z}\left( \Phi _{\mathcal{Z}_{1}}^{\ast }\right) 
}{\left\vert \mathcal{Z}_{1}\right\vert }\frac{\Phi _{\mathcal{Z}_{2}}}{%
\left\vert \mathcal{Z}_{2}\right\vert }\sum_{x\in \mathcal{Z}_{1},y\in 
\mathcal{Z}_{2}}\frac{\gamma ^{d}f\left( \gamma \left( x-y\right) +\gamma
z\right) }{\left\vert \mathcal{Z}_{1}\right\vert \left\vert \mathcal{Z}%
_{2}\right\vert }\right) \text{ },  \label{sdsdssd}
\end{eqnarray}%
where $\{\alpha _{x}\}_{x\in \mathbb{Z}^{d}}$ is the set of (translation) $%
\ast $-automorphisms of $\mathcal{U}$ defined by (\ref{transl}) and%
\begin{equation}
\mathfrak{R}_{\Phi }^{f,\gamma }\doteq \sum_{\mathcal{Z}_{1},\mathcal{Z}%
_{2}\in \mathcal{P}_{\mathrm{f}}:\mathcal{Z}_{1}\cap \mathcal{Z}%
_{2}\supseteq \left\{ 0\right\} }\frac{\Phi _{\mathcal{Z}_{1}}^{\ast }\Phi _{%
\mathcal{Z}_{2}}}{\left\vert \mathcal{Z}_{1}\right\vert +\left\vert \mathcal{%
Z}_{2}\right\vert }\sum_{x\in \mathcal{Z}_{1},y\in \mathcal{Z}_{2}}\frac{%
\gamma ^{d}f\left( \gamma \left( x-y\right) \right) }{\left\vert \mathcal{Z}%
_{1}\right\vert \left\vert \mathcal{Z}_{2}\right\vert }\ .
\label{sdasdasdad}
\end{equation}%
Equation (\ref{sdsdssd}) results from arguments that are very similar to the
ones used to get (\ref{fefeffefef}). See also (\ref{ti interaction}). Note
from Lemma \ref{propuni} and the triangle inequality that 
\begin{equation}
\Vert \mathfrak{R}_{\Phi }^{f,\gamma }\Vert _{\mathcal{U}}\leq \mathfrak{F}%
_{\gamma }\left( \Phi ,f\right) \qquad \text{and thus,}\qquad \lim_{\gamma
\rightarrow 0^{+}}\Vert \mathfrak{R}_{\Phi }^{f,\gamma }\Vert _{\mathcal{U}%
}=0\ .  \label{sdasdasdadbis}
\end{equation}%
We therefore only need to study the right-hand side of (\ref{sdsdssd}) more
precisely. This is done by using a cyclic representation of
translation-invariant states to have the spectral theorem at our disposal.
\medskip

\noindent \underline{Step 2:} Any state $\rho \in E$ induces a cyclic
representation of the $C^{\ast }$-algebra $\mathcal{U}$ \cite[Theorem 2.3.16]%
{BrattelliRobinsonI}: For any $\rho \in E$, there exist a Hilbert space $%
\mathcal{H}_{\rho }$, a representation $\pi _{\rho }:\mathcal{U}\rightarrow 
\mathcal{B}(\mathcal{H}_{\rho })$ of $\mathcal{U}$ on $\mathcal{H}_{\rho }$,
and a norm-one vector $\Omega _{\rho }\in \mathcal{H}_{\rho }$, which is
cyclic with respect to $\pi _{\rho }(\mathcal{U})$, such that 
\begin{equation*}
\rho (A)=\left\langle \Omega _{\rho },\pi _{\rho }\left( A\right) \Omega
_{\rho }\right\rangle _{\mathcal{H}_{\rho }}\ ,\qquad A\in \mathcal{U}\ .
\end{equation*}%
Recall that any state $\rho $ is translation-invariant, i.e., $\rho \in
E_{1} $, iff $\rho \circ \alpha _{x}=\rho $ for any $x\in \mathbb{Z}^{d}$.
See Equation (\ref{periodic invariant states}). Since the mapping $x\mapsto
\alpha _{x}$ from $\left( \mathbb{Z}^{d},+\right) $ to the group of $\ast $%
--automorphisms of $\mathcal{U}$ is a group homomorphism, if $\rho $ is
translation-invariant, then there is a uniquely defined family $%
\{U_{x}\}_{x\in \mathbb{Z}^{d}}$ of unitary operators in $\mathcal{B}(%
\mathcal{H}_{\rho })$ with invariant vector $\Omega _{\rho }$, i.e., $\Omega
_{\rho }=U_{x}\Omega _{\rho }$ for any $x\in \mathbb{Z}^{d}$, and such that 
\begin{equation*}
\pi _{\rho }(\alpha _{x}(A))=U_{x}\pi _{\rho }(A)U_{x}^{\ast }\ ,\qquad A\in 
\mathcal{U},\ x\in \mathbb{Z}^{d}\ .
\end{equation*}%
In particular, $U_{x+y}=U_{x}U_{y}$ for any $x,y\in \mathbb{Z}^{d}$ and $%
U_{x}^{\ast }=U_{-x}$. See for instance \cite[Corollary 2.3.17]%
{BrattelliRobinsonI}. Since $\left( \mathbb{Z}^{d},+\right) $ is abelian,
the normal operators 
\begin{equation*}
U_{(1,0,\ldots ,0)},U_{(0,1,\ldots ,0)},\ldots ,U_{(0,0,\ldots ,1)}\in 
\mathcal{B}(\mathcal{H}_{\rho })
\end{equation*}%
commute with each other and their joint spectrum is contained in the $d$%
-dimensional torus%
\begin{equation*}
\mathrm{T}_{d}\doteq \left\{ \left( z_{1},\ldots ,z_{d}\right) \in \mathbb{C}%
^{d}:\,\left\vert z_{i}\right\vert =1,\,i=1,\ldots ,d\right\} \ .
\end{equation*}%
The spectral theorem\footnote{%
For any family $\{A_{j},A_{j}^{\ast }\}_{j\in J}$ of commuting bounded
operators on some Hilbert space, the real and imaginary parts of these
operators commute, i.e., $\{\func{Re}\{A_{j}\},\func{Im}\{A_{j}\}\}_{j\in J}$
is also a commuting family. The spectral theorem for such general commuting
families $\{A_{j},A_{j}^{\ast }\}_{j\in J}$ of (normal) operators directly
follows from the one for (commuting) self-adjoint operators, as in \cite[%
Chap. 6, Theorem 2 and compare with Sect. 5]{spectral thm}.} ensures the
existence of a projection-valued measure $\mathrm{P}$ on the Borel $\sigma $%
-algebra of $\mathrm{T}_{d}$ (with respect to the usual Euclidean metric for 
$\mathbb{C}^{d}$), such that%
\begin{equation}
U_{z}=\int_{\mathrm{T}_{d}}x_{1}^{z_{1}}\cdots x_{d}^{z_{d}}\mathrm{d}%
\mathrm{P}\left( x\right) =\int_{\Theta _{d}}\mathrm{e}^{i\theta \cdot z}%
\mathrm{d}\left( \mathrm{F}_{\ast }\mathrm{P}\right) \left( \theta \right) \
,\qquad z\in \mathbb{Z}^{d}\ ,  \label{ssdsssd}
\end{equation}%
where $\Theta _{d}\doteq \lbrack -\pi ,\pi )^{d}$ and $\mathrm{F}_{\ast }%
\mathrm{P}$ is the pushforward of $\mathrm{P}$ under the (Borel measurable)
mapping $\mathrm{F}:\mathrm{T}_{d}\rightarrow \Theta _{d}$ defined by%
\begin{equation}
\mathrm{F}(x_{1},\dots ,x_{d})\doteq \left( \func{arg}\left( x_{1}\right)
,\dots ,\func{arg}\left( x_{d}\right) \right) \ ,\qquad \left( x_{1},\dots
,x_{d}\right) \in \mathrm{T}_{d}\ .  \label{sdsdsdsdsdssd}
\end{equation}%
By the measurable functional calculus, the mapping 
\begin{equation*}
f\mapsto \int_{\Theta _{d}}f(\theta )\mathrm{d}(\mathrm{F}_{\ast }\mathrm{P}%
)\left( \theta \right)
\end{equation*}%
is a $\ast $-homomorphism from the (commutative) $C^{\ast }$-algebra of
bounded Borel-measurable functions on $\Theta _{d}$ to $\mathcal{B}(\mathcal{%
H}_{\rho })$. In particular, for any pair $f,g$ of bounded Borel-measurable
functions on $\Theta _{d}$, one has%
\begin{equation}
\int_{\Theta _{d}}f(\theta )g(\theta )\mathrm{d}(\mathrm{F}_{\ast }\mathrm{P}%
)\left( \theta \right) =\int_{\Theta _{d}}f(\theta )\mathrm{d}(\mathrm{F}%
_{\ast }\mathrm{P})\left( \theta \right) \int_{\Theta _{d}}g(\theta )\mathrm{%
d}(\mathrm{F}_{\ast }\mathrm{P})\left( \theta \right) \ .  \label{sdsdsdsd}
\end{equation}%
See, e.g., \cite[Section 3 of Chapter 5, in particular Equation (5) and
Theorem 1]{spectral thm}.\medskip

\noindent \underline{Step 3:} Using the reflection-symmetry of $f$ and a
cyclic representation (Step 2), for any translation-invariant state $\rho
\in E_{1}$, we infer from Equations (\ref{sdsdssd})--(\ref{sdasdasdad}) that%
\begin{equation}
\rho \left( \mathfrak{e}_{\mathcal{K}_{\gamma }\left( \Phi ,f\right)
}\right) +\rho (\mathfrak{R}_{\Phi }^{f,\gamma })=\sum_{\mathcal{Z}_{1},%
\mathcal{Z}_{2}\in \mathcal{P}_{\mathrm{f}}:\mathcal{Z}_{1}\cap \mathcal{Z}%
_{2}\supseteq \left\{ 0\right\} }\left\langle \frac{\pi _{\rho }\left( \Phi
_{\mathcal{Z}_{1}}\right) }{\left\vert \mathcal{Z}_{1}\right\vert }\Omega
_{\rho },\sum_{x\in \mathcal{Z}_{1},y\in \mathcal{Z}_{2}}\frac{B_{\gamma
}\left( y-x\right) }{\left\vert \mathcal{Z}_{1}\right\vert \left\vert 
\mathcal{Z}_{2}\right\vert }\frac{\pi _{\rho }\left( \Phi _{\mathcal{Z}%
_{2}}\right) }{\left\vert \mathcal{Z}_{2}\right\vert }\Omega _{\rho
}\right\rangle _{\mathcal{H}_{\rho }},  \label{sdasdasdad2}
\end{equation}%
where, for any $a\in \mathfrak{L}$ and $\gamma \in (0,1)$, we define the
operator%
\begin{equation}
B_{\gamma }\left( a\right) \doteq \sum_{z\in \mathfrak{L}}\gamma ^{d}f\left(
\gamma a+\gamma z\right) U_{z}\in \mathcal{B}\left( \mathcal{H}_{\rho
}\right) \ .  \label{operator B}
\end{equation}%
Note that $B_{\gamma }(a)$ is well-defined because of (\ref{estimate 1}) and
the triangle inequality. The next step is a study of the operator (\ref%
{operator B}), since we already know that, as $\gamma \rightarrow 0^{+}$, $%
\rho (\mathfrak{R}_{\Phi }^{f,\gamma })\rightarrow 0$ uniformly in $\rho \in
E$, thanks to Equation (\ref{sdasdasdadbis}). \medskip

\noindent \underline{Step 4:} Using Equations (\ref{ssdsssd})--(\ref%
{sdsdsdsdsdssd}), the operator (\ref{operator B}) is equal for $a\in 
\mathfrak{L}$ and $\gamma \in (0,1)$ to%
\begin{equation}
B_{\gamma }\left( a\right) =\int_{\Theta _{d}}\sum_{z\in \mathfrak{L}}\gamma
^{d}f\left( \gamma a+\gamma z\right) \mathrm{e}^{i\theta \cdot z}\mathrm{d}%
\left( \mathrm{F}_{\ast }\mathrm{P}\right) \left( \theta \right) \ .
\label{B}
\end{equation}%
Defining the function $g$ on the lattice $\mathfrak{L}$ by%
\begin{equation*}
g\left( y\right) \doteq f\left( y\right) \mathrm{e}^{\gamma ^{-1}i\theta
\cdot y}\mathrm{e}^{-i\theta \cdot a}\ ,\qquad y\in \mathfrak{L}\ ,
\end{equation*}%
for any $a\in \mathfrak{L}$,\ $\gamma \in (0,1)$\ and $\theta \in \Theta
_{d} $, we obviously have the equalities%
\begin{equation*}
g\left( \gamma a+\gamma z\right) =f\left( \gamma a+\gamma z\right) \mathrm{e}%
^{i\theta \cdot z}\ ,\qquad \ z\in \mathfrak{L}\ ,
\end{equation*}%
as well as 
\begin{equation*}
\hat{g}(k)=\hat{f}\left( k-\gamma ^{-1}\theta \right) \mathrm{e}^{-i\theta
\cdot a}\ ,\qquad \ k\in \mathbb{R}^{d}\ .
\end{equation*}%
Since $f\in \mathfrak{D}_{0}$, we can thus infer from Equation (\ref{B}),
combined with the Poisson summation formula applied to $g$ (Proposition \ref%
{poissonsum}), that%
\begin{equation}
B_{\gamma }\left( a\right) =\int_{\Theta _{d}}\hat{f}\left( -\gamma
^{-1}\theta \right) \mathrm{e}^{-i\theta \cdot a}\mathrm{d}\left( \mathrm{F}%
_{\ast }\mathrm{P}\right) \left( \theta \right) +R_{f,a,\gamma }\ ,
\label{operator B sympatoch}
\end{equation}%
for any $a\in \mathfrak{L}$ and\ $\gamma \in (0,1)$, where%
\begin{equation}
R_{f,a,\gamma }\doteq \int_{\Theta _{d}}\sum_{z\in \mathfrak{L}\backslash
\{0\}}\hat{f}\left( \gamma ^{-1}\left( 2\pi z-\theta \right) \right) \mathrm{%
e}^{i(2\pi z-\theta )\cdot a}\mathrm{d}\left( \mathrm{F}_{\ast }\mathrm{P}%
\right) \left( \theta \right) \ .  \label{reste}
\end{equation}%
By Lemma \ref{colsumlim}, observe that, for any $\varepsilon \in \mathbb{R}%
^{+}$, there is a constant $M_{\varepsilon }\in \mathbb{R}^{+}$ such that,
for any $\gamma \in (0,1)$, $\theta \in (-\pi ,\pi ]^{d}$ and $f\in 
\mathfrak{D}_{\varepsilon }$,%
\begin{equation*}
\sum_{k\in \mathfrak{L}\backslash \{0\}}\left\vert \hat{f}\left( \gamma
^{-1}\left( 2\pi z-\theta \right) \right) \right\vert \leq \gamma
^{2}M_{\varepsilon }\left\Vert f\right\Vert _{\mathfrak{D}_{\varepsilon }}\ .
\end{equation*}%
It then directly follows from (\ref{reste}) that%
\begin{equation}
\left\Vert R_{f,a,\gamma }\right\Vert _{\mathcal{B}\left( \mathcal{H}_{\rho
}\right) }\leq \gamma ^{2}M_{\varepsilon }\left\Vert f\right\Vert _{%
\mathfrak{D}_{\varepsilon }}\ ,\qquad f\in \mathfrak{D}_{\varepsilon },\
\varepsilon \in \mathbb{R}^{+}\ .  \label{bound symaptoch2}
\end{equation}%
We are now in a position to study the Kac limit of the energy density given
by Equations (\ref{sdasdasdad2})--(\ref{operator B}). \medskip

\noindent \underline{Step 5:} The function $f\in \mathfrak{D}_{0,+}$ being
reflection-symmetric, $\hat{f}$ is reflection-symmetric. It follows from
Equations (\ref{sdasdasdad2}) and (\ref{operator B sympatoch}) that%
\begin{eqnarray}
&&\rho \left( \mathfrak{e}_{\mathcal{K}_{\gamma }\left( \Phi ,f\right)
}\right) +\rho (\mathfrak{R}_{\Phi }^{f,\gamma })+\zeta \left( \gamma \right)
\label{dsdsdsd} \\
&=&\sum_{\mathcal{Z}_{1},\mathcal{Z}_{2}\in \mathcal{P}_{\mathrm{f}}:%
\mathcal{Z}_{1}\cap \mathcal{Z}_{2}\supseteq \left\{ 0\right\} }\left\langle 
\frac{\pi _{\rho }\left( \Phi _{\mathcal{Z}_{1}}\right) }{\left\vert 
\mathcal{Z}_{1}\right\vert }\Omega _{\rho },\int_{\Theta _{d}}\hat{f}\left(
\gamma ^{-1}\theta \right) \sum_{x\in \mathcal{Z}_{1},y\in \mathcal{Z}_{2}}%
\frac{\mathrm{e}^{i\theta \cdot \left( x-y\right) }}{\left\vert \mathcal{Z}%
_{1}\right\vert \left\vert \mathcal{Z}_{2}\right\vert }\mathrm{d}\left( 
\mathrm{F}_{\ast }\mathrm{P}\right) \left( \theta \right) \frac{\pi _{\rho
}\left( \Phi _{\mathcal{Z}_{2}}\right) }{\left\vert \mathcal{Z}%
_{2}\right\vert }\Omega _{\rho }\right\rangle _{\mathcal{H}_{\rho }}  \notag
\end{eqnarray}%
for some complex-valued function $\zeta $ satisfying%
\begin{equation}
\left\vert \zeta \left( \gamma \right) \right\vert \leq \gamma
^{2}M_{\varepsilon }\left\Vert f\right\Vert _{\mathfrak{D}_{\varepsilon
}}\left\Vert \Phi \right\Vert _{\mathcal{W}_{1}}^{2}\ ,\qquad f\in \mathfrak{%
D}_{\varepsilon },\ \varepsilon \in \mathbb{R}^{+}\ ,  \label{dsdsdsd2}
\end{equation}%
thanks to Equations (\ref{iteration0}) and (\ref{bound symaptoch2}). On the
other hand, we infer from \cite[Equation (4.17)]{BruPedra2} that 
\begin{equation}
\Delta _{A}\left( \rho \right) =\left\Vert P_{\rho }\pi _{\rho }\left(
A\right) \Omega _{\rho }\right\Vert _{\mathcal{H}_{\rho }}^{2}\ ,\qquad \rho
\in E_{1}\ ,  \label{sdsdsddsdd}
\end{equation}%
where $P_{\rho }$ is the orthogonal projection on the subspace of elements
of $\mathcal{H}_{\rho }$ that are invariant with respect to all unitary
operators $U_{x}$ for $x\in \mathbb{Z}^{d}$. In other words, by Equation (%
\ref{ssdsssd}) and other well-known properties of the measurable spectral
calculus, this projection can be written as 
\begin{equation}
P_{\rho }\doteq \int_{\Theta _{d}}h\left( \theta \right) \mathrm{d}\left( 
\mathrm{F}_{\ast }\mathrm{P}\right) \left( \theta \right) \ ,
\label{sdsdsddsdd2}
\end{equation}%
where $h:\Theta _{d}\rightarrow \mathbb{R}$ is the function defined by $%
h(0)\doteq 1$ and $h(\theta )\doteq 0$ for all $\theta \in \Theta
_{d}\backslash \{0\}$. Equality (\ref{sdsdsddsdd}) is in fact a direct
consequence of (\ref{Limit of Space-Averages}) and (\ref{space averaging})
together with the von Neumann ergodic theorem \cite[Theorem 4.2]{BruPedra2}.
Now, for every (nonempty) finite subset $\mathcal{Z}\in \mathcal{P}_{\mathrm{%
f}}$ with $0\in \mathcal{Z}$, observe that%
\begin{equation*}
\lim_{\gamma \rightarrow 0^{+}}\hat{f}\left( \gamma ^{-1}\theta \right)
\sum_{x\in \mathcal{Z}_{1},y\in \mathcal{Z}_{2}}\frac{\mathrm{e}^{i\theta
\cdot \left( x-y\right) }}{\left\vert \mathcal{Z}_{1}\right\vert \left\vert 
\mathcal{Z}_{2}\right\vert }=\hat{f}\left( 0\right) h\left( \theta \right)
,\qquad \theta \in \Theta _{d}\ .
\end{equation*}%
By Lebesgue's dominated convergence theorem, it follows that the operators 
\begin{equation*}
\int_{\Theta _{d}}\hat{f}\left( \gamma ^{-1}\theta \right) \sum_{x\in 
\mathcal{Z}_{1},y\in \mathcal{Z}_{2}}\frac{\mathrm{e}^{i\theta \cdot \left(
x-y\right) }}{\left\vert \mathcal{Z}_{1}\right\vert \left\vert \mathcal{Z}%
_{2}\right\vert }\mathrm{d}\left( \mathrm{F}_{\ast }\mathrm{P}\right) \left(
\theta \right) \in \mathcal{B}\left( \mathcal{H}_{\rho }\right) ,\qquad
\gamma \in \left( 0,1\right) \ ,
\end{equation*}%
converge strongly to the operator 
\begin{equation*}
\int_{\Theta _{d}}\hat{f}\left( 0\right) h\left( \theta \right) \mathrm{d}%
\left( \mathrm{F}_{\ast }\mathrm{P}\right) \left( \theta \right) =\hat{f}%
\left( 0\right) P_{\rho }\ ,
\end{equation*}%
as $\gamma \rightarrow 0^{+}$. Together with Equations (\ref{equation
utilise}), (\ref{sdasdasdadbis}) and (\ref{dsdsdsd})--(\ref{sdsdsddsdd}),
the theorem then follows.
\end{proof}

\subsubsection{Multiple-Field Case}

Recall that Kac interactions of Definition \ref{definition Kac interaction}
are generalized by Definition \ref{definition Kac interaction copy(1)} to
accommodate possibly infinitely many mean-field components in the Kac limit.
This generalization uses Borel measures of bounded variation on either $%
\mathbb{S}\times \mathbb{D}_{\varepsilon }$, $\mathbb{S}\times \mathbb{D}%
_{\varepsilon ,+}$ or $\mathbb{S}\times \mathbb{C}_{\varepsilon ,+}$,
corresponding to the spaces $\mathcal{D}_{\varepsilon }$, $\mathcal{D}%
_{\varepsilon ,+}\subseteq \mathcal{D}_{\varepsilon }$ and $\mathcal{C}%
_{\varepsilon ,+}\subseteq \mathcal{D}_{\varepsilon }$, respectively, for
any fixed $\varepsilon \in \mathbb{R}^{+}$. See also Equation (\ref{D0}). In
the same way we obtain the energy density of the Kac interaction in the
limit $\gamma \rightarrow 0^{+}$ for the simple-field case (Theorem \ref%
{propuni copy(1)}), we get an analogous result for the general
(multiple-field) case. To this end, we need a standard result of measure
theory, referring to the disintegration of measures:

\begin{theorem}[Disintegration of measures]
\label{disintegration theorem}\mbox{ }\newline
Let $\mathcal{X},\mathcal{Y}$ be complete separable metric spaces.\ Take $%
\mathfrak{b}$ a finite (positive) Borel measure on $\mathcal{X}\times 
\mathcal{Y}$ and let $\pi _{\ast }\mathfrak{b}$ be the pushforward of $%
\mathfrak{b}$ through the projection $\pi :\mathcal{X}\times \mathcal{Y}%
\rightarrow \mathcal{X}$ defined by%
\begin{equation*}
\pi \left( x,y\right) \doteq x\ ,\qquad x\in \mathcal{X},\ y\in \mathcal{Y}\
.
\end{equation*}%
Then, there exists a $\pi _{\ast }\mathfrak{b}$-a.e. uniquely defined family 
$\{\mu _{x}\}_{x\in \mathcal{X}}$ of probability measures on $\mathcal{Y}$,
such that, for any Borel set $\mathcal{A}\subseteq \mathcal{Y}$, the mapping 
$x\mapsto \mu _{x}(\mathcal{A})$ is Borel measurable and, for any measurable
function $g:\mathcal{X}\times \mathcal{Y}\rightarrow \lbrack 0,\infty ]$,%
\begin{equation*}
\int_{\mathcal{X}\times \mathcal{Y}}g\left( x,y\right) \mathfrak{b}\left( 
\mathrm{d}\left( x,y\right) \right) =\int_{\mathcal{X}}\left( \int_{\mathcal{%
Y}}g\left( x,y\right) \mu _{x}\left( \mathrm{d}y\right) \right) (\pi _{\ast }%
\mathfrak{b})\left( \mathrm{d}x\right) \ .
\end{equation*}
\end{theorem}

\begin{proof}
This theorem is a straightforward application of \cite[Theorem 5.3.1]%
{desintegrationreference}.
\end{proof}

\noindent This disintegration theorem is useful in the sequel by allowing us
to extend Theorem \ref{propuni copy(1)} to Kac interactions of Definition %
\ref{definition Kac interaction copy(1)}.

To explain how we use this result, note first that, for any $\varepsilon \in 
\mathbb{R}^{+}$, the set $\mathbb{S}\times \mathbb{D}_{\varepsilon }$ is a
complete and separable metric space with respect to the restriction of the
norm of $\mathcal{W}_{1}\times \mathcal{D}_{\varepsilon }$, which is a
separable Banach space. Then, for any $\varepsilon \in \mathbb{R}^{+}$ and
any finite (positive) measure $\mathfrak{b}\in \mathcal{D}_{\varepsilon }$,
we use Theorem \ref{disintegration theorem} to define the Borel measure $%
\mathfrak{a}_{\mathfrak{b}}\in \mathcal{S}_{1}$ by 
\begin{eqnarray}
\mathfrak{a}_{\mathfrak{b}}\left( \mathcal{A}\right) &\doteq &\int_{\mathcal{%
A}}\left( \int_{\mathbb{D}_{\varepsilon }}\hat{f}\left( 0\right) \mu _{\Psi
}\left( \mathrm{d}f\right) \right) (\pi _{\ast }\mathfrak{b)}\left( \mathrm{d%
}\Psi \right)  \label{ab} \\
&=&\int_{\mathcal{A\times }\mathbb{D}_{\varepsilon }}\hat{f}\left( 0\right) 
\mathfrak{b}\left( \mathrm{d}(\Psi ,f)\right)  \notag
\end{eqnarray}%
for all Borel sets $\mathcal{A}\subseteq \mathbb{S}$, with $\pi _{\ast }%
\mathfrak{b}$ being the pushforward of $\mathfrak{b}$ under the projection
mapping $\pi :\mathbb{S}\times \mathbb{D}_{\varepsilon }\rightarrow \mathbb{S%
}$ defined by%
\begin{equation}
\pi \left( \Psi ,f\right) \doteq \Psi \ ,\qquad \Psi \in \mathbb{S},\ f\in 
\mathbb{D}_{\varepsilon }\ .  \label{projection mapping}
\end{equation}%
Note that this measure is well-defined because the mapping $f\mapsto \hat{f}%
\left( 0\right) $ from $\mathfrak{D}_{\varepsilon }$ to $\mathbb{R}_{0}^{+}$
is continuous. Observe also that the mapping $\mathfrak{b}\mapsto \mathfrak{a%
}_{\mathfrak{b}}$ from $\mathcal{D}_{\varepsilon }$ to $\mathcal{S}_{1}$ is
(real) linear.

We then obtain the following theorem:

\begin{theorem}[Energy density in the Kac limit]
\label{propuni copy(4)}\mbox{ }\newline
For any $\mathfrak{b}\in \mathcal{D}_{0}$, the family $(e_{\Phi ^{\mathfrak{b%
},\gamma }})_{\gamma \in (0,1)}$ of energy density functionals converges
pointwise to the functional $\Delta _{\mathfrak{a}_{\mathfrak{b}}}$ in the
Kac limit, i.e., 
\begin{equation*}
\lim_{\gamma \rightarrow 0^{+}}e_{\Phi ^{\mathfrak{b},\gamma }}\left( \rho
\right) =\Delta _{\mathfrak{a}_{\mathfrak{b}}}(\rho )\ ,\qquad \rho \in
E_{1}\ .
\end{equation*}
\end{theorem}

\begin{proof}
Take a finite measure $\mathfrak{b}\in \mathcal{D}_{0}$ and observe from
Equation (\ref{equation utilise}) and Definition \ref{definition Kac
interaction copy(1)}, along with the continuity of states and basic
properties of the Bochner integral, that%
\begin{equation}
e_{\Phi ^{\mathfrak{b},\gamma }}\left( \rho \right) =\int_{\mathbb{S}\times 
\mathbb{D}_{\varepsilon }}e_{\mathcal{K}_{\gamma }\left( \Psi ,f\right)
}\left( \rho \right) \mathfrak{b}\left( \mathrm{d}\left( \Psi ,f\right)
\right) \ ,\qquad \rho \in E_{1}\ .  \label{kjkjk}
\end{equation}%
See also Lemma \ref{lemma Kac norm} (i). So, since $\mathfrak{b}$ is a
finite measure, it suffices to invoke Equation (\ref{inequality a la con}),
Theorem \ref{propuni copy(1)} and Lemma \ref{lemma Kac norm} (i), as well as
Lebesgue's dominated convergence theorem to arrive at 
\begin{equation*}
\lim_{\gamma \rightarrow 0^{+}}e_{\Phi ^{\mathfrak{b},\gamma }}\left( \rho
\right) =\int_{\mathbb{S}\times \mathbb{D}_{\varepsilon }}\hat{f}\left(
0\right) \Delta _{\mathfrak{e}_{\Psi }}\left( \rho \right) \mathfrak{b}%
\left( \mathrm{d}\left( \Psi ,f\right) \right) \ ,\qquad \rho \in E_{1}\ .
\end{equation*}%
Combined with Theorem \ref{disintegration theorem}, this last equality in
turn implies the pointwise convergence, as $\gamma \rightarrow 0^{+}$, of
the family $(e_{\Phi ^{\mathfrak{b},\gamma }})_{\gamma \in (0,1)}$ to the
functional $\Delta _{\mathfrak{a}_{\mathfrak{b}}}$, because, for any
translation-invariant state $\rho \in E_{1}$, 
\begin{eqnarray}
\int_{\mathbb{S}\times \mathbb{D}_{\varepsilon }}\hat{f}\left( 0\right)
\Delta _{\mathfrak{e}_{\Psi }}\left( \rho \right) \mathfrak{b}\left( \mathrm{%
d}\left( \Psi ,f\right) \right) &=&\int_{\mathbb{S}}\Delta _{\mathfrak{e}%
_{\Psi }}\left( \rho \right) \left( \int_{\mathbb{D}_{\varepsilon }}\hat{f}%
\left( 0\right) \mu _{\Psi }\left( \mathrm{d}f\right) \right) (\pi _{\ast }%
\mathfrak{b})\left( \mathrm{d}\Psi \right)  \notag \\
&=&\int_{\mathbb{S}}\Delta _{\mathfrak{e}_{\Psi }}\left( \rho \right) 
\mathfrak{a}_{\mathfrak{b}}\left( \mathrm{d}\Psi \right) \doteq \Delta _{%
\mathfrak{a}_{\mathfrak{b}}}\left( \rho \right) \ ,  \label{sssdsdsd}
\end{eqnarray}%
by Equations (\ref{Free-energy density long range0}) and (\ref{ab}).
\end{proof}

\subsection{Thermodynamics in the Kac Limit -- Repulsive Case\label%
{Thermodynamic in the Kac Limit -- Repulsive Case}}

In this section we study the convergence of Fermi systems associated with
Kac interactions towards purely repulsive mean-field models in the Kac
limit, after taking the thermodynamic limit. As explained in Section \ref%
{Section purely attrac}, purely repulsive mean-field models refer here to
translation-invariant mean-field models $\mathfrak{m}=(\Phi ,\mathfrak{%
\mathfrak{a}})\in \mathcal{M}_{1}$, with $\mathfrak{\mathfrak{a}}=\mathfrak{a%
}_{+}$ being a finite positive measure on $\mathbb{S}$. The attractive case
is studied in the next section.\ Similar to the study of the Kac limit of
energy densities done in Section \ref{Kac Limit of Energy Densities}, we
divide this study in two steps, by starting with the pedagogical case of
models with a single mean-field interaction.

\subsubsection{Simple-Field Case}

In order to study the convergence of Fermi systems associated with Kac
interactions towards repulsive mean-field models, we use an estimate on
energy densities of Kac interactions with respect to space-averaging
functionals on translation-invariant states (Section \ref{Section space
averaging}). It is obtained from a slight modification of the proof of
Theorem \ref{propuni copy(1)} and corresponds to the following assertion:

\begin{proposition}[Energy densities of Kac interactions and space-averaging
functionals]
\label{propuni copy(2)}\mbox{ }\newline
For any parameter $\varepsilon \in \mathbb{R}^{+}$, there is a constant $%
M_{\varepsilon }>0$ such that, for any $\gamma \in (0,1)$, each function$\
f\in \mathfrak{D}_{\varepsilon ,+}$, any interaction $\Phi \in \mathcal{W}%
_{1}$ and $\rho \in E_{1}$,%
\begin{equation*}
\hat{f}\left( 0\right) \Delta _{\mathfrak{e}_{\Phi }}\left( \rho \right)
\leq e_{\mathcal{K}_{\gamma }\left( \Phi ,f\right) }\left( \rho \right) +%
\mathfrak{F}_{\gamma }\left( \Phi ,f\right) +\gamma ^{2}M_{\varepsilon
}\left\Vert f\right\Vert _{\mathfrak{D}_{\varepsilon }}\left\Vert \Phi
\right\Vert _{\mathcal{W}_{1}}^{2}\ ,
\end{equation*}%
where $\hat{f}$ denotes the Fourier\ transform of $f$ and $\mathfrak{F}%
_{\gamma }$ is the mapping from $\mathcal{W}_{1}\times \mathfrak{D}_{0}$ to $%
\mathbb{R}_{0}^{+}$ defined by Equation (\ref{technical function}).
\end{proposition}

\begin{proof}
Using all mathematical objects described in the proof of Theorem \ref%
{propuni copy(1)}, the identity 
\begin{equation*}
h(\theta )g(\theta )=h(\theta )g(0)\ ,\qquad \theta \in \Theta _{d}\ ,
\end{equation*}%
for any function $g:\Theta _{d}\rightarrow \mathbb{C}$, as well as Equation (%
\ref{sdsdsdsd}), we obtain from (\ref{equation utilise}) and (\ref%
{sdsdsddsdd})--(\ref{sdsdsddsdd2}) that%
\begin{align*}
\hat{f}\left( 0\right) \Delta _{\mathfrak{e}_{\Phi }}\left( \rho \right) &
=\left\Vert \int_{\Theta _{d}}\sqrt{\hat{f}\left( 0\right) }h\left( \theta
\right) \mathrm{d}\left( \mathrm{F}_{\ast }\mathrm{P}\right) \left( \theta
\right) \sum_{\mathcal{Z}\in \mathcal{P}_{\mathrm{f}}:\mathcal{Z}\supseteq
\left\{ 0\right\} }\frac{\pi _{\rho }\left( \Phi _{\mathcal{Z}}\right) }{%
\left\vert \mathcal{Z}\right\vert }\Omega _{\rho }\right\Vert _{\mathcal{H}%
_{\rho }}^{2} \\
& =\left\Vert \sum_{\mathcal{Z}\in \mathcal{P}_{\mathrm{f}}:\mathcal{Z}%
\supseteq \left\{ 0\right\} }\int_{\Theta _{d}}h\left( \theta \right) \sqrt{%
\hat{f}\left( \gamma ^{-1}\theta \right) }\sum_{x\in \mathcal{Z}}\frac{%
\mathrm{e}^{i\theta \cdot x}}{\left\vert \mathcal{Z}\right\vert }\mathrm{d}%
\left( \mathrm{F}_{\ast }\mathrm{P}\right) \left( \theta \right) \frac{\pi
_{\rho }\left( \Phi _{\mathcal{Z}}\right) }{\left\vert \mathcal{Z}%
\right\vert }\Omega _{\rho }\right\Vert _{\mathcal{H}_{\rho }}^{2} \\
& =\left\Vert P_{\rho }\sum_{\mathcal{Z}\in \mathcal{P}_{\mathrm{f}}:%
\mathcal{Z}\supseteq \left\{ 0\right\} }\int_{\Theta _{d}}\sqrt{\hat{f}%
\left( \gamma ^{-1}\theta \right) }\sum_{x\in \mathcal{Z}}\frac{\mathrm{e}%
^{i\theta \cdot x}}{\left\vert \mathcal{Z}\right\vert }\mathrm{d}\left( 
\mathrm{F}_{\ast }\mathrm{P}\right) \left( \theta \right) \frac{\pi _{\rho
}\left( \Phi _{\mathcal{Z}}\right) }{\left\vert \mathcal{Z}\right\vert }%
\Omega _{\rho }\right\Vert _{\mathcal{H}_{\rho }}^{2}\ ,
\end{align*}%
which, together with Equation (\ref{dsdsdsd}), directly implies the upper
bound%
\begin{equation*}
\hat{f}\left( 0\right) \Delta _{\mathfrak{e}_{\Phi }}\left( \rho \right)
\leq \rho \left( \mathfrak{e}_{\mathcal{K}_{\gamma }\left( \Phi ,f\right)
}\right) +\rho (\mathfrak{R}_{\Phi }^{f,\gamma })+\zeta \left( \gamma \right)
\end{equation*}%
for any $\gamma \in (0,1)$. Now, it suffices to invoke Equations (\ref%
{sdasdasdadbis}) and (\ref{dsdsdsd2}) to get the assertion.
\end{proof}

Proposition \ref{propuni copy(2)} gives in particular an upper bound for the
space-averaging functional on trans%
%TCIMACRO{\TeXButton{\-}{\-}}%
%BeginExpansion
\-%
%EndExpansion
lation-invariant states via an energy density functional of a short range
interaction in $\mathcal{W}_{1}$. This upper bound is \emph{uniform} with
respect to translation-invariant states, by (\ref{ssdsdsdsdsddsdsds}) and
Lemma \ref{propuni}. This is a strong property which directly implies that
the infinite volume pressure $\mathrm{P}_{\mathcal{K}_{\gamma }\left( \Phi
,f\right) }$ and the equilibrium states of $\mathit{M}_{\mathcal{K}_{\gamma
}\left( \Phi ,f\right) }$ associated with some Kac interaction $\mathcal{K}%
_{\gamma }\left( \Phi ,f\right) $, as respectively defined by (\ref{pressure
short range}) and (\ref{minimizer short range}), converge in the Kac limit $%
\gamma \rightarrow 0^{+}$ to those of the expected purely repulsive
mean-field model $\mathfrak{m}\in \mathcal{M}_{1}$. See (\ref{pressure long
range}) and (\ref{definition equilibirum state}) for the definitions of the
infinite volume pressure and (generalized) equilibrium states of any
mean-field model. More precisely, one gets the following result:

\begin{theorem}[From short-range interactions to repulsive mean-field models]

\label{propuni copy(3)}\mbox{ }\newline
Take interactions $\Phi \in \mathcal{W}_{1}^{\mathbb{R}}$, $\Psi \in \mathbb{%
S}$ (see (\ref{unit sphere})) and a positive definite function $f\in 
\mathfrak{D}_{0,+}$, whose Fourier\ transform is denoted by $\hat{f}$. Let $%
\delta _{\Psi }\in \mathcal{S}_{1}$ be the Dirac measure on $\Psi \in 
\mathbb{S}$.\newline
\emph{(i)}\ Convergence of infinite-volume pressures: 
\begin{equation*}
\lim_{\gamma \rightarrow 0^{+}}\mathrm{P}_{\Phi +\mathcal{K}_{\gamma }\left(
\Psi ,f\right) }=\mathrm{P}_{(\Phi ,\hat{f}\left( 0\right) \delta _{\Psi
})}^{\sharp }=\mathrm{P}_{(\Phi ,\hat{f}\left( 0\right) \delta _{\Psi
})}^{\flat }\ .
\end{equation*}%
\emph{(ii)}\ Convergence of equilibrium states: Weak$^{\ast }$ accumulation
points of any net of equilibrium states $\omega _{\gamma }\in \mathit{M}%
_{\Phi +\mathcal{K}_{\gamma }\left( \Psi ,f\right) }$ as $\gamma \rightarrow
0^{+}$ are generalized equilibrium states of the purely repulsive mean-field
model $(\Phi ,\hat{f}\left( 0\right) \delta _{\Psi })$, i.e., they belong to
the weak$^{\ast }$-compact convex set $\mathit{\Omega }_{(\Phi ,\hat{f}%
\left( 0\right) \delta _{\Psi })}^{\sharp }=\mathit{\Omega }_{(\Phi ,\hat{f}%
\left( 0\right) \delta _{\Psi })}^{\flat }$.
\end{theorem}

\begin{proof}
Before starting, recall that $\mathrm{P}_{\mathfrak{m}}^{\sharp }=\mathrm{P}%
_{\mathfrak{m}}^{\flat }$ and $\mathit{\Omega }_{\mathfrak{m}}^{\sharp }=%
\mathit{\Omega }_{\mathfrak{m}}^{\flat }$ for any purely repulsive
mean-field model $\mathfrak{m}\in \mathcal{M}_{1}$, thanks to \cite[Theorem
2.25]{BruPedra2}. By Equations (\ref{pressure free energy}) and (\ref{map
free energy}), for any $\Phi \in \mathcal{W}_{1}^{\mathbb{R}}$, $\Psi \in 
\mathbb{S}$ and $f\in \mathfrak{D}_{0}$, 
\begin{equation}
\mathrm{P}_{\Phi +\mathcal{K}_{\gamma }\left( \Psi ,f\right) }=-\inf_{\rho
\in E_{1}}\left\{ f_{\Phi }\left( \rho \right) +e_{\mathcal{K}_{\gamma
}\left( \Psi ,f\right) }\left( \rho \right) \right\} \ ,\qquad \gamma \in
\left( 0,1\right) \ ,  \label{sdsdsd}
\end{equation}%
while from (\ref{BCS main theorem 1eq})--(\ref{Free-energy density long
range}), 
\begin{equation}
\mathrm{P}_{(\Phi ,\hat{f}\left( 0\right) \delta _{\Psi })}=-\inf_{\rho \in
E_{1}}\{f_{\Phi }\left( \rho \right) +\hat{f}\left( 0\right) \Delta _{%
\mathfrak{e}_{\Psi }}\left( \rho \right) \}\ .  \label{sdsdsd1}
\end{equation}%
Assertion (i) is therefore a direct consequence of (\ref{sdsdsd})--(\ref%
{sdsdsd1}) combined with Theorem \ref{propuni copy(1)}, Proposition \ref%
{propuni copy(2)} and Lemma \ref{propuni}. The proof of Assertion (ii) is
similar:\ For any net of equilibrium states $\omega _{\gamma }\in \mathit{M}%
_{\Phi +\mathcal{K}_{\gamma }\left( \Psi ,f\right) }$, we combine (\ref%
{sdsdsd}) with Proposition \ref{propuni copy(2)} to deduce that, for any $%
\gamma \in \left( 0,1\right) $,%
\begin{equation*}
f_{\Phi }\left( \omega _{\gamma }\right) +\Delta _{\mathfrak{e}_{\Psi
}}\left( \omega _{\gamma }\right) \leq \inf_{\rho \in E_{1}}\left\{ f_{\Phi
}\left( \rho \right) +e_{\mathcal{K}_{\gamma }\left( \Psi ,f\right) }\left(
\rho \right) \right\} +\mathfrak{F}_{\gamma }\left( \Phi ,f\right) +\gamma
^{2}M_{\varepsilon }\left\Vert f\right\Vert _{\mathfrak{D}_{\varepsilon
}}\left\Vert \Phi \right\Vert _{\mathcal{W}_{1}}^{2}\ ,
\end{equation*}%
which, combined with Equations (\ref{definition equilibirum state}) and (\ref%
{sdsdsd1}), Lemma \ref{propuni} and Assertion (i), implies the second
statement of the theorem.
\end{proof}

If one considers the simple example given by (\ref{N interaction0})--(\ref{N
interaction2}), then Theorem \ref{propuni copy(3)} and Remark \ref{remark
idiote} imply that, for any translation-invariant interaction $\Phi \in 
\mathcal{W}_{1}^{\mathbb{R}}$ and positive definite function $f\in \mathfrak{%
D}_{0,+}$,%
\begin{eqnarray}
&&\lim_{\gamma \rightarrow 0^{+}}\underset{L\rightarrow \infty }{\lim }\frac{%
1}{\beta |\Lambda _{L}|}\ln \mathrm{Trace}\left( \exp \left\{ -\beta \left(
U_{L}^{\Phi }+\sum_{x,y\in \Lambda _{L},\mathrm{s},\mathrm{t}\in \mathrm{S}%
}\gamma ^{d}f\left( \gamma \left( x-y\right) \right) a_{y,\mathrm{t}}^{\ast
}a_{y,\mathrm{t}}a_{x,\mathrm{s}}^{\ast }a_{x,\mathrm{s}}\right) \right\}
\right)  \notag \\
&=&\underset{L\rightarrow \infty }{\lim }\frac{1}{\beta |\Lambda _{L}|}\ln 
\mathrm{Trace}\left( \exp \left\{ -\beta \left( U_{L}^{\Phi }+\frac{\hat{f}%
\left( 0\right) }{\left\vert \Lambda _{L}\right\vert }\sum_{x,y\in \Lambda
_{L},\mathrm{s},\mathrm{t}\in \mathrm{S}}a_{y,\mathrm{t}}^{\ast }a_{y,%
\mathrm{t}}a_{x,\mathrm{s}}^{\ast }a_{x,\mathrm{s}}\right) \right\} \right)
\ ,  \label{dddfdfdf}
\end{eqnarray}%
while the equilibrium states of this Kac interaction can be approximated by
generalized equilibrium states of the corresponding mean-field model. Note
that the thermodynamics of this repulsive mean-field model can be \textbf{%
explicitly computed} when $\Phi $ defines a quasi-free fermion system, by
using the thermodynamic game (cf. Sections \ref{Section thermo game} and \ref%
{Section effective theories}).

This application of Theorem \ref{propuni copy(3)} is \textbf{nontrivial},
albeit simple, keeping in mind that the results hold true for all
translation-invariant interactions $\Phi \in \mathcal{W}_{1}^{\mathbb{R}}$.
In particular, $\mathcal{W}_{1}^{\mathbb{R}}$ encodes -- by far -- all
short-range lattice fermion models used in condensed matter physics, the
norm on $\mathcal{W}_{1}^{\mathbb{R}}$ being quite weak. E.g., 
\begin{equation*}
\left\Vert \Phi \right\Vert _{\mathcal{W}_{1}}\doteq \sum\limits_{\Lambda
\in \mathcal{P}_{\mathrm{f}},\;\Lambda \supseteq \{0\}}\left\vert \Lambda
\right\vert ^{-1}\left\Vert \Phi _{\Lambda }\right\Vert _{\mathcal{U}}\leq
\sum\limits_{\Lambda \in \mathcal{P}_{\mathrm{f}},\;\Lambda \supseteq
\{0\}}\left\Vert \Phi _{\Lambda }\right\Vert _{\mathcal{U}}\ .
\end{equation*}%
See Equation (\ref{iteration0}).

Regarding the simple example given by (\ref{N interaction0})--(\ref{N
interaction2}), $f$ is a pair potential characterizing an interparticle
interaction whose range is tuned by the parameter $\gamma \in (0,1)$. The
main limitation at this point concerning the choice of $f$ is the fact that
this function has to be \emph{positive definite}. Conditions of this type
are relaxed, from Section \ref{section attractive} on, and very general
results are proven in Section \ref{section general}. Remark, however, that
it is current in theoretical physics to use a positive definite function for
such two-body interaction potential. This property of $f$ is reminiscent of
a superstability condition, which is important in the bosonic case \cite[%
Section 2.2 and Appendix G]{BruZagrebnov8}.

In the next section we extend our results to approximate -- via (Kac)\
short-range interactions of $\mathcal{W}_{1}$ -- models of $\mathcal{M}_{1}$
with possibly infinitely many repulsive mean-field components.

\subsubsection{Multiple-Field Case}

In order to approximate a model with possibly infinitely many mean-field
repulsions, we use the Kac interactions $\Phi ^{\mathfrak{b},\gamma }\in 
\mathcal{W}_{1}^{\mathbb{R}}$ of Definition \ref{definition Kac interaction
copy(1)} for positive measures $\mathfrak{b}\in \mathcal{D}_{0,+}$ and $%
\gamma \in (0,1)$. In this context, we need to adapt Proposition \ref%
{propuni copy(2)}. For any positive measure $\mathfrak{b}\in \mathcal{D}_{0}$%
, recall that we use Theorem \ref{disintegration theorem} to define the
positive measure $\mathfrak{a}_{\mathfrak{b}}\in \mathcal{S}_{1}$ by (\ref%
{ab}). Then, we have the following estimate:

\begin{proposition}[Energy densities of Kac interactions and space-averaging
functionals]
\label{propuni copy(7)}\mbox{ }\newline
For any parameter $\varepsilon \in \mathbb{R}^{+}$, there is a constant $%
M_{\varepsilon }>0$ such that, for any $\gamma \in (0,1)$, positive measure $%
\mathfrak{b}\in \mathcal{D}_{\varepsilon ,+}$, interaction $\Phi \in 
\mathcal{W}_{1}$ and $\rho \in E_{1}$, 
\begin{equation*}
\Delta _{\mathfrak{a}_{\mathfrak{b}}}\left( \rho \right) \leq e_{\Phi ^{%
\mathfrak{b},\gamma }}\left( \rho \right) +\int_{\mathbb{S}\times \mathbb{D}%
_{\varepsilon ,+}}\mathfrak{F}_{\gamma }\left( \Phi ,f\right) \mathfrak{b}%
\left( \mathrm{d}\left( \Psi ,f\right) \right) +\gamma ^{2}M_{\varepsilon }%
\mathfrak{b}\left( \mathbb{S}\times \mathbb{D}_{\varepsilon ,+}\right)
\end{equation*}%
with $\mathfrak{F}_{\gamma }$ being the mapping from $\mathcal{W}_{1}\times 
\mathfrak{D}_{0}$ to $\mathbb{R}_{0}^{+}$ defined by Equation (\ref%
{technical function}) for any fixed $\gamma \in (0,1)$.
\end{proposition}

\begin{proof}
The assertion is a consequence of Proposition \ref{propuni copy(2)}, along
with Equations (\ref{kjkjk}) and (\ref{sssdsdsd}). Note that the above
integral is well-defined. See for instance Lemma \ref{propuni}.
\end{proof}

We can now generalize Theorem \ref{propuni copy(3)} to models with possibly
infinitely many long-range repulsions, by replacing Proposition \ref{propuni
copy(2)} with Proposition \ref{propuni copy(7)} and by using basically the
same arguments:

\begin{theorem}[From short-range interactions to repulsive mean-field models]

\label{propuni copy(5)}\mbox{ }\newline
Take an interaction $\Phi \in \mathcal{W}_{1}^{\mathbb{R}}$ and a positive
measure $\mathfrak{b}\in \mathcal{D}_{0,+}$.\newline
\emph{(i)}\ Convergence of infinite-volume pressures: 
\begin{equation*}
\lim_{\gamma \rightarrow 0^{+}}\mathrm{P}_{\Phi +\Phi ^{\mathfrak{b},\gamma
}}=\mathrm{P}_{(\Phi ,\mathfrak{a}_{\mathfrak{b}})}^{\sharp }=\mathrm{P}%
_{(\Phi ,\mathfrak{a}_{\mathfrak{b}})}^{\flat }\ .
\end{equation*}%
\emph{(ii)}\ Convergence of equilibrium states: Weak$^{\ast }$ accumulation
points of any net of equilibrium states $\omega _{\gamma }\in \mathit{M}%
_{\Phi +\Phi ^{\mathfrak{b},\gamma }}$ as $\gamma \rightarrow 0^{+}$ are
generalized equilibrium states of the purely repulsive model $(\Phi ,%
\mathfrak{a}_{\mathfrak{b}})$, i.e., they belong to the weak$^{\ast }$%
-compact convex set $\mathit{\Omega }_{(\Phi ,\mathfrak{a}_{\mathfrak{b}%
})}^{\sharp }=\mathit{\Omega }_{(\Phi ,\mathfrak{a}_{\mathfrak{b}})}^{\flat
} $.
\end{theorem}

\begin{proof}
Recall again that $\mathrm{P}_{\mathfrak{m}}^{\sharp }=\mathrm{P}_{\mathfrak{%
m}}^{\flat }$ and $\mathit{\Omega }_{\mathfrak{m}}^{\sharp }=\mathit{\Omega }%
_{\mathfrak{m}}^{\flat }$ for any purely repulsive model $\mathfrak{m}\in 
\mathcal{M}_{1}$, thanks to \cite[Theorem 2.25]{BruPedra2}. By Equations (%
\ref{pressure free energy}) and (\ref{map free energy}), for any $\Phi \in 
\mathcal{W}_{1}^{\mathbb{R}}$ and positive measure $\mathfrak{b}\in \mathcal{%
D}_{0}$,%
\begin{equation}
\mathrm{P}_{\Phi +\Phi ^{\mathfrak{b},\gamma }}=-\inf_{\rho \in
E_{1}}\left\{ f_{\Phi }\left( \rho \right) +e_{\Phi ^{\mathfrak{b},\gamma
}}\left( \rho \right) \right\} \ ,  \label{sdsdsdbis}
\end{equation}%
while from (\ref{BCS main theorem 1eq})--(\ref{Free-energy density long
range}), 
\begin{equation}
\mathrm{P}_{(\Phi ,\mathfrak{a}_{\mathfrak{b}})}=-\inf_{\rho \in
E_{1}}\left\{ f_{\Phi }\left( \rho \right) +\Delta _{\mathfrak{a}_{\mathfrak{%
b}}}\left( \rho \right) \right\} \ .  \label{sdsdsd1bis}
\end{equation}%
Assertion (i) thus results from Equations Equations (\ref{sdsdsdbis})--(\ref%
{sdsdsd1bis}), together with Proposition \ref{propuni copy(7)}, Theorem \ref%
{propuni copy(4)} and Lemma \ref{propuni}, while the proof of (ii) requires
the additional use of Equations (\ref{definition equilibirum state}), all
combined with Lebesgue's dominated convergence theorem. We omit the details.
\end{proof}

Since\ the positive measure $\mathfrak{b}\in \mathcal{D}_{0,+}$ is general
in Theorem \ref{propuni copy(5)}, it is easy to check that all purely
repulsive mean-field models in $\mathcal{M}_{1}$ can be approximated in the
sense of Theorem \ref{propuni copy(5)} via a Kac interaction of Definition %
\ref{definition Kac interaction copy(1)}.

\subsection{Thermodynamics in the Kac Limit -- Attractive Case\label{section
attractive}}

In the present section we study the convergence of Fermi systems associated
with Kac interactions towards purely attractive mean-field models in the Kac
limit, after taking the thermodynamic limit. As explained in Section \ref%
{Section purely attrac}, purely attractive mean-field models refer to
translation-invariant mean-field models $\mathfrak{m}=(\Phi ,\mathfrak{%
\mathfrak{a}})\in \mathcal{M}_{1}$, with $\mathfrak{a}_{-}=-\mathfrak{%
\mathfrak{a}}$ being a finite positive measure on $\mathbb{S}$. Similar to
the repulsive case (Section \ref{Thermodynamic in the Kac Limit -- Repulsive
Case}), we split this study in two steps, by starting with the pedagogical
case of models with a single mean-field interaction.

\subsubsection{Simple-Field Case}

The upper bound given by Proposition \ref{propuni copy(2)} for the purely
repulsive case does not hold true anymore in the attractive case. In fact,
Theorem \ref{propuni copy(1)} and Equations (\ref{sdsdsd})--(\ref{sdsdsd1})
only lead to the inequality%
\begin{equation}
\mathrm{P}_{(\Phi ,-\hat{f}\left( 0\right) \delta _{\Psi })}^{\sharp }=%
\mathrm{P}_{(\Phi ,-\hat{f}\left( 0\right) \delta _{\Psi })}^{\flat }\leq
\inf_{\gamma \in \left( 0,1\right) }\mathrm{P}_{\Phi +\mathcal{K}_{\gamma
}\left( \Psi ,-f\right) }^{\sharp }\ ,\qquad \Psi \in \mathbb{S},\ \Phi \in 
\mathcal{W}_{1},\ f\in \mathfrak{D}_{0,+}\ ,  \label{inequality pressure}
\end{equation}%
with $\delta _{\Psi }\in \mathcal{S}_{1}$ being the Dirac measure on $\Psi $%
. In the present case, Proposition \ref{propuni copy(2)} is replaced by the
monotonicity of energy densities of Kac interactions with respect to $\gamma 
$, allowing us to use Lemma \ref{colsumlim copy(1)}, which is reminiscent of
Dini's theorem. To this end, we need to restrict the choice of functions $%
f\in \mathfrak{D}_{0,+}$ to the subcone $\mathfrak{C}_{0,+}\subseteq 
\mathfrak{D}_{0,+}$, which is defined by (\ref{D frac0}). More precisely, to
tackle the attractive case, Proposition \ref{propuni copy(2)} is replaced by
the following result:

\begin{proposition}[Monotonicity of energy density functionals]
\label{propuni copy(8)}\mbox{ }\newline
For $\Psi \in \mathcal{W}_{1}$, $f\in \mathfrak{C}_{0,+}$ and $\eta \in 
\mathbb{R}^{+}$, there exists $\gamma _{0}\in \left( 0,1\right) $ such that,
for all $\gamma _{1},\gamma _{2}\in \left( 0,\gamma _{0}\right) $ with $%
\gamma _{1}\geq \gamma _{2}$, 
\begin{equation*}
e_{\mathcal{K}_{\gamma _{2}}\left( \Phi ,-f\right) }\geq e_{\mathcal{K}%
_{\gamma _{1}}\left( \Phi ,-f\right) }-\eta \ .
\end{equation*}
\end{proposition}

\begin{proof}
By Lemma \ref{lemma Kac norm} (i), Equation (\ref{inequality a la con}) and
the density of $\mathcal{W}_{0}$ in $\mathcal{W}_{1}$, it suffices to
consider the special case $\Psi \in \mathcal{W}_{0}$ to prove the
proposition. Using the characteristic function $\chi _{r}$ of the ball of
radius $r\in \mathbb{R}^{+}$ centered at zero in the torus $\Theta
_{d}\doteq \lbrack -\pi ,\pi )^{d}$, we observe from (\ref{dsdsdsd}) that,
for any $\Psi \in \mathcal{W}_{0}$, $r\in \mathbb{R}^{+}$, $\gamma \in
\left( 0,1\right) $ and $f\in \mathfrak{C}_{0,+}$, 
\begin{eqnarray}
&&\rho \left( \mathfrak{e}_{\mathcal{K}_{\gamma }\left( \Psi ,f\right)
}\right) +\rho (\mathfrak{R}_{\Psi }^{f,\gamma })+\zeta \left( \gamma
\right) +\mathfrak{X}_{\Psi }^{f,\gamma }\left( r,\rho \right) +\mathfrak{Y}%
_{\Psi }^{f,\gamma }\left( r,\rho \right)  \label{ssdsdsdsd} \\
&=&\sum_{\mathcal{Z}_{1},\mathcal{Z}_{2}\in \mathcal{P}_{\mathrm{f}}:%
\mathcal{Z}_{1}\cap \mathcal{Z}_{2}\supseteq \left\{ 0\right\} }\left\langle 
\frac{\pi _{\rho }\left( \Psi _{\mathcal{Z}_{1}}\right) }{\left\vert 
\mathcal{Z}_{1}\right\vert }\Omega _{\rho },\int_{\Theta _{d}}\hat{f}\left(
\gamma ^{-1}\theta \right) \chi _{r}\left( \theta \right) \mathrm{d}\left( 
\mathrm{F}_{\ast }\mathrm{P}\right) \left( \theta \right) \frac{\pi _{\rho
}\left( \Psi _{\mathcal{Z}_{2}}\right) }{\left\vert \mathcal{Z}%
_{2}\right\vert }\Omega _{\rho }\right\rangle _{\mathcal{H}_{\rho }},  \notag
\end{eqnarray}%
where 
\begin{eqnarray*}
\mathfrak{X}_{\Psi }^{f,\gamma }\left( r,\rho \right) &\doteq &\sum_{%
\mathcal{Z}_{1},\mathcal{Z}_{2}\in \mathcal{P}_{\mathrm{f}}:\mathcal{Z}%
_{1}\cap \mathcal{Z}_{2}\supseteq \left\{ 0\right\} }\left\langle \frac{\pi
_{\rho }\left( \Psi _{\mathcal{Z}_{1}}\right) }{\left\vert \mathcal{Z}%
_{1}\right\vert }\Omega _{\rho },\int_{\Theta _{d}}\hat{f}\left( \gamma
^{-1}\theta \right) \chi _{r}\left( \theta \right) \right. \\
&&\qquad \qquad \qquad \qquad \qquad \qquad \left. \sum_{x\in \mathcal{Z}%
_{1},y\in \mathcal{Z}_{2}}\frac{\left( 1-\mathrm{e}^{i\theta \cdot \left(
x-y\right) }\right) }{\left\vert \mathcal{Z}_{1}\right\vert \left\vert 
\mathcal{Z}_{2}\right\vert }\mathrm{d}\left( \mathrm{F}_{\ast }\mathrm{P}%
\right) \left( \theta \right) \frac{\pi _{\rho }\left( \Psi _{\mathcal{Z}%
_{2}}\right) }{\left\vert \mathcal{Z}_{2}\right\vert }\Omega _{\rho
}\right\rangle _{\mathcal{H}_{\rho }} \\
\mathfrak{Y}_{\Psi }^{f,\gamma }\left( r,\rho \right) &\doteq &\sum_{%
\mathcal{Z}_{1},\mathcal{Z}_{2}\in \mathcal{P}_{\mathrm{f}}:\mathcal{Z}%
_{1}\cap \mathcal{Z}_{2}\supseteq \left\{ 0\right\} }\left\langle \frac{\pi
_{\rho }\left( \Psi _{\mathcal{Z}_{1}}\right) }{\left\vert \mathcal{Z}%
_{1}\right\vert }\Omega _{\rho },\int_{\Theta _{d}}\hat{f}\left( \gamma
^{-1}\theta \right) \left( \chi _{r}\left( \theta \right) -1\right) \right.
\\
&&\qquad \qquad \qquad \qquad \qquad \qquad \left. \sum_{x\in \mathcal{Z}%
_{1},y\in \mathcal{Z}_{2}}\frac{\mathrm{e}^{i\theta \cdot \left( x-y\right) }%
}{\left\vert \mathcal{Z}_{1}\right\vert \left\vert \mathcal{Z}%
_{2}\right\vert }\mathrm{d}\left( \mathrm{F}_{\ast }\mathrm{P}\right) \left(
\theta \right) \frac{\pi _{\rho }\left( \Psi _{\mathcal{Z}_{2}}\right) }{%
\left\vert \mathcal{Z}_{2}\right\vert }\Omega _{\rho }\right\rangle _{%
\mathcal{H}_{\rho }}.
\end{eqnarray*}%
For any $\varepsilon \in \mathbb{R}^{+}$ and finite-range interaction $\Psi
\in \mathcal{W}_{0}$, there is a constant $C_{\varepsilon ,\Psi }\in \mathbb{%
R}^{+}$ such that, for any $f\in \mathfrak{C}_{\varepsilon ,+}$ and $\gamma
\in \left( 0,1\right) $, 
\begin{equation}
\left\vert \mathfrak{X}_{\Psi }^{f,\gamma }\left( r,\rho \right) \right\vert
\leq C_{\varepsilon ,\Psi }r\left\Vert f\right\Vert _{\mathfrak{D}%
_{\varepsilon }}\ ,\qquad \left\vert \mathfrak{Y}_{\Psi }^{f,\gamma }\left(
r,\rho \right) \right\vert \leq C_{\varepsilon ,\Psi }\left( \frac{\gamma }{r%
}\right) ^{2d}\left\Vert f\right\Vert _{\mathfrak{D}_{\varepsilon }}\ .
\label{ssdsdsdsd2}
\end{equation}%
Observe meanwhile that, for any function $f\in \mathfrak{C}_{0,+}$, the
family of functions $(g_{\gamma })_{\gamma \in \left( 0,1\right) }$ defined
on the torus $\Theta _{d}$ by 
\begin{equation*}
g_{\gamma }\left( \theta \right) \doteq \hat{f}\left( \gamma ^{-1}\theta
\right) \chi _{r}\left( \theta \right) \ ,\qquad \theta \in \Theta _{d}\ ,
\end{equation*}%
is monotonically increasing with respect to $\gamma \in \left( 0,1\right) $.
It suffices now to use Equations (\ref{sdasdasdadbis}), (\ref{dsdsdsd2}) and
(\ref{ssdsdsdsd})--(\ref{ssdsdsdsd2}) to get the assertion.
\end{proof}

Proposition \ref{propuni copy(8)} combined with Lemma \ref{colsumlim copy(1)}
implies that the infinite volume pressure $\mathrm{P}_{\mathcal{K}_{\gamma
}\left( \Phi ,-f\right) }$ and the equilibrium states of $\mathit{M}_{%
\mathcal{K}_{\gamma }\left( \Phi ,-f\right) }$ associated with some Kac
interaction $\mathcal{K}_{\gamma }\left( \Phi ,-f\right) $, as respectively
defined by (\ref{pressure short range}) and (\ref{minimizer short range}),
converges in the Kac limit $\gamma \rightarrow 0^{+}$ to those of some
purely attractive mean-field model $\mathfrak{m}\in \mathcal{M}_{1}$. See (%
\ref{pressure long range}) and (\ref{definition equilibirum state}) for the
definitions of the infinite volume pressure and equilibrium states of any
mean-field model. More precisely, one gets the following result:

\begin{theorem}[From short-range interactions to attractive mean-field models%
]
\label{propuni copy(6)0}\mbox{ }\newline
Take interactions $\Phi \in \mathcal{W}_{1}^{\mathbb{R}}$, $\Psi \in \mathbb{%
S}\cap \mathcal{W}_{0}$ and a positive definite function $f\in \mathfrak{C}%
_{0,+}$, whose Fourier\ transform is denoted by $\hat{f}$. Let $\delta
_{\Psi }\in \mathcal{S}_{1}$ denote the Dirac measure on $\Psi \in \mathbb{S}
$.\newline
\emph{(i)}\ Convergence of infinite-volume pressures: 
\begin{equation*}
\lim_{\gamma \rightarrow 0^{+}}\mathrm{P}_{\Phi +\mathcal{K}_{\gamma }\left(
\Psi ,-f\right) }=\mathrm{P}_{(\Phi ,-\hat{f}\left( 0\right) \delta _{\Psi
})}^{\sharp }=\mathrm{P}_{(\Phi ,-\hat{f}\left( 0\right) \delta _{\Psi
})}^{\flat }\ .
\end{equation*}%
\emph{(ii)}\ Convergence of equilibrium states: Weak$^{\ast }$ accumulation
points of any net of equilibrium states $\omega _{\gamma }\in \mathit{M}%
_{\Phi +\mathcal{K}_{\gamma }\left( \Psi ,-f\right) }$ as $\gamma
\rightarrow 0^{+}$ are equilibrium states of $(\Phi ,-\hat{f}\left( 0\right)
\delta _{\Psi })$, i.e., they belong to the weak$^{\ast }$-compact convex
set $\mathit{\Omega }_{(\Phi ,-\hat{f}\left( 0\right) \delta _{\Psi
})}^{\sharp }=\mathit{\Omega }_{(\Phi ,-\hat{f}\left( 0\right) \delta _{\Psi
})}^{\flat }$.
\end{theorem}

\begin{proof}
Assertion (i) is a direct consequence of Proposition \ref{propuni copy(8)},
Equations (\ref{sdsdsd})--(\ref{sdsdsd1}) and Theorem \ref{propuni copy(1)}
together with Lemma \ref{colsumlim copy(1)}. To prove the second assertion,
fix $\Phi \in \mathcal{W}_{1}^{\mathbb{R}}$, $\Psi \in \mathbb{S}\cap 
\mathcal{W}_{0}$ and a positive definite function $f\in \mathfrak{C}_{0,+}$%
.\ For any $\gamma \in \left( 0,1\right) $, pick a minimizer $\omega
_{\gamma }\in \mathit{M}_{\Phi +\mathcal{K}_{\gamma }\left( \Psi ,-f\right)
} $ of the variational problem%
\begin{equation*}
\inf_{\rho \in E_{1}}\left\{ f_{\Phi }\left( \rho \right) -e_{\mathcal{K}%
_{\gamma }\left( \Psi ,-f\right) }\left( \rho \right) \right\} =-\mathrm{P}%
_{\Phi +\mathcal{K}_{\gamma }\left( \Psi ,-f\right) }\ .
\end{equation*}%
Observe that, for any $\gamma \in \left( 0,1\right) $, $\omega _{\gamma }$
can be equivalently seen as a tangent functional at $\Phi \in \mathcal{W}%
_{1}^{\mathbb{R}}$ to the convex continuous function $\Phi \mapsto \mathrm{P}%
_{\Phi +\mathcal{K}_{\gamma }\left( \Psi ,-f\right) }$ from the real Banach
space $\mathcal{W}_{1}^{\mathbb{R}}$ to $\mathbb{R}$. See \cite[Section 2.6]%
{BruPedra2} for more details. In particular, for any $\Phi _{1}\in \mathcal{W%
}_{1}^{\mathbb{R}}$, 
\begin{equation*}
\mathrm{P}_{\Phi +\Phi _{1}+\mathcal{K}_{\gamma }\left( \Psi ,-f\right) }-%
\mathrm{P}_{\Phi +\mathcal{K}_{\gamma }\left( \Psi ,-f\right) }\geq -\omega
_{\gamma }\left( \mathfrak{e}_{\Phi _{1}}\right) \ .
\end{equation*}%
By weak$^{\ast }$ compactness of $E_{1}$, the net $(\omega _{\gamma
})_{\gamma \in \left( 0,1\right) }$ has weak$^{\ast }$-converging subnets
and any weak$^{\ast }$ accumulation point of the net is the limit of such a
subnet. By Assertion (i), each weak$^{\ast }$ accumulation point can be seen
as a tangent functional at $\Phi \in \mathcal{W}_{1}^{\mathbb{R}}$ to the
convex continuous function $\tilde{\Phi}\mapsto \mathrm{P}_{(\tilde{\Phi},-%
\hat{f}\left( 0\right) \delta _{\Psi })}^{\sharp }$ from the real Banach
space $\mathcal{W}_{1}^{\mathbb{R}}$ to $\mathbb{R}$. By \cite[Theorem 2.28]%
{BruPedra2}, Assertion (ii) then follows.
\end{proof}

If one considers the simple example given by (\ref{BCS interaction})--(\ref%
{BCS interaction1}), then Theorem \ref{propuni copy(6)0} and Remark \ref%
{remark idiote} imply that, for any translation-invariant interaction $\Phi
\in \mathcal{W}_{1}^{\mathbb{R}}$ and positive definite function $f\in 
\mathfrak{C}_{0,+}$,%
\begin{eqnarray}
&&\lim_{\gamma \rightarrow 0^{+}}\underset{L\rightarrow \infty }{\lim }\frac{%
1}{\beta |\Lambda _{L}|}\ln \mathrm{Trace}\left( \exp \left\{ -\beta \left(
U_{L}^{\Phi }-\sum_{x,y\in \Lambda _{L}}\gamma ^{d}f\left( \gamma \left(
x-y\right) \right) a_{y,\downarrow }^{\ast }a_{y,\uparrow }^{\ast
}a_{x,\uparrow }a_{x,\downarrow }\right) \right\} \right)  \notag \\
&=&\underset{L\rightarrow \infty }{\lim }\frac{1}{\beta |\Lambda _{L}|}\ln 
\mathrm{Trace}\left( \exp \left\{ -\beta \left( U_{L}^{\Phi }-\frac{\hat{f}%
\left( 0\right) }{\left\vert \Lambda _{L}\right\vert }\sum_{x,y\in \Lambda
_{L}}a_{y,\downarrow }^{\ast }a_{y,\uparrow }^{\ast }a_{x,\uparrow
}a_{x,\downarrow }\right) \right\} \right) \ ,  \label{dddfdfdfdddfdfdf}
\end{eqnarray}%
while the equilibrium states of this Kac interaction can be approximated by
equilibrium states of the corresponding mean-field model. Note that the
thermodynamics of this attractive mean-field model can again be \textbf{%
explicitly computed} when $\Phi $ defines a quasi-free fermion system, by
using the thermodynamic game (cf. Sections \ref{Section thermo game} and \ref%
{Section effective theories}).

This application of Theorem \ref{propuni copy(6)0} is again \textbf{%
nontrivial}, albeit simple, keeping in mind that the results hold true for
all translation-invariant interactions $\Phi \in \mathcal{W}_{1}^{\mathbb{R}%
} $. Regarding the simple example given by (\ref{BCS interaction})--(\ref%
{BCS interaction1}), $f$ encodes the hopping strength of Cooper pairs,
characterizing the BCS interaction whose range is tuned by the parameter $%
\gamma \in (0,1)$. The limit mean-field models are also important in
theoretical physics. For instance, the reduced BCS Hamiltonian or the strong
coupling BCS Hamiltonian, which are mean-field models, qualitatively display
most of the basic properties of real conventional type I superconductors.
See, e.g., \cite[Chapter VII, Section 4]{Thou}. \ Concerning the choice of $%
f $, the main limitation at this point is the fact that this function has to
be \emph{positive definite} with \emph{scaling-monotone} Fourier transform,
see (\ref{secaling monotone}). This kind of condition is relaxed in Section %
\ref{section general}, where very general results are proven. Observe,
however, that functions of this type are used in theoretical physics, like
in the original BCS\ theory of superconductivity \cite{BCS}.

In the next section we extend our results to approximate -- via (Kac)\
short-range interactions of $\mathcal{W}_{1}$ -- models of $\mathcal{M}_{1}$
with possibly infinitely many attractive mean-field components.

\subsubsection{Multiple-Field Case}

In order to approximate a model with possibly infinitely many mean-field
attractions, we use the Kac interactions $\Phi ^{-\mathfrak{b},\gamma }\in 
\mathcal{W}_{1}^{\mathbb{R}}$ of Definition \ref{definition Kac interaction
copy(1)} for positive measures $\mathfrak{b}\in \mathcal{C}_{0,+}$ and $%
\gamma \in (0,1)$. In this context, we need to adapt Proposition \ref%
{propuni copy(8)}. For any positive measure $\mathfrak{b}\in \mathcal{D}%
_{0,+}$, recall that we use Theorem \ref{disintegration theorem} to define
the positive measure $\mathfrak{a}_{\mathfrak{b}}\in \mathcal{S}_{1}$ by (%
\ref{ab}). Then, we have the following property of energy densities of Kac
interactions:

\begin{proposition}[Monotonicity of energy density functionals]
\label{propuni copy(9)}\mbox{ }\newline
For any positive measure $\mathfrak{b}\in \mathcal{C}_{0,+}$ and $\eta \in 
\mathbb{R}^{+}$, there exists $\gamma _{0}\in \left( 0,1\right) $ such that,
for all $\gamma _{1},\gamma _{2}\in \left( 0,\gamma _{0}\right) $ with $%
\gamma _{1}\geq \gamma _{2}$, 
\begin{equation*}
e_{\Phi ^{-\mathfrak{b},\gamma _{2}}}\geq e_{\Phi ^{-\mathfrak{b},\gamma
_{1}}}-\eta \ .
\end{equation*}
\end{proposition}

\begin{proof}
By Lemma \ref{definition Kac interaction copy(2)} and Equation (\ref%
{inequality a la con}), we can assume without loss of generality that 
\begin{equation*}
\mathfrak{b}\doteq \sum_{j=1}^{n}\delta _{\left( \Psi _{j},f_{j}\right) }
\end{equation*}%
for some finite sequence $\left( \Psi _{1},f_{1}\right) ,\ldots ,\left( \Psi
_{n},f_{n}\right) $ in $\mathbb{S}\times (\mathbb{D}_{\varepsilon }\cap 
\mathfrak{C}_{0,+})$. Then, by Proposition \ref{propuni copy(8)} and the
linearity of the mapping $\Phi \mapsto e_{\Phi }\left( \rho \right) $ from $%
\mathcal{W}_{1}$ to $\mathbb{C}$ at fixed $\rho \in E_{1}$, the assertion
follows.
\end{proof}

We can now generalize Theorem \ref{propuni copy(6)0} to models with possibly
infinitely many mean-field attractions by replacing Proposition \ref{propuni
copy(8)} with Proposition \ref{propuni copy(9)} and by using basically the
same arguments as in the simple-field case:

\begin{theorem}[From short-range interactions to attractive mean-field models%
]
\label{propuni copy(6)}\mbox{ }\newline
Take an interaction $\Phi \in \mathcal{W}_{1}^{\mathbb{R}}$ and a positive
measure $\mathfrak{b}\in \mathcal{C}_{0,+}$.\newline
\emph{(i)}\ Convergence of infinite-volume pressures: 
\begin{equation*}
\lim_{\gamma \rightarrow 0^{+}}\mathrm{P}_{\Phi +\Phi ^{-\mathfrak{b},\gamma
}}=\mathrm{P}_{(\Phi ,-\mathfrak{a}_{\mathfrak{b}})}^{\sharp }=\mathrm{P}%
_{(\Phi ,-\mathfrak{a}_{\mathfrak{b}})}^{\flat }\ .
\end{equation*}%
\emph{(ii)}\ Convergence of equilibrium states: Weak$^{\ast }$ accumulation
points of any net of equilibrium states $\omega _{\gamma }\in \mathit{M}%
_{\Phi +\Phi ^{-\mathfrak{b},\gamma }}$ as $\gamma \rightarrow 0^{+}$ are
equilibrium states of $(\Phi ,-\mathfrak{a}_{\mathfrak{b}})$, i.e., they
belong to the weak$^{\ast }$-compact convex set $\mathit{\Omega }_{(\Phi ,-%
\mathfrak{a}_{\mathfrak{b}})}^{\sharp }=\mathit{\Omega }_{(\Phi ,-\mathfrak{a%
}_{\mathfrak{b}})}^{\flat }$.
\end{theorem}

\begin{proof}
Assertion (i) is a direct consequence of Proposition \ref{propuni copy(9)},
Equations (\ref{sdsdsdbis})--(\ref{sdsdsd1bis}) and Theorem \ref{propuni
copy(4)} together with Lemma \ref{colsumlim copy(1)}. The proof of the
second assertion is the same as the one of Theorem \ref{propuni copy(6)0}
(ii). We omit the details.
\end{proof}

It is easy to check that all purely attractive mean-field models can be
approximated in the sense of Theorem \ref{propuni copy(6)} via a Kac
interaction of Definition \ref{definition Kac interaction copy(1)} for
positive measures $\mathfrak{b}\in \mathcal{C}_{0,+}$.

\subsection{Thermodynamics in the Kac Limit -- General Case\label{section
general}}

\subsubsection{Mixed Case\label{section mixed}}

In this section we generalize the results on purely attractive or repulsive
mean-field models to all mean-field models in $\mathcal{M}_{1}$, with
possibly infinitely many mean-field attractions and repulsions. We first
generalize Inequality (\ref{inequality pressure}), which relates the
conventional and non-conventional pressures $\mathrm{P}^{\sharp }$ and $%
\mathrm{P}^{\flat }$\ of mean-field models with the pressures of the
corresponding Kac interactions. Recall that $\mathrm{P}^{\sharp }$ and $%
\mathrm{P}^{\flat }$ are defined by (\ref{pressure long range})--(\ref%
{Free-energy density long range}) and (\ref{convex functional g_m})--(\ref%
{Pressure bemol}), respectively.\ In fact, in the general case, we have the
following inequalities:

\begin{proposition}[Bounds on Kac pressures]
\label{propuni copy(10)}\mbox{ }\newline
Take an interaction $\Phi \in \mathcal{W}_{1}^{\mathbb{R}}$ and two positive
measures $\mathfrak{b}_{+}\in \mathcal{D}_{0,+}$ and $\mathfrak{b}_{-}\in 
\mathcal{C}_{0,+}$. Then, for any sequences $(\gamma _{+,n})_{n\in \mathbb{N}%
}$ and $(\gamma _{-,n})_{n\in \mathbb{N}}$ of real numbers in the interval $%
\left( 0,1\right) $ converging to zero,%
\begin{equation*}
\mathrm{P}_{(\Phi ,\mathfrak{a}_{\mathfrak{b}_{+}}-\mathfrak{a}_{\mathfrak{b}%
_{-}})}^{\sharp }\leq \liminf_{n\rightarrow \infty }\mathrm{P}_{\Phi +\Phi ^{%
\mathfrak{b}_{+},\gamma _{+,n}}+\Phi ^{-\mathfrak{b}_{-},\gamma _{-,n}}}\leq
\limsup_{n\rightarrow \infty }\mathrm{P}_{\Phi +\Phi ^{\mathfrak{b}%
_{+},\gamma _{+,n}}+\Phi ^{-\mathfrak{b}_{-},\gamma _{-,n}}}\leq \mathrm{P}%
_{(\Phi ,\mathfrak{a}_{\mathfrak{b}_{+}}-\mathfrak{a}_{\mathfrak{b}%
_{-}})}^{\flat }\ ,
\end{equation*}%
where, for any measure $\mathfrak{b}\in \mathcal{D}_{0}$, the measure $%
\mathfrak{a}_{\mathfrak{b}}\in \mathcal{S}_{1}$ is defined by Equation (\ref%
{ab}).
\end{proposition}

\begin{proof}
Fix once and for all $\Phi \in \mathcal{W}_{1}^{\mathbb{R}}$ and two
positive measures $\mathfrak{b}_{+}\in \mathcal{D}_{0,+}$ and $\mathfrak{b}%
_{-}\in \mathcal{C}_{0,+}$. On the one hand, using Proposition \ref{propuni
copy(7)}\ and Equation (\ref{sdsdsdbis}), for any parameters $\gamma
_{-},\gamma _{+}\in \left( 0,1\right) $, 
\begin{equation}
\mathrm{P}_{\Phi +\Phi ^{\mathfrak{b}_{+},\gamma _{+}}+\Phi ^{-\mathfrak{b}%
_{-},\gamma _{-}}}\leq -\inf h_{\gamma _{-}}\left( E_{1}\right) +\int_{%
\mathbb{S}\times \mathbb{D}_{\varepsilon ,+}}\mathfrak{F}_{\gamma
_{+}}\left( \Phi ,f\right) \mathfrak{b}\left( \mathrm{d}\left( \Psi
,f\right) \right) +\gamma _{+}^{2}M_{\varepsilon }\mathfrak{b}\left( \mathbb{%
S}\times \mathbb{D}_{\varepsilon ,+}\right)  \label{bornesuperieur}
\end{equation}%
with $h_{\gamma _{-}}$ being the functional defined on the space $E_{1}$ of
translation-invariant states by 
\begin{equation}
h_{\gamma _{-}}\left( \rho \right) \doteq f_{\Phi +\Phi ^{-\mathfrak{b}%
_{-},\gamma _{-}}}\left( \rho \right) +\Delta _{\mathfrak{a}_{\mathfrak{b}%
_{+}}}\left( \rho \right) \ ,\qquad \rho \in E_{1}\ .  \label{function h}
\end{equation}%
Note that this functional is generally not lower semicontinuous in the weak$%
^{\ast }$ topology, because $\Delta _{\mathfrak{a}_{\mathfrak{b}_{+}}}$ is
only upper semicontinuous for non-trivial measures $\mathfrak{a}_{\mathfrak{b%
}_{+}}\neq 0$. By \cite[Theorem 1.4]{BruPedra-convex}, observe that 
\begin{equation}
\inf h_{\gamma _{-}}\left( E_{1}\right) =\inf \Gamma (h_{\gamma _{-}})\left(
E_{1}\right) \ ,  \label{tyrtyryt1}
\end{equation}%
where $\Gamma (h)$ denotes the so-called $\Gamma $-regularization of a
functional $h$ on $E_{1}$, defined by (\ref{gamma regulirisation}). In
contrast with $h_{\gamma _{-}}$, the functional $\Gamma (h_{\gamma _{-}})$
is weak$^{\ast }$-lower semicontinuous and, moreover, it can be explicitly
computed, as done in the proof of \cite[Theorem 2.21]{BruPedra2}: 
\begin{equation}
\Gamma (h_{\gamma _{-}})\left( \rho \right) =f_{\Phi +\Phi ^{-\mathfrak{b}%
_{-},\gamma _{-}}}\left( \rho \right) +\int_{\mathbb{S}}\left\vert \rho
\left( \mathfrak{e}_{\Psi }\right) \right\vert ^{2}\mathfrak{a}_{\mathfrak{b}%
_{+}}\left( \mathrm{d}\Psi \right) \ ,\qquad \rho \in E_{1}\ .
\label{tyrtyryt2}
\end{equation}%
We now invoke Proposition \ref{propuni copy(9)}, Theorem \ref{propuni
copy(4)} and Lemma \ref{colsumlim copy(1)} to deduce from (\ref{tyrtyryt1}%
)--(\ref{tyrtyryt2}) that 
\begin{equation}
\lim_{\gamma _{-}\rightarrow 0^{+}}\inf h_{\gamma _{-}}\left( E_{1}\right)
=\inf f_{(\Phi ,\mathfrak{a}_{\mathfrak{b}_{+}}-\mathfrak{a}_{\mathfrak{b}%
_{-}})}^{\flat }\left( E_{1}\right) =-\mathrm{P}_{(\Phi ,\mathfrak{a}_{%
\mathfrak{b}_{+}}-\mathfrak{a}_{\mathfrak{b}_{-}})}^{\flat }\ .
\label{function hlimit}
\end{equation}%
By Inequality (\ref{bornesuperieur}), Lemma \ref{propuni} and Lebesgue's
dominated convergence theorem, it follows that%
\begin{equation*}
\limsup_{n\rightarrow \infty }\mathrm{P}_{\Phi +\Phi ^{\mathfrak{b}%
_{+},\gamma _{+,n}}+\Phi ^{-\mathfrak{b}_{-},\gamma _{-,n}}}\leq \mathrm{P}%
_{(\Phi ,\mathfrak{a}_{\mathfrak{b}_{+}}-\mathfrak{a}_{\mathfrak{b}%
_{-}})}^{\flat }
\end{equation*}%
for any sequences $(\gamma _{+,n})_{n\in \mathbb{N}}$ and $(\gamma
_{-,n})_{n\in \mathbb{N}}$ of real numbers in the interval $\left(
0,1\right) $ converging to zero.

On the other hand, using Equation (\ref{sdsdsdbis}), for any $\gamma
_{-},\gamma _{+}\in \left( 0,1\right) $, 
\begin{equation*}
\mathrm{P}_{\Phi +\Phi ^{\mathfrak{b}_{+},\gamma _{+}}+\Phi ^{-\mathfrak{b}%
_{-},\gamma _{-}}}\geq -f_{\Phi +\Phi ^{\mathfrak{b}_{+},\gamma _{+}}+\Phi
^{-\mathfrak{b}_{-},\gamma _{-}}}\left( \rho \right) \ ,\qquad \rho \in
E_{1}\ .
\end{equation*}%
By Theorem \ref{propuni copy(4)}, it follows that, for any $\rho \in E_{1}$
and all sequences $(\gamma _{+,n})_{n\in \mathbb{N}}$ and $(\gamma
_{-,n})_{n\in \mathbb{N}}$ of real numbers in the interval $\left(
0,1\right) $ converging to zero, 
\begin{equation*}
\liminf_{n\rightarrow \infty }\mathrm{P}_{\Phi +\Phi ^{\mathfrak{b}%
_{+},\gamma _{+,n}}+\Phi ^{-\mathfrak{b}_{-},\gamma _{-,n}}}\geq
-\lim_{n\rightarrow \infty }f_{\Phi +\Phi ^{\mathfrak{b}_{+},\gamma
_{+,n}}+\Phi ^{-\mathfrak{b}_{-},\gamma _{-,n}}}\left( \rho \right)
=-f_{(\Phi ,\mathfrak{a}_{\mathfrak{b}_{+}}-\mathfrak{a}_{\mathfrak{b}%
_{-}})}^{\sharp }\left( \rho \right) \ ,
\end{equation*}%
which, combined with Equation (\ref{BCS main theorem 1eq}), in turn implies
that%
\begin{equation*}
\liminf_{n\rightarrow \infty }\mathrm{P}_{\Phi +\Phi ^{\mathfrak{b}%
_{+},\gamma _{+,n}}+\Phi ^{-\mathfrak{b}_{-},\gamma _{-,n}}}\geq \mathrm{P}%
_{(\Phi ,\mathfrak{a}_{\mathfrak{b}_{+}}-\mathfrak{a}_{\mathfrak{b}%
_{-}})}^{\sharp }\ .
\end{equation*}
\end{proof}

In other words, Proposition \ref{propuni copy(10)} tells us that the limit
of Kac pressures or, at least, its accumulation points belong to the
interval $[\mathrm{P}_{\mathfrak{m}}^{\sharp },\mathrm{P}_{\mathfrak{m}%
}^{\flat }]$ for some mean-field model $\mathfrak{m}\in \mathcal{M}_{1}$. It
also means that the pressures associated with attractive and repulsive Kac
interactions \textbf{may not} converge to its natural mean-field
approximation, in contrast with the intuitive guess\footnote{%
Such kind of discrepancy between quantum models in the Kac limit and their
expected mean-field approximations already appears in a different context
for the Kac limit\ of a Bose gas in presence of two-body interactions, see 
\cite{Smedt-Zagrebnov}. However, in this previous case \cite{Smedt-Zagrebnov}%
, there is no competition between repulsive and attractive long-range
forces, but instead a perturbation with a scaled external field. See also
Section \ref{Kac-dffdfd} for more details.}. Below, we show that the
extremes of the interval $[\mathrm{P}_{\mathfrak{m}}^{\sharp },\mathrm{P}_{%
\mathfrak{m}}^{\flat }]$ can be attained by a convenient choice of the
sequences $(\gamma _{+,n})_{n\in \mathbb{N}}$ and $(\gamma _{-,n})_{n\in 
\mathbb{N}}$. We start with the lower boundary of the interval:

\begin{theorem}[From short-range interactions to conventional mean-field
models]
\label{propuni copy(11)}\mbox{ }\newline
Take an interaction $\Phi \in \mathcal{W}_{1}^{\mathbb{R}}$ and two positive
measures $\mathfrak{b}_{+}\in \mathcal{D}_{0,+}$ and $\mathfrak{b}_{-}\in 
\mathcal{C}_{0,+}$.\newline
\emph{(i)}\ Convergence of infinite-volume pressures: 
\begin{equation*}
\lim_{\gamma _{+}\rightarrow 0^{+}}\lim_{\gamma _{-}\rightarrow 0^{+}}%
\mathrm{P}_{\Phi +\Phi ^{\mathfrak{b}_{+},\gamma _{+}}+\Phi ^{-\mathfrak{b}%
_{-},\gamma _{-}}}=\mathrm{P}_{(\Phi ,\mathfrak{a}_{\mathfrak{b}_{+}}-%
\mathfrak{a}_{\mathfrak{b}_{-}})}^{\sharp }\ .
\end{equation*}%
\emph{(ii)}\ Convergence of equilibrium states: For all $\gamma _{+}\in
\left( 0,1\right) $, take any weak$^{\ast }$ accumulation point $\omega
_{\gamma _{+}}$ of any net $(\omega _{\gamma _{+},\gamma _{-}})_{\gamma
_{-}\in \left( 0,1\right) }\subseteq \mathit{M}_{\Phi +\Phi ^{\mathfrak{b}%
_{+},\gamma _{+}}+\Phi ^{-\mathfrak{b}_{-},\gamma _{-}}}$ as $\gamma
_{-}\rightarrow 0^{+}$. Pick any weak$^{\ast }$ accumulation point $\omega $
of the net $(\omega _{\gamma _{+}})_{\gamma _{+}\in \left( 0,1\right) }$, as 
$\gamma _{+}\rightarrow 0^{+}$. Then, $\omega \in \mathit{\Omega }_{(\Phi ,%
\mathfrak{a}_{\mathfrak{b}_{+}}-\mathfrak{a}_{\mathfrak{b}_{-}})}^{\sharp }$.
\end{theorem}

\begin{proof}
Pick $\Phi \in \mathcal{W}_{1}^{\mathbb{R}}$ and two positive measures $%
\mathfrak{b}_{+}\in \mathcal{D}_{0,+}$ and $\mathfrak{b}_{-}\in \mathcal{C}%
_{0,+}$. Fix $\gamma _{+}\in \left( 0,1\right) $. Then, by Theorem \ref%
{propuni copy(6)}, 
\begin{equation}
\lim_{\gamma _{-}\rightarrow 0^{+}}\mathrm{P}_{\Phi +\Phi ^{\mathfrak{b}%
_{+},\gamma _{+}}+\Phi ^{-\mathfrak{b}_{-},\gamma _{-}}}=\mathrm{P}_{(\Phi
+\Phi ^{\mathfrak{b}_{+},\gamma _{+}},-\mathfrak{a}_{\mathfrak{b}%
_{-}})}^{\sharp }=-\inf_{\rho \in E_{1}}\left\{ f_{\Phi }\left( \rho \right)
+e_{\Phi ^{\mathfrak{b}_{+},\gamma _{+}}}\left( \rho \right) -\Delta _{%
\mathfrak{a}_{\mathfrak{b}_{-}}}\left( \rho \right) \right\} \ .
\label{ghjghj}
\end{equation}%
In the same way we prove Theorem \ref{propuni copy(5)}, we use Proposition %
\ref{propuni copy(7)}, Theorem \ref{propuni copy(4)}, Lemma \ref{propuni}
and Lebesgue's dominated convergence theorem as well as Equations (\ref%
{sdsdsd1bis}) and (\ref{ghjghj}) to get Assertion (i). Finally, for any $%
\gamma _{+}\in \left( 0,1\right) $, take any weak$^{\ast }$ accumulation
point $\omega _{\gamma _{+}}$ of any net $(\omega _{\gamma _{+},\gamma
_{-}})_{\gamma _{-}\in \left( 0,1\right) }\subseteq \mathit{M}_{\Phi +\Phi ^{%
\mathfrak{b}_{+},\gamma _{+}}+\Phi ^{-\mathfrak{b}_{-},\gamma _{-}}}$ as $%
\gamma _{-}\rightarrow 0^{+}$. Then, pick any weak$^{\ast }$ accumulation
point $\omega $ of the net $(\omega _{\gamma _{+}})_{\gamma _{+}\in \left(
0,1\right) }$, as $\gamma _{+}\rightarrow 0^{+}$. Similar to the proof of
Theorem \ref{propuni copy(6)0} (ii), one shows that $\omega $ can be seen as
a tangent functional at $\Phi \in \mathcal{W}_{1}^{\mathbb{R}}$ to the
convex continuous function $\tilde{\Phi}\mapsto \mathrm{P}_{(\tilde{\Phi},%
\mathfrak{a}_{\mathfrak{b}_{+}}-\mathfrak{a}_{\mathfrak{b}_{-}})}^{\sharp }$
from the real Banach space $\mathcal{W}_{1}^{\mathbb{R}}$ to $\mathbb{R}$.
Then, by \cite[Theorem 2.28]{BruPedra2}, the assertion follows.
\end{proof}

\begin{corollary}[From short-range interactions to conventional mean-field
models]
\label{propuni copy(13)}\mbox{ }\newline
Take an interaction $\Phi \in \mathcal{W}_{1}^{\mathbb{R}}$ and two positive
measures $\mathfrak{b}_{+}\in \mathcal{D}_{0,+}$ and $\mathfrak{b}_{-}\in 
\mathcal{C}_{0,+}$. There exist two sequences $(\gamma _{+,n})_{n\in \mathbb{%
N}}$ and $(\gamma _{-,n})_{n\in \mathbb{N}}$ of real numbers in the interval 
$\left( 0,1\right) $ converging to zero and a sequence $(\omega _{n})_{n\in 
\mathbb{N}}$ of equilibrium states $\omega _{n}\in \mathit{M}_{\Phi +\Phi ^{%
\mathfrak{b}_{+},\gamma _{+,n}}+\Phi ^{-\mathfrak{b}_{-},\gamma _{-,n}}}$
such that 
\begin{equation*}
\lim_{n\rightarrow \infty }\mathrm{P}_{\Phi +\Phi ^{\mathfrak{b}_{+},\gamma
_{+,n}}+\Phi ^{-\mathfrak{b}_{-},\gamma _{-,n}}}=\mathrm{P}_{(\Phi ,%
\mathfrak{a}_{\mathfrak{b}_{+}}-\mathfrak{a}_{\mathfrak{b}_{-}})}^{\sharp }
\end{equation*}%
and $(\omega _{n})_{n\in \mathbb{N}}$ weak$^{\ast }$ converges to a
generalized equilibrium state of $\mathit{\Omega }_{(\Phi ,\mathfrak{a}_{%
\mathfrak{b}_{+}}-\mathfrak{a}_{\mathfrak{b}_{-}})}^{\sharp }$.
\end{corollary}

\begin{proof}
It suffices to combine Theorem \ref{propuni copy(11)} with \cite[Chapter 2,
4 Theorem]{topology}. Note that the weak$^{\ast }$-topology is metrizable on
the weak$^{\ast }$--compact convex set of all states on $\mathcal{U}$,
because $\mathcal{U}$ is separable. See, e.g., \cite[Theorem 10.10]%
{BruPedra2}.
\end{proof}

We show now that the upper bound of Proposition \ref{propuni copy(10)} is
also attained by a convenient choice of the sequences $(\gamma _{+,n})_{n\in 
\mathbb{N}}$ and $(\gamma _{-,n})_{n\in \mathbb{N}}$:

\begin{theorem}[From short-range interactions to non-conventional mean-field
models]
\label{propuni copy(12)bis}\mbox{ }\newline
Take an interaction $\Phi \in \mathcal{W}_{1}^{\mathbb{R}}$ and two positive
measures $\mathfrak{b}_{+}\in \mathcal{D}_{0,+}$ and $\mathfrak{b}_{-}\in 
\mathcal{C}_{0,+}$.\newline
\emph{(i)}\ Convergence of infinite-volume pressures: 
\begin{equation*}
\lim_{\gamma _{-}\rightarrow 0^{+}}\lim_{\gamma _{+}\rightarrow 0^{+}}%
\mathrm{P}_{\Phi +\Phi ^{\mathfrak{b}_{+},\gamma _{+}}+\Phi ^{-\mathfrak{b}%
_{-},\gamma _{-}}}=\mathrm{P}_{(\Phi ,\mathfrak{a}_{\mathfrak{b}_{+}}-%
\mathfrak{a}_{\mathfrak{b}_{-}})}^{\flat }\ .
\end{equation*}%
\emph{(ii)}\ Convergence of equilibrium states: For all $\gamma _{-}\in
\left( 0,1\right) $, take any weak$^{\ast }$ accumulation point $\omega
_{\gamma _{-}}$ of any net $(\omega _{\gamma _{+},\gamma _{-}})_{\gamma
_{+}\in \left( 0,1\right) }\subseteq \mathit{M}_{\Phi +\Phi ^{\mathfrak{b}%
_{+},\gamma _{+}}+\Phi ^{-\mathfrak{b}_{-},\gamma _{-}}}$ as $\gamma
_{+}\rightarrow 0^{+}$. Pick any weak$^{\ast }$ accumulation point $\omega $
of the net $(\omega _{\gamma _{-}})_{\gamma _{-}\in \left( 0,1\right) }$, as 
$\gamma _{-}\rightarrow 0^{+}$. Then, $\omega \in \mathit{\Omega }_{(\Phi ,%
\mathfrak{a}_{\mathfrak{b}_{+}}-\mathfrak{a}_{\mathfrak{b}_{-}})}^{\flat }$.
\end{theorem}

\begin{proof}
Pick $\Phi \in \mathcal{W}_{1}^{\mathbb{R}}$ and two positive measures $%
\mathfrak{b}_{+}\in \mathcal{D}_{0,+}$ and $\mathfrak{b}_{-}\in \mathcal{C}%
_{0,+}$. Fix $\gamma _{-}\in \left( 0,1\right) $. By Theorem \ref{propuni
copy(5)}, 
\begin{equation}
\lim_{\gamma _{+}\rightarrow 0^{+}}\mathrm{P}_{\Phi +\Phi ^{\mathfrak{b}%
_{+},\gamma _{+}}+\Phi ^{-\mathfrak{b}_{-},\gamma _{-}}}=\mathrm{P}_{(\Phi
+\Phi ^{-\mathfrak{b}_{-},\gamma _{-}},\mathfrak{a}_{\mathfrak{b}%
_{+}})}^{\sharp }  \label{sdsdsdsdsdsdsdsd}
\end{equation}%
and weak$^{\ast }$ accumulation points of any net of equilibrium states $%
\omega _{\gamma _{-}}\in \mathit{M}_{\Phi +\Phi ^{\mathfrak{b}_{+},\gamma
_{+}}+\Phi ^{-\mathfrak{b}_{-},\gamma _{-}}}$ as $\gamma _{+}\rightarrow
0^{+}$ are generalized equilibrium states of $(\Phi +\Phi ^{-\mathfrak{b}%
_{-},\gamma _{-}},\mathfrak{a}_{\mathfrak{b}_{+}})$, i.e., they belong to
the weak$^{\ast }$-compact convex set $\mathit{\Omega }_{(\Phi +\Phi ^{-%
\mathfrak{b}_{-},\gamma _{-}},\mathfrak{a}_{\mathfrak{b}_{+}})}^{\sharp }$.
Observe that 
\begin{equation*}
\mathrm{P}_{(\Phi +\Phi ^{-\mathfrak{b}_{-},\gamma _{-}},\mathfrak{a}_{%
\mathfrak{b}_{+}})}^{\sharp }=-\inf h_{\gamma _{-}}\left( E_{1}\right)
\end{equation*}%
with $h_{\gamma _{-}}$ being the functional defined on $E_{1}$ by (\ref%
{function h}). By (\ref{function hlimit}), we deduce the first assertion.
Finally, for any $\gamma _{-}\in \left( 0,1\right) $, take any weak$^{\ast }$
accumulation point $\omega _{\gamma _{-}}$ of any net $(\omega _{\gamma
_{+},\gamma _{-}})_{\gamma _{+}\in \left( 0,1\right) }\subseteq \mathit{M}%
_{\Phi +\Phi ^{\mathfrak{b}_{+},\gamma _{+}}+\Phi ^{-\mathfrak{b}_{-},\gamma
_{-}}}$ as $\gamma _{+}\rightarrow 0^{+}$. Then, pick any weak$^{\ast }$
accumulation point $\omega $ of the net $(\omega _{\gamma _{-}})_{\gamma
_{-}\in \left( 0,1\right) }$, as $\gamma _{-}\rightarrow 0^{+}$. Similar to
the proof of Theorem \ref{propuni copy(6)} (ii), one shows that $\omega $
can be seen as a tangent functional at $\Phi \in \mathcal{W}_{1}^{\mathbb{R}%
} $ to the convex continuous function $\tilde{\Phi}\mapsto \mathrm{P}_{(%
\tilde{\Phi},\mathfrak{a}_{\mathfrak{b}_{+}}-\mathfrak{a}_{\mathfrak{b}%
_{-}})}^{\flat }$ from the real Banach space $\mathcal{W}_{1}^{\mathbb{R}}$
to $\mathbb{R}$. Then, by \cite[Theorems 1.4 and 1.8]{BruPedra-convex}, the
assertion follows, since $\mathit{\Omega }_{(\Phi ,\mathfrak{a}_{\mathfrak{b}%
_{+}}-\mathfrak{a}_{\mathfrak{b}_{-}})}^{\flat }$ is a weak$^{\ast }$%
-compact convex set.
\end{proof}

\begin{corollary}[From short-range interactions to non-conventional
mean-field models]
\label{propuni copy(14)}\mbox{ }\newline
Take an interaction $\Phi \in \mathcal{W}_{1}^{\mathbb{R}}$ and two positive
measures $\mathfrak{b}_{+}\in \mathcal{D}_{0,+}$ and $\mathfrak{b}_{-}\in 
\mathcal{C}_{0,+}$. There exist two sequences $(\gamma _{+,n})_{n\in \mathbb{%
N}}$ and $(\gamma _{-,n})_{n\in \mathbb{N}}$ of real numbers in the interval 
$\left( 0,1\right) $ converging to zero and a sequence $(\omega _{n})_{n\in 
\mathbb{N}}$ of equilibrium states $\omega _{n}\in \mathit{M}_{\Phi +\Phi ^{%
\mathfrak{b}_{+},\gamma _{+,n}}+\Phi ^{-\mathfrak{b}_{-},\gamma _{-,n}}}$
such that 
\begin{equation*}
\lim_{n\rightarrow \infty }\mathrm{P}_{\Phi +\Phi ^{\mathfrak{b}_{+},\gamma
_{+,n}}+\Phi ^{-\mathfrak{b}_{-},\gamma _{-,n}}}=\mathrm{P}_{(\Phi ,%
\mathfrak{a}_{\mathfrak{b}_{+}}-\mathfrak{a}_{\mathfrak{b}_{-}})}^{\flat }
\end{equation*}%
and $(\omega _{n})_{n\in \mathbb{N}}$ weak$^{\ast }$ converges to a
non-conventional equilibrium state of $\mathit{\Omega }_{(\Phi ,\mathfrak{a}%
_{\mathfrak{b}_{+}}-\mathfrak{a}_{\mathfrak{b}_{-}})}^{\flat }$.
\end{corollary}

\begin{proof}
Exactly like for Corollary \ref{propuni copy(13)}, one combines Theorem \ref%
{propuni copy(12)bis} with \cite[Chapter 2, 4 Theorem]{topology}.
\end{proof}

We conclude by showing that the limit of Kac pressures can attain \textbf{all%
} the values of the interval $[\mathrm{P}_{\mathfrak{m}}^{\sharp },\mathrm{P}%
_{\mathfrak{m}}^{\flat }]$ for the corresponding mean-field model $\mathfrak{%
m}\in \mathcal{M}_{1}$.

\begin{theorem}[{Reachability of all pressures in $[\mathrm{P}^{\sharp },%
\mathrm{P}^{\flat }]$}]
\label{propuni copy(12)}\mbox{ }\newline
Take an interaction $\Phi \in \mathcal{W}_{1}^{\mathbb{R}}$ and two positive
measures $\mathfrak{b}_{+}\in \mathcal{D}_{0,+}$ and $\mathfrak{b}_{-}\in 
\mathcal{C}_{0,+}$. Let $\mathfrak{m}\doteq (\Phi ,\mathfrak{a}_{\mathfrak{b}%
_{+}}-\mathfrak{a}_{\mathfrak{b}_{-}})\in \mathcal{M}_{1}$. For any $\mathrm{%
p}\in \lbrack \mathrm{P}_{\mathfrak{m}}^{\sharp },\mathrm{P}_{\mathfrak{m}%
}^{\flat }]$, there are two sequences $(\gamma _{+,n})_{n\in \mathbb{N}}$
and $(\gamma _{-,n})_{n\in \mathbb{N}}$ of real numbers in the interval $%
\left( 0,1\right) $ converging to zero, such that 
\begin{equation*}
\lim_{n\rightarrow \infty }\mathrm{P}_{\Phi +\Phi ^{\mathfrak{b}_{+},\gamma
_{+,n}}+\Phi ^{-\mathfrak{b}_{-},\gamma _{-,n}}}=\mathrm{p}\ .
\end{equation*}
\end{theorem}

\begin{proof}
Fix all parameters of the theorem. By Corollaries \ref{propuni copy(13)} and %
\ref{propuni copy(14)}, without loss of generality, we can take $\mathrm{p}%
\in (\mathrm{P}_{\mathfrak{m}}^{\sharp },\mathrm{P}_{\mathfrak{m}}^{\flat })$
and, for any $n\in \mathbb{N}$, there exist real numbers $\gamma
_{+,n}^{\sharp },\gamma _{-,n}^{\sharp },\gamma _{+,n}^{\flat },\gamma
_{-,n}^{\flat }$ in the interval $\left( 0,1\right) $, converging to zero as 
$n\rightarrow \infty $, such that 
\begin{equation}
\mathrm{P}_{\Phi +\Phi ^{\mathfrak{b}_{+},\gamma _{+,n}^{\sharp }}+\Phi ^{-%
\mathfrak{b}_{-},\gamma _{-,n}^{\sharp }}}<\mathrm{p}<\mathrm{P}_{\Phi +\Phi
^{\mathfrak{b}_{+},\gamma _{+,n}^{\flat }}+\Phi ^{-\mathfrak{b}_{-},\gamma
_{-,n}^{\flat }}}\ ,\qquad n\in \mathbb{N}\ .  \label{sddsds}
\end{equation}%
Note from Lemma \ref{lemma Kac norm} (ii) and Lebesgue's dominated
convergence theorem that the mappings $\gamma \mapsto \Phi ^{\pm \mathfrak{b}%
_{\pm },\gamma }$ from $\left( 0,1\right) $ to $\mathcal{W}_{1}$ are
continuous. Hence, by continuity of the pressure $\Phi \mapsto \mathrm{P}%
_{\Phi }$ on $\mathcal{W}_{1}$ and Inequality (\ref{sddsds}), there are
sequences $(\gamma _{+,n})_{n\in \mathbb{N}}$ and $(\gamma _{-,n})_{n\in 
\mathbb{N}}$ of real numbers in the interval $\left( 0,1\right) $ converging
to zero such that 
\begin{equation*}
\mathrm{P}_{\Phi +\Phi ^{\mathfrak{b}_{+},\gamma _{+,n}}+\Phi ^{-\mathfrak{b}%
_{-},\gamma _{-,n}}}=\mathrm{p}\ ,\qquad n\in \mathbb{N}\ .
\end{equation*}
\end{proof}

\subsubsection{Relaxing Positive-Definiteness and Scaling-Monotonicity \label%
{General Case}}

The general results of Section \ref{section mixed} use two functions $%
f_{+}\in \mathfrak{D}_{0,+}$ and $f_{-}\in \mathfrak{C}_{0,+}$ in the simple
field case, or two positive measures $\mathfrak{b}_{+}\in \mathcal{D}_{0,+}$
and $\mathfrak{b}_{-}\in \mathcal{C}_{0,+}$ in the multiple-field case. The
objects $f_{+}$ and $\mathfrak{b}_{+}$ encode repulsions, while $f_{-}$ and $%
\mathfrak{b}_{-}$ represent attractions of fermions on the lattice $%
\mathfrak{L}$. Therefore, up to this point, the repulsions and attractions
are defined from positive definite, $2d$ times continuously differentiable,
reflexion-symmetric, real-valued functions on $\mathbb{R}^{d}$, whose
derivatives of order $l\leq 2d$ decay faster than $\left\vert x\right\vert
^{-(d+\varepsilon +l)}$, for some $\varepsilon >0$, as $\left\vert
x\right\vert \rightarrow \infty $. Note additionally that the attractive
part also requires the scaling-monotone property of Fourier transforms, as
defined by (\ref{secaling monotone}). In the present section we show that,
by only imposing sufficient regularity on general functions $f\in \mathfrak{D%
}_{0}$, the case of Section \ref{section mixed} is recovered and we arrive
at a fully general result for sufficiently regular interactions.

Heuristically, the main part of the argument is as follows: A (not
necessarily positive definite) function $f\in \mathfrak{D}_{0}$ can be
written as a difference $f=f_{+}-f_{-}$, where $f_{+}\in \mathfrak{D}_{0,+}$
and $f_{-}\in \mathfrak{C}_{0,+}$. In fact, one may add and subtract a
positive definite and scaling-monotone function $\xi _{f}\in \mathfrak{C}%
_{0,+}$ to $f$ in such a way that $f_{+}=f+\xi _{f}\in \mathfrak{D}_{0,+}$
and $f_{-}=\xi _{f}\in \mathfrak{C}_{0,+}$. A similar fact is true in the
multiple-field case. This argument can be rigorously carried out by adding
stronger regularity and asymptotic properties to functions defining Kac
interactions.\ In fact, it suffices to take $\kappa =3d+1$ in the
definitions (\ref{D fracepsilon}) and (\ref{norm definition function kac})
of $\mathfrak{D}_{\varepsilon ,\kappa }$ and its norm. Recall that $%
\mathfrak{D}_{\varepsilon ,3d+1}$ is a vector subspace of $\mathfrak{D}%
_{\varepsilon }\equiv \mathfrak{D}_{\varepsilon ,2d}$ and 
\begin{equation*}
\left\Vert f\right\Vert _{\mathfrak{D}_{\varepsilon ,2d}}\leq \left\Vert
f\right\Vert _{\mathfrak{D}_{\varepsilon ,3d+1}}\ ,\qquad f\in \mathfrak{D}%
_{\varepsilon ,3d+1}\ .
\end{equation*}

For any parameter $\varepsilon \in \mathbb{R}^{+}$ and function $f\in 
\mathfrak{D}_{\varepsilon ,3d+1}$,%
\begin{equation}
|\hat{f}\left( k\right) |\leq \kappa !\left\Vert f\right\Vert _{\mathfrak{D}%
_{\varepsilon ,\kappa }}\min \left\{ 1,\min_{j\in \left\{ 1,\ldots
,d\right\} }|k_{j}|^{-\kappa }\right\} \int_{\mathbb{R}^{d}}\left(
1+\left\vert x\right\vert \right) ^{-(d+\varepsilon )}\mathrm{d}^{d}x\leq
C_{\varepsilon }\left\Vert f\right\Vert _{\mathfrak{D}_{\varepsilon ,\kappa
}}\hat{\xi}\left( k\right)  \label{jkjkjkjk}
\end{equation}%
for some constant $C_{\varepsilon }\in \mathbb{R}^{+}$ depending only upon $%
\varepsilon $, where $\kappa =3d+1$ and%
\begin{equation*}
\hat{\xi}\left( k\right) \doteq \left( 1+\left\vert k\right\vert ^{2}\right)
^{-\frac{3d+1}{2}}\ ,\qquad k\in \mathbb{R}^{d}\ .
\end{equation*}%
Now, we define $\xi \in \bigcap_{\varepsilon \in \mathbb{R}^{+}}\mathfrak{D}%
_{\varepsilon ,2d}$ to be the function whose Fourier transform is $\hat{\xi}$%
. Note that $\xi \in \mathfrak{C}_{0,+}$, see Inequality (\ref{secaling
monotone}). As a consequence, we infer from (\ref{jkjkjkjk}) that, for any $%
\varepsilon \in \mathbb{R}^{+}$ and $f\in \mathfrak{D}_{\varepsilon ,3d+1}$, 
\begin{equation}
f_{+}\doteq f+C_{\varepsilon }\left\Vert f\right\Vert _{\mathfrak{D}%
_{\varepsilon ,3d+1}}\xi \in \mathfrak{D}_{0,+,2d}\qquad \text{and}\qquad
f_{-}\doteq C_{\varepsilon }\left\Vert f\right\Vert _{\mathfrak{D}%
_{\varepsilon ,3d+1}}\xi \in \mathfrak{C}_{0,+,2d}  \label{sdsdsdsdd}
\end{equation}%
for some constant $C_{\varepsilon }\in \mathbb{R}^{+}$ depending only upon $%
\varepsilon $. We now extend this argument to the multiple field case and
arrive at the following lemma:

\begin{lemma}[Positive decompositions of measures]
\label{lemmaextra}\mbox{ }\newline
For any fixed $\varepsilon \in \mathbb{R}^{+}$ and $\mathfrak{b}\in \mathcal{%
D}_{\varepsilon ,3d+1}$, 
\begin{equation*}
\mathfrak{Z}_{\mathfrak{b}}\doteq \left\{ (\mathfrak{\bar{b}}_{+},\mathfrak{%
\bar{b}}_{-})\in \mathcal{D}_{\varepsilon ,+,2d}\times \mathcal{C}%
_{\varepsilon ,+,2d}:\mathfrak{a}_{\mathfrak{\bar{b}}_{+}}-\mathfrak{a}_{%
\mathfrak{\bar{b}}_{-}}=\mathfrak{a}_{\mathfrak{b}},\text{ }\Phi ^{\mathfrak{%
b},\gamma }=\Phi ^{\mathfrak{\bar{b}}_{+}-\mathfrak{\bar{b}}_{-},\gamma
}\quad \text{for}\quad \gamma \in (0,1)\right\} \neq \emptyset \ .
\end{equation*}
\end{lemma}

\begin{proof}
Fix the parameter $\varepsilon \in \mathbb{R}^{+}$. We consider first the
case of a positive measure $\mathfrak{b}\in \mathcal{D}_{\varepsilon ,3d+1}$%
. Let $\mathfrak{d}:\mathbb{S}\times \mathbb{D}_{\varepsilon
,3d+1}\rightarrow \mathbb{R}_{0}^{+}$ be the (continuous, bounded) positive
valued function defined by 
\begin{equation*}
\mathfrak{d}(\Psi ,f)\doteq \left\Vert f+2C_{\varepsilon }\xi \right\Vert _{%
\mathfrak{D}_{\varepsilon ,2d}}\text{ },\qquad \Psi \in \mathbb{S},\ f\in 
\mathbb{D}_{\varepsilon ,3d+1}\ ,
\end{equation*}%
with the constant $C_{\varepsilon }\in \mathbb{R}^{+}$ defined by (\ref%
{jkjkjkjk}). This function is used to define the (positive)\ Borel measure $%
\mathfrak{db}$ of finite variation by%
\begin{equation*}
\mathfrak{db}(\mathcal{A})\doteq \int_{\mathcal{A}}\mathfrak{d}(\Psi ,f)%
\mathfrak{b}\left( \mathrm{d}(\Psi ,f)\right)
\end{equation*}%
for any Borel set $\mathcal{A}\subseteq \mathbb{S}\times \mathbb{D}%
_{\varepsilon ,3d+1}$. Let $\mathfrak{\bar{b}}_{+}\doteq \mathbf{\xi }_{\ast
}^{+}(\mathfrak{db)}$ be the pushforward of $\mathfrak{db}$ through the
(continuous) mapping $\mathbf{\xi }^{+}:\mathbb{S}\times \mathbb{D}%
_{\varepsilon ,3d+1}\rightarrow \mathbb{S}\times \mathbb{D}_{\varepsilon
,2d} $ defined by 
\begin{equation*}
\mathbf{\xi }^{+}\left( \Psi ,f\right) \doteq \left( \Psi ,\left\Vert
f+2C_{\varepsilon }\xi \right\Vert _{\mathfrak{D}_{\varepsilon
,2d}}^{-1}\left( f+2C_{\varepsilon }\xi \right) \right) \ ,\qquad \Psi \in 
\mathbb{S},\ f\in \mathbb{D}_{\varepsilon ,3d+1}\ .
\end{equation*}%
Remark that $f+2C_{\varepsilon }\xi \neq 0$ for every $f\in \mathbb{D}%
_{\varepsilon ,3d+1}$. Define further $\mathfrak{\bar{b}}_{-}\in \mathcal{C}%
_{\varepsilon ,+,2d}$ by 
\begin{equation*}
\mathfrak{\bar{b}}_{-}\doteq 2C_{\varepsilon }\left\Vert \xi \right\Vert _{%
\mathfrak{D}_{\varepsilon ,2d}}\mathbf{\xi }_{\ast }^{-}\mathfrak{b}\ ,
\end{equation*}%
where $\mathbf{\xi }_{\ast }^{-}\mathfrak{b}$ is the pushforward of $%
\mathfrak{b}$ through the (continuous) mapping $\mathbf{\xi }^{-}:\mathbb{S}%
\times \mathbb{D}_{\varepsilon ,3d+1}\rightarrow \mathbb{S}\times \mathbb{D}%
_{\varepsilon ,2d}$ defined by 
\begin{equation*}
\mathbf{\xi }^{-}\left( \Psi ,f\right) \doteq \left( \Psi ,\left\Vert \xi
\right\Vert _{\mathfrak{D}_{\varepsilon ,2d}}^{-1}\xi \right) \ ,\qquad \Psi
\in \mathbb{S},\ f\in \mathbb{D}_{\varepsilon ,3d+1}\ .
\end{equation*}%
It is easy to check that $(\mathfrak{\bar{b}}_{+},\mathfrak{\bar{b}}_{-})\in 
\mathfrak{Z}_{\mathfrak{b}}$. To prove the general case, remark that, for
any $\mathfrak{b}\in \mathcal{D}_{\varepsilon ,3d+1}$, one has $\mathfrak{a}%
_{\mathfrak{\breve{b}}}=-\mathfrak{a}_{\mathfrak{b}}$ and $\Phi ^{\mathfrak{%
\breve{b}},\gamma }=-\Phi ^{\mathfrak{b},\gamma }$ for any $\gamma \in (0,1)$%
, where $\mathfrak{\breve{b}}$ is the pushforward of $\mathfrak{b}$ under
the continuous mapping $(\Psi ,f)\mapsto (\Psi ,-f)$ from $\mathbb{S}\times 
\mathbb{D}_{\varepsilon }$ to itself. Hence, for a general (not necessarily
positive) $\mathfrak{b}\in \mathcal{D}_{\varepsilon ,3d+1}$, one may
consider any decomposition $\mathfrak{b}=\mathfrak{b}_{+}-\mathfrak{b}_{-}$
with $\mathfrak{b}_{+},\mathfrak{b}_{-}\in \mathcal{D}_{\varepsilon ,+,3d+1}$%
. (Take, for instance, the Jordan decomposition of $\mathfrak{b}$.) Then,
replace $\mathfrak{b}$ with the positive measure $\mathfrak{b}_{+}+\mathfrak{%
\breve{b}}_{-}$. This change in the measure $\mathfrak{b}$ does not affect
the corresponding objects $\mathfrak{a}_{\mathfrak{b}}$ and $\Phi ^{%
\mathfrak{b},\gamma }$ for any $\gamma \in (0,1)$. Recall that the mappings $%
\mathfrak{b}\mapsto \mathfrak{a}_{\mathfrak{b}}$\ and $\mathfrak{b}\mapsto
\Phi ^{\mathfrak{b},\gamma }$ at fixed $\gamma \in (0,1)$ are (real) linear.
\end{proof}

By means of Lemma \ref{lemmaextra}, one can apply \textbf{all} results of
Section \ref{section mixed} to Kac and mean-field models associated with
general elements $\mathfrak{b}\in \mathcal{D}_{0,3d+1}$ and get the
following result:

\begin{proposition}[From short-range interactions to mean-field models - I]
\label{propuni copy(17)}\mbox{ }\newline
Let $\Phi \in \mathcal{W}_{1}^{\mathbb{R}}$ and $\mathfrak{b}\in \mathcal{D}%
_{0,3d+1}$. Take any decomposition $(\mathfrak{\bar{b}}_{+},\mathfrak{\bar{b}%
}_{-})\in \mathfrak{Z}_{\mathfrak{b}}\neq \emptyset $ of $\mathfrak{b}$.
(See Lemma \ref{lemmaextra}.) Then, Proposition \ref{propuni copy(10)},
Theorems \ref{propuni copy(11)}, \ref{propuni copy(12)bis} and \ref{propuni
copy(12)}, as well as Corollaries \ref{propuni copy(13)} and \ref{propuni
copy(14)} hold true with $\mathfrak{b}_{\pm }=\mathfrak{\bar{b}}_{\pm }$ and
mean-field interaction $\mathfrak{a}_{\mathfrak{b}_{+}}-\mathfrak{a}_{%
\mathfrak{b}_{-}}=\mathfrak{a}_{\mathfrak{b}}$.
\end{proposition}

\begin{proof}
The proposition directly follows from Lemma \ref{lemmaextra}, along with
Proposition \ref{propuni copy(10)}, Theorems \ref{propuni copy(11)}, \ref%
{propuni copy(12)bis} and \ref{propuni copy(12)}, as well as Corollaries \ref%
{propuni copy(13)} and \ref{propuni copy(14)}, from Section \ref{section
mixed}.
\end{proof}

Remark that, by Proposition \ref{propuni copy(17)}, if the conventional and
non-conventional pressures of the mean-field model $\mathfrak{m}=(\Phi ,%
\mathfrak{a}_{\mathfrak{b}})$ are the same for some $\Phi \in \mathcal{W}%
_{1}^{\mathbb{R}}$ and $\mathfrak{b}\in \mathcal{D}_{0,3d+1}$, then we
obtain the convergence of the Kac pressure $\mathrm{P}_{\Phi +\Phi ^{%
\mathfrak{b},\gamma }}$ to the pressure associated with the corresponding
mean-field model. Mutatis mutandis for the convergence of equilibrium
states. More precisely, we have the following:

\begin{theorem}[From short-range interactions to mean-field models - II]
\label{propuni copy(15)}\mbox{ }\newline
Take any interaction $\Phi \in \mathcal{W}_{1}^{\mathbb{R}}$ and a positive
measure $\mathfrak{b}\in \mathcal{D}_{0}$, such that $\mathrm{P}_{(\Phi ,%
\mathfrak{a}_{\mathfrak{b}})}^{\sharp }=\mathrm{P}_{(\Phi ,\mathfrak{a}_{%
\mathfrak{b}})}^{\flat }$.\newline
\emph{(i)}\ Convergence of infinite-volume pressures: 
\begin{equation*}
\lim_{\gamma \rightarrow 0^{+}}\mathrm{P}_{\Phi +\Phi ^{\mathfrak{b},\gamma
}}=\mathrm{P}_{(\Phi ,\mathfrak{a}_{\mathfrak{b}})}^{\sharp }=\mathrm{P}%
_{(\Phi ,\mathfrak{a}_{\mathfrak{b}})}^{\flat }\ .
\end{equation*}%
\emph{(ii)}\ Convergence of equilibrium states: If $\mathrm{P}_{(\Psi ,%
\mathfrak{a}_{\mathfrak{b}})}^{\sharp }=\mathrm{P}_{(\Psi ,\mathfrak{a}_{%
\mathfrak{b}})}^{\flat }$ for all $\Psi \in \mathcal{W}_{1}^{\mathbb{R}}$,
then weak$^{\ast }$ accumulation points of any net of equilibrium states $%
\omega _{\gamma }\in \mathit{M}_{\Phi +\Phi ^{\mathfrak{b},\gamma }}$ as $%
\gamma \rightarrow 0^{+}$ are generalized equilibrium states of the
mean-field model $(\Phi ,\mathfrak{a}_{\mathfrak{b}})$, i.e., they belong to
the weak$^{\ast }$-compact convex set $\mathit{\Omega }_{(\Phi ,\mathfrak{a}%
_{\mathfrak{b}})}^{\sharp }=\mathit{\Omega }_{(\Phi ,\mathfrak{a}_{\mathfrak{%
b}})}^{\flat }$.
\end{theorem}

\begin{proof}
The first part of the theorem is a direct consequence of the part of Theorem %
\ref{propuni copy(17)} referring to Proposition \ref{propuni copy(10)} with $%
\gamma _{+}=\gamma _{-}=\gamma $. The second part follows from the first
one, via a simple adaptation of the proof of the second part of Theorem \ref%
{propuni copy(6)0}.
\end{proof}

Recall that $\mathrm{P}^{\sharp }$ and $\mathrm{P}^{\flat }$ are defined by (%
\ref{pressure long range})--(\ref{Free-energy density long range}) and (\ref%
{convex functional g_m})--(\ref{Pressure bemol}), respectively. As already
explained above, $\mathrm{P}_{(\Psi ,\mathfrak{a}_{\mathfrak{b}})}^{\sharp }=%
\mathrm{P}_{(\Psi ,\mathfrak{a}_{\mathfrak{b}})}^{\flat }$ for all $\Psi \in 
\mathcal{W}_{1}^{\mathbb{R}}$ whenever the mean-field interaction $\mathfrak{%
a}_{\mathfrak{b}}\in \mathcal{S}_{1}$ is purely attractive, or purely
repulsive. Thus, Theorem \ref{propuni copy(15)} is a stronger version of
Theorems \ref{propuni copy(5)} and \ref{propuni copy(6)}, because $\mathfrak{%
b}_{\pm }\in \mathcal{D}_{0}$ are not restricted anymore to be in $\mathcal{D%
}_{0,+}$, nor in $\mathcal{C}_{0,+}$, i.e., the requirement that functions
have to be of positive type and have scaling monotone Fourier transforms can
be relaxed. Recall also that a sufficient condition on mean-field models
that have non-trivial attractive and repulsive components for $\mathrm{P}%
^{\sharp }=\mathrm{P}^{\flat }$ to hold true is given by Lemma \ref{theorem
structure of omega copy(1)}.

\section{Illustration of the General Results: Two-Body Potentials\label%
{Illustration}}

For the reader's convenience, we illustrate our general abstract results in
a particular example, which however generalizes the model (\ref{SR}),
presented in the introduction for $\mathrm{S}=\{\uparrow ,\downarrow \}$, in
the sense that it also includes long-range attractions, as described in
Sections \ref{section attractive} and \ref{section general}. This is an
important generalization, because attractive interactions have always been
an issue in the Kac limit, and, what is more, the competition between
repulsions and attractions can lead to unconventional mean-field effective
models in the Kac limit (Section \ref{section mixed}). Note that this
section is based on the short paper \cite{Kacproceeding}, which we summarize
below.

\subsection{Short-Range Model with Two-Body Interaction Potentials}

Using the cubic boxes $\Lambda _{L}\doteq \{\mathbb{Z}\cap \left[ -L,L\right]
\}^{d}$ of volume $|\Lambda _{L}|$, we define the translation-invariant
local Hamiltonians 
\begin{equation}
H_{\Lambda _{L}}\left( \gamma _{-},\gamma _{+}\right) \doteq T_{\Lambda
_{L}}-H_{\Lambda _{L},-}+H_{\Lambda _{L},+}\ ,  \label{SRbis}
\end{equation}%
for two parameters $\gamma _{-},\gamma _{+}\in \left( 0,1\right) $ and
length $L\in \mathbb{N}_{0}$, where 
\begin{eqnarray*}
T_{\Lambda _{L}} &\doteq &\underset{x,y\in \Lambda _{L},\ \mathrm{s}\in
\{\uparrow ,\downarrow \}}{\sum }h\left( x-y\right) a_{x,\mathrm{s}}^{\ast
}a_{y,\mathrm{s}}\ , \\
H_{\Lambda _{L},-} &\doteq &\sum_{x,y\in \Lambda _{L}}\gamma
_{-}^{d}f_{-}\left( \gamma _{-}\left( x-y\right) \right) a_{y,\uparrow
}^{\ast }a_{y,\downarrow }^{\ast }a_{x,\downarrow }a_{x,\uparrow }\ , \\
H_{\Lambda _{L},+} &\doteq &\sum\limits_{x,y\in \Lambda _{L},\mathrm{s},%
\mathrm{t}\in \{\uparrow ,\downarrow \}}\gamma _{+}^{d}f_{+}\left( \gamma
_{+}\left( x-y\right) \right) a_{y,\mathrm{t}}^{\ast }a_{y,\mathrm{t}}a_{x,%
\mathrm{s}}^{\ast }a_{x,\mathrm{s}}\ .
\end{eqnarray*}%
For $\gamma _{-}=0$, the Hamiltonian is nothing else than the Hamiltonian $%
H_{L}^{\mathrm{SR}}$ defined by (\ref{SR}), presented in the introduction
for $\mathrm{S}=\{\uparrow ,\downarrow \}$. Compare also (\ref{SRbis}) with (%
\ref{N interaction2}) and (\ref{BCS interaction1}). Here, $h$ is some
reflection-symmetric real-valued function on $\mathbb{Z}^{d}$, i.e.,%
\begin{equation*}
h(-z)=h(z)\in \mathbb{R}\text{ },\qquad z\in \mathbb{Z}^{d}\ .
\end{equation*}%
It represents the kinetic part of the model. Usually, $h\left( x\right)
=v\left( \left\vert x\right\vert \right) $ for some function $v:\mathbb{R}%
_{0}^{+}\rightarrow \mathbb{R}$. Instead, one could have used the condition%
\begin{equation*}
h(-z)=\overline{h(z)}\in \mathbb{C}\text{ },\qquad z\in \mathbb{Z}^{d},
\end{equation*}%
on the one-particle hopping strength with minor changes. This slightly more
general situation allows, for instance, for an external magnetic potential
in the model. For simplicity we stick to the real case and assume
additionally that $h$ is finitely supported. This encompasses the case of
the discrete Laplacian, which corresponds to the choice 
\begin{equation*}
h\left( z\right) =\left\{ 
\begin{array}{lll}
0 & \text{for} & \left\vert z\right\vert >1 \\ 
-1 & \text{for} & \left\vert z\right\vert =1 \\ 
2d & \text{for} & z=0%
\end{array}%
\right. ,\qquad z\in \mathbb{Z}^{d}\ .
\end{equation*}

As explained in all the paper, the reflection-symmetric real-valued function 
$f_{+}$ is a (non-zero) pair potential encoding interparticle forces, whose
range is tuned by the parameter $\gamma _{+}\in (0,1)$. The (non-zero)
reflection-symmetric real-valued function $f_{-}$ encodes the hopping
strength of Cooper pairs. This model thus implements a BCS interaction whose
range is tuned by the parameter $\gamma _{-}\in (0,1)$. Similar to the
one-particle hopping strength $h$, with minor modifications in order to
include external magnetic forces, one could have used a complex-valued
function $f_{-}$ satisfying%
\begin{equation*}
f_{-}(-z)=\overline{f_{-}(z)}\in \mathbb{C}\text{ },\qquad z\in \mathbb{Z}%
^{d}\ .
\end{equation*}%
Again for simplicity we stick to the real case. As is usual in theoretical
physics, $f_{-},f_{+}$ are assumed to be fast decaying and positive
definite, i.e., the Fourier\ transforms $\hat{f}_{-},\hat{f}_{+}$ of $%
f_{-},f_{+}$ are positive functions on $\mathbb{R}^{d}$. This choice for $%
f_{+}$ is reminiscent of a superstability condition, which is essential in
the bosonic case \cite[Section 2.2 and Appendix G]{BruZagrebnov8}. For
simplicity, we assume that $f_{-},f_{+}\in C_{0}^{2d}\left( \mathbb{R}^{d},%
\mathbb{R}\right) $, that is, they are both compactly supported and
sufficiently regular. For technical reasons, we also assume that the Fourier
transform of the Cooper pair hopping strength is scaling monotone, in the
sense of (\ref{secaling monotone}), that is,%
\begin{equation*}
\hat{f}_{-}(\gamma ^{-1}k)\leq \hat{f}_{-}\left( k\right) \ ,\qquad k\in 
\mathbb{R}^{d}\ ,\text{ }\gamma \left( 0,1\right) \text{ }.
\end{equation*}%
All these properties are already explained in Section \ref{sect banach refl
inv functions}, see in particular Equations (\ref{D fracepsilon}) and (\ref%
{D frac0}) as well as the explicit examples given there (like the
Yukawa-type potential).

By (\ref{pressure free energy})--(\ref{map free energy}), the pressure of
the model can be represented in the thermodynamic limit as a variational
problem over translation-invariant states: First, for $\gamma _{-},\gamma
_{+}\in \left( 0,1\right) $, the energy density functional 
\begin{equation*}
\mathfrak{e}(\gamma _{-},\gamma _{+}):E_{1}\rightarrow \mathbb{R}
\end{equation*}%
is defined by 
\begin{equation*}
\mathfrak{e}(\gamma _{-},\gamma _{+})\left( \rho \right) \doteq
\lim\limits_{L\rightarrow \infty }\frac{1}{\left\vert \Lambda
_{L}\right\vert }\rho \left( H_{\Lambda _{L}}\left( \gamma _{-},\gamma
_{+}\right) \right) <\infty
\end{equation*}%
for any translation-invariant state $\rho \in E_{1}$. By (\ref{ssssssssss}%
)--(\ref{eq:enpersite}), this limit exists and it can be split into three
parts:\ 
\begin{equation}
\mathfrak{e}(\gamma _{-},\gamma _{+})=\underset{\text{repulsive interaction
term }(+)}{\underbrace{\mathfrak{e}_{+}}}-\underset{\text{attractive
interaction term }(-)}{\underbrace{\mathfrak{e}_{-}}}+\underset{\text{%
kinetic term}}{\underbrace{\mathfrak{e}_{0}}},  \label{energy density0}
\end{equation}%
where, for any translation-invariant state $\rho \in E_{1}$, 
\begin{equation}
\mathfrak{e}_{\pm }\left( \rho \right) \doteq \lim\limits_{L\rightarrow
\infty }\frac{1}{\left\vert \Lambda _{L}\right\vert }\rho \left( H_{\Lambda
_{L},\pm }\right) <\infty \qquad \text{and}\qquad \mathfrak{e}_{0}\left(
\rho \right) \doteq \lim\limits_{L\rightarrow \infty }\frac{1}{\left\vert
\Lambda _{L}\right\vert }\rho \left( T_{\Lambda _{L}}\right) <\infty .
\label{energy density}
\end{equation}

Then, by (\ref{map free energy}), using the specific notation of this
section, for any inverse temperature $\beta \in (0,\infty )$ and $\gamma
_{-},\gamma _{+}\in \left( 0,1\right) $, the free energy density functional $%
\mathfrak{f}_{\beta }(\gamma _{-},\gamma _{+}):E_{1}\rightarrow \mathbb{R}$
is defined by%
\begin{equation}
\mathfrak{f}_{\beta }(\gamma _{-},\gamma _{+})\doteq \mathfrak{e}(\gamma
_{-},\gamma _{+})-\beta ^{-1}\mathfrak{s}\text{ },  \label{sdssdd00}
\end{equation}%
where $\mathfrak{s}:E_{1}\rightarrow \mathbb{R}_{0}^{+}$ is the entropy
density functional defined as the thermodynamic limit (\ref{entropy density}%
) of the von Neumann entropy per unit volume. With these definitions, the
thermodynamic limit of the (grand-canonical) pressure equals 
\begin{equation}
P_{\beta }\left( \gamma _{-},\gamma _{+}\right) \doteq \underset{%
L\rightarrow \infty }{\lim }\frac{1}{\beta |\Lambda _{L}|}\ln \mathrm{Tr}(%
\mathrm{e}^{-\beta H_{\Lambda _{L}}\left( \gamma _{-},\gamma _{+}\right)
})=-\inf \mathfrak{f}_{\beta }(\gamma _{-},\gamma _{+})\left( E_{1}\right)
<\infty  \label{sdssdd01}
\end{equation}%
for any inverse temperature $\beta \in (0,\infty )$ and parameters $\gamma
_{-},\gamma _{+}\in \left( 0,1\right) $. See Equation (\ref{pressure free
energy})--(\ref{map free energy}). The free energy density functional $%
\mathfrak{f}_{\beta }(\gamma _{-},\gamma _{+})$ is weak$^{\ast }$-lower
semicontinuous and the (space homogeneous) infinite volume equilibrium
states of the short-range model are defined as being the minimizers of this
functional. They form thus the set (\ref{definition equilibirum state}),
which is denoted in this particular case by 
\begin{equation*}
\mathit{\Omega }_{\beta }\left( \gamma _{-},\gamma _{+}\right) \doteq
\left\{ \omega \in E_{1}:\mathfrak{f}_{\beta }(\gamma _{-},\gamma
_{+})\left( \omega \right) =-P_{\beta }\left( \gamma _{-},\gamma _{+}\right)
\right\}
\end{equation*}%
for any $\beta \in (0,\infty )$ and $\gamma _{-},\gamma _{+}\in \left(
0,1\right) $.

As explained in Section \ref{Translation Invariant Equilibrium States
copy(1)}, $\mathit{\Omega }_{\beta }(\gamma _{-},\gamma _{+})$ is a
(non-empty) convex weak$^{\ast }$-compact subset of the space $E_{1}$ of
translation-invariant states. It can be directly related to the limit of
Gibbs states associated with the local Hamiltonians $H_{\Lambda _{L}}(\gamma
_{-},\gamma _{+})$, $L\in \mathbb{N}_{0}$: The set of all states on $%
\mathcal{U}$ being weak$^{\ast }$-compact, the sequence of Gibbs states of
the local Hamiltonians, seen as periodic states on $\mathcal{U}$, has weak$%
^{\ast }$-convergent subsequences. However, it is not clear that such limits
always belong to the set $E_{1}$ of translation-invariant states. In fact,
if a weak$^{\ast }$-convergent sequence of Gibbs states has a
translation-invariant state $\omega $ as its limit, then the state $\omega $
must belong to $\mathit{\Omega }_{\beta }(\gamma _{-},\gamma _{+})$. This
condition can be ensured by imposing periodic boundary conditions, as
explained in \cite[Chapter 3]{BruPedra2}. In particular, in this case, the
weak$^{\ast }$-accumulation points of Gibbs states are equilibrium states,
i.e., minimizers of the free energy density functional. See \cite[Theorem
3.13]{BruPedra2}.

\subsection{Mean-Field Approximations}

Recall now that the Kac, or van der Waals, limit refers to the limits $%
\gamma _{\pm }\rightarrow 0^{+}$ of the short-range model that is already in
infinite volume, i.e., $\gamma _{\pm }\rightarrow 0^{+}$ after the
thermodynamic limit $L\rightarrow \infty $. For small parameters $\gamma
_{\pm }\ll 1$, the short-range model defined in finite volume by (\ref{SRbis}%
) has interparticle ($+$) and BCS ($-$) interactions whose ranges ($\mathcal{%
O}(\gamma _{\pm }^{-1})$) are very large as compared to lattice constant
(here, $\mathcal{O}(1)$), but the interaction strength is small as $\gamma
_{\pm }^{d}$, in such a way that the first Born approximation\footnote{%
I.e., $\int_{\mathbb{R}^{d}}\gamma _{\pm }^{d}f_{\pm }\left( \gamma _{\pm
}x\right) \mathrm{d}x=\int_{\mathbb{R}^{d}}f_{\pm }\left( x\right) \mathrm{d}%
x\doteq \hat{f}_{\pm }(0)$.} to the scattering lengths of the interparticle
and BCS potentials remains constant, as is usual. One therefore expects to
have some effective mean-field, or long-range, model in the limits $\gamma
_{\pm }\rightarrow 0^{+}$.

The effective local Hamiltonians for the long-range limit of the above
short-range model should be 
\begin{equation}
H_{\Lambda _{L}}^{\sharp }\left( \eta _{-},\eta _{+}\right) \doteq
T_{\Lambda _{L}}+\underset{\text{mean-field repulsion }(+)}{\underbrace{%
\frac{\eta _{+}}{\left\vert \Lambda _{L}\right\vert }\sum\limits_{x,y\in
\Lambda _{L},\mathrm{s},\mathrm{t}\in \{\uparrow ,\downarrow \}}a_{y,\mathrm{%
t}}^{\ast }a_{y,\mathrm{t}}a_{x,\mathrm{s}}^{\ast }a_{x,\mathrm{s}}}}-%
\underset{\text{mean-field attraction }(-)}{\underbrace{\frac{\eta _{-}}{%
\left\vert \Lambda _{L}\right\vert }\sum_{x,y\in \Lambda _{L}}a_{y,\uparrow
}^{\ast }a_{y,\downarrow }^{\ast }a_{x,\downarrow }a_{x,\uparrow }}}
\label{MF}
\end{equation}%
for all $L\in \mathbb{N}_{0}$ and some positive parameters $\eta _{-},\eta
_{+}\in \mathbb{R}_{0}^{+}$. Compare these local Hamiltonians with (\ref{MF0}%
) and (\ref{SRbis}). They refer to a mean-field model, as defined in Section %
\ref{Long-Range Models}. See, e.g., (\ref{dddfdfdf}) and (\ref%
{dddfdfdfdddfdfdf}).

As a consequence, we can define from (\ref{space averaging}) the
space-averaging functionals associated with the mean-field repulsions ($+$)
and attractions ($-$), which are respectively denoted by $\Delta _{\pm
}:E_{1}\rightarrow \mathbb{R}$ and equals 
\begin{equation}
\Delta _{\pm }\left( \rho \right) \doteq \lim\limits_{L\rightarrow \infty }%
\frac{1}{|\Lambda _{L}|^{2}}\sum\limits_{x,y\in \Lambda _{L}}\rho \left(
\alpha _{y}\left( A_{\pm }^{\ast }\right) \alpha _{x}\left( A_{\pm }\right)
\right) \in \left[ |\rho (A_{\pm })|^{2},\Vert A_{\pm }\Vert _{\mathcal{U}%
}^{2}\right] \ ,  \label{sdssdsd}
\end{equation}%
for any translation-invariant state $\rho \in E_{1}$, where 
\begin{equation}
A_{-}\doteq a_{0,\downarrow }a_{0,\uparrow }\qquad \text{and}\qquad
A_{+}\doteq a_{0,\uparrow }^{\ast }a_{0,\uparrow }+a_{0,\downarrow }^{\ast
}a_{0,\downarrow }\ .  \label{sdssdsd1}
\end{equation}%
Compare this functional with Equations (\ref{Limit of Space-Averages})--(\ref%
{Ergodicity2}) and recall the ergodic property of extreme points of the
convex weak$^{\ast }$-compact space $E_{1}$ of translation-invariant states.

For any inverse temperature $\beta \in (0,\infty )$ and $\eta _{-},\eta
_{+}\in \mathbb{R}_{0}^{+}$, the (conventional) free energy density
functional of the mean-field model, as defined by (\ref{Free-energy density
long range}), is well-defined and denoted here by 
\begin{equation}
\mathfrak{f}_{\beta }^{\sharp }(\eta _{-},\eta _{+})\doteq \eta _{+}\Delta
_{+}-\eta _{-}\Delta _{-}+\mathfrak{e}_{0}-\beta ^{-1}\mathfrak{s}\ .
\label{sdssdd0}
\end{equation}%
Compare this definition with Equations (\ref{energy density0})--(\ref%
{sdssdd00}), the corresponding one for the short-range model. By (\ref%
{pressure long range})--(\ref{Free-energy density long range}), the
thermodynamic limit of the (grand-canonical) pressure of the mean-field
model equals 
\begin{equation}
P_{\beta }^{\sharp }\left( \eta _{-},\eta _{+}\right) \doteq \underset{%
L\rightarrow \infty }{\lim }\frac{1}{\beta |\Lambda _{L}|}\ln \mathrm{Tr}(%
\mathrm{e}^{-\beta H_{\Lambda _{L}}^{\sharp }\left( \eta _{-},\eta
_{+}\right) })=-\inf \mathfrak{f}_{\beta }^{\sharp }(\eta _{-},\eta
_{+})\left( E_{1}\right) <\infty  \label{sdssddbis}
\end{equation}%
for any inverse temperature $\beta \in (0,\infty )$ and parameters $\eta
_{-},\eta _{+}\in \mathbb{R}_{0}^{+}$.

As explained in Section \ref{Translation Invariant Equilibrium States
copy(2)}, the free energy density functional $\mathfrak{f}_{\beta }^{\sharp
}(\eta _{-},\eta _{+})$ is generally not weak$^{\ast }$-lower semicontinuous
on the convex weak$^{\ast }$-compact space $E_{1}$ of translation-invariant
states because of repulsive mean-field interactions. See Equation (\ref%
{sdsdsdssd}), that is here, 
\begin{equation*}
\mathfrak{f}_{\beta }^{\sharp }(\eta _{-},\eta _{+})=\underset{\text{upper
semicont.}}{\underbrace{\eta _{+}\Delta _{+}}}\underset{\text{lower semicont.%
}}{+\quad \underbrace{\left( -\eta _{-}\Delta _{-}+\mathfrak{e}_{0}-\beta
^{-1}\mathfrak{s}\right) }}\ .
\end{equation*}%
In particular, in contrast with the short-range model, this functional does
not necessarily have minimizers and this leads to the definition of
(conventional) equilibrium states given via (\ref{definition equilibirum
state}), that is in this special case, 
\begin{multline}
\mathit{\Omega }_{\beta }^{\sharp }(\eta _{-},\eta _{+})\doteq 
%TCIMACRO{\TeXButton{\Big{\{}}{\Big{\{}}}%
%BeginExpansion
\Big{\{}%
%EndExpansion
\omega \in E_{1}:\exists (\rho _{n})_{n\in \mathbb{N}}\subseteq E_{1}\mathrm{%
\ }\text{weak}^{\ast }\text{ converging to}\ \omega \\
\text{so\ that\ }\lim_{n\rightarrow \infty }\mathfrak{f}_{\beta }^{\sharp
}(\eta _{-},\eta _{+})(\rho _{n})=-P_{\beta }^{\sharp }\left( \eta _{-},\eta
_{+}\right) 
%TCIMACRO{\TeXButton{\Big{\}}}{\Big{\}}}}%
%BeginExpansion
\Big{\}}%
%EndExpansion
\ .  \notag
\end{multline}

Again, this last set can be directly related to the limit of Gibbs states
associated with the local Hamiltonians $H_{\Lambda _{L}}^{\sharp }(\eta
_{-},\eta _{+})$, $L\in \mathbb{N}_{0}$: If a weak$^{\ast }$-convergent
subsequence of Gibbs states has a translation-invariant state $\omega $ as
its limit, then the state $\omega $ must belong to $\mathit{\Omega }_{\beta
}^{\sharp }(\eta _{-},\eta _{+})$. Exactly as in the short-range case, this
condition can be ensured by imposing periodic boundary conditions, as
explained in \cite[Chapter 3]{BruPedra2}. In particular, in this case, the
weak$^{\ast }$-accumulation points of Gibbs states are conventional
equilibrium states, i.e., minimizers of the conventional free energy density
functional defined by (\ref{sdssdd0}). See again \cite[Theorem 3.13]%
{BruPedra2}.

The set $\mathit{\Omega }_{\beta }^{\sharp }(\eta _{-},\eta _{+})$ gathers
all conventional equilibrium states, being associated with approximating
minimizers of the free energy density functional (\ref{sdssdd0}).
Non-conventional equilibrium states are instead minimizers of the
non-conventional free energy density functional (\ref{convex functional g_m}%
), which is in the present case denoted by 
\begin{equation*}
\mathfrak{f}_{\beta }^{\flat }\left( \eta _{-},\eta _{+}\right) \left( \rho
\right) \doteq \quad \underset{\text{convex cont.}}{\underbrace{\eta
_{+}\left\vert \rho (a_{0,\uparrow }^{\ast }a_{0,\uparrow }+a_{0,\downarrow
}^{\ast }a_{0,\downarrow })\right\vert ^{2}}}\underset{\text{affine lower
semicont.}}{\quad +\quad \underbrace{\left( -\eta _{-}\Delta _{-}\left( \rho
\right) +\mathfrak{e}_{0}\left( \rho \right) -\beta ^{-1}\mathfrak{s}\left(
\rho \right) \right) }}
\end{equation*}%
for any translation-invariant state $\rho \in E_{1}$. In particular, this
leads to the set (\ref{definition equilibirum state nonconventional}) of
non-conventional equilibrium states, that is in this case,%
\begin{equation*}
\mathit{\Omega }_{\beta }^{\flat }\left( \eta _{-},\eta _{+}\right) \doteq
\left\{ \omega \in E_{1}:\mathfrak{f}_{\beta }^{\flat }(\eta _{-},\eta
_{+})(\omega )=\inf \,\mathfrak{f}_{\beta }^{\flat }(\eta _{-},\eta
_{+})(E_{1})\doteq -P_{\beta }^{\flat }\left( \eta _{-},\eta _{+}\right)
\right\} \ ,
\end{equation*}%
where $P_{\beta }^{\flat }\left( \eta _{-},\eta _{+}\right) $ is nothing
else than the non-conventional pressure defined in the general case by
Equation (\ref{Pressure bemol}).

\subsection{Thermodynamic Game\label{Thermodynamic Game}}

The thermodynamic game explained in Section \ref{Section thermo game} can be
explicitly given in this case. Applying the concept of approximating
(self-adjoint, short-range) interactions (\ref{approx interaction}) and
Hamiltonians (\ref{approx hamil}) to the present example, we introduce
so-called approximating local short-range Hamiltonians for the mean-field
model: 
\begin{eqnarray}
\tilde{H}_{\Lambda _{L}}\left( \eta _{-},\eta _{+},c_{-},c_{+}\right)
&\doteq &T_{\Lambda _{L}}+\eta _{+}^{1/2}\left( \bar{c}_{+}+c_{+}\right)
\sum_{x\in \Lambda _{L},\mathrm{s}\in \{\uparrow ,\downarrow \}}a_{x,\mathrm{%
s}}^{\ast }a_{x,\mathrm{s}}  \label{qppr} \\
&&-\eta _{-}^{1/2}\sum_{x\in \Lambda _{L}}\left( \bar{c}_{-}a_{x,\uparrow
}^{\ast }a_{x,\downarrow }^{\ast }+c_{-}a_{x,\downarrow }a_{x,\uparrow
}\right)  \notag
\end{eqnarray}%
for $c_{-},c_{+}\in \mathbb{C}$, $L\in \mathbb{N}_{0}$ and $\eta _{-},\eta
_{+}\in \mathbb{R}_{0}^{+}$. Given an inverse temperature $\beta \in
(0,\infty )$, we define the function $\tilde{P}_{\beta }:\mathbb{C}%
^{2}\times (\mathbb{R}_{0}^{+})^{2}\rightarrow \mathbb{R}$ by the infinite
volume pressure 
\begin{equation*}
\text{$\tilde{P}_{\beta }\left( c_{-},c_{+},\eta _{+},\eta _{-}\right) $}%
\doteq \underset{L\rightarrow \infty }{\lim }\frac{1}{\beta |\Lambda _{L}|}%
\ln \mathrm{Tr}(\mathrm{e}^{-\beta \tilde{H}_{\Lambda _{L}}\left( \eta
_{-},\eta _{+},c_{-},c_{+}\right) })<\infty \ ,\quad c_{-},c_{+}\in \mathbb{C%
},\ \eta _{-},\eta _{+}\in \mathbb{R}_{0}^{+}\ ,
\end{equation*}%
which exists, thanks to (\ref{variational problem approx}). Then, for fixed $%
\beta \in (0,\infty )$ and $\eta _{-},\eta _{+}\in \mathbb{R}_{0}^{+}$, the
approximating free energy density, as defined by (\ref{approximating free
energy density functional}), is equal to%
\begin{equation*}
\mathfrak{h}_{\beta }\left( c_{-},c_{+}\right) \doteq -\left\vert
c_{+}\right\vert ^{2}+\left\vert c_{-}\right\vert ^{2}-\text{$\tilde{P}%
_{\beta }\left( c_{-},c_{+},\eta _{+},\eta _{-}\right) $}\ ,\qquad
c_{-},c_{+}\in \mathbb{C}\ .
\end{equation*}%
Given an inverse temperature $\beta \in (0,\infty )$ and fixed parameters $%
\eta _{-},\eta _{+}\in \mathbb{R}_{0}^{+}$, this function is seen as the
payoff function of a two-person zero-sum game, the thermodynamic game
associated with the mean-field model. See Section \ref{Section thermo game}
or \cite{Kacproceeding} for more pedagogical explanations. Note from (\ref%
{pressures}) that 
\begin{equation}
P_{\beta }^{\sharp }\left( \eta _{-},\eta _{+}\right) =-\inf_{c_{-}\in 
\mathbb{C}}\sup_{c_{+}\in \mathbb{C}}\mathfrak{h}_{\beta }\left(
c_{-},c_{+}\right) \qquad \text{and}\qquad P_{\beta }^{\flat }\left( \eta
_{-},\eta _{+}\right) =-\sup_{c_{+}\in \mathbb{C}}\inf_{c_{-}\in \mathbb{C}}%
\mathfrak{h}_{\beta }\left( c_{-},c_{+}\right) \ .  \label{pressure bis}
\end{equation}%
The $\sup $ and the $\inf $ of this last equation do not commute in general,
but a sufficient condition for the $\sup $ and $\inf $ to commute is given
by Lemma \ref{theorem structure of omega copy(1)}.

Notice that the approximating Hamiltonians (\ref{qppr}) are \textbf{quadratic%
} in the annihilation and creation operators. It can thus be diagonalized by
a so-called Bogoliubov transformation and the pressures $P_{\beta }^{\flat }$
and $P_{\beta }^{\sharp }$, as well as the payoff function $\mathfrak{h}%
_{\beta }$ of the thermodynamic game, can be analytically and numerically
studied. Additionally, the sets $\mathit{\Omega }_{\beta }^{\sharp }(\eta
_{-},\eta _{+})$ and $\mathit{\Omega }_{\beta }^{\flat }\left( \eta
_{-},\eta _{+}\right) $ of all equilibrium states can be \textbf{explicitly
determined}, thanks to Theorem \ref{theorem structure of omega}.

\subsection{The Kac Limit and Mean-Field Approximations\label{Kac-dffdfd}}

By Theorem \ref{propuni copy(1)}, the energy densities (\ref{energy density}%
) associated with the short-range attractions and repulsions converge
pointwise to a mean-field one associated with the elements $A_{\pm }$ (\ref%
{sdssdsd1}):\ 
\begin{equation}
\lim_{\gamma _{\pm }\rightarrow 0^{+}}\mathfrak{e}_{\pm }\left( \rho \right)
\doteq \lim_{\gamma _{\pm }\rightarrow 0^{+}}\lim\limits_{L\rightarrow
\infty }\frac{1}{\left\vert \Lambda _{L}\right\vert }\rho \left( H_{\Lambda
_{L},\pm }\right) =\hat{f}_{\pm }(0)\Delta _{\pm }\left( \rho \right) \doteq
\lim\limits_{L\rightarrow \infty }\frac{\hat{f}_{\pm }(0)}{|\Lambda _{L}|^{2}%
}\sum\limits_{x,y\in \Lambda _{L}}\rho \left( \alpha _{y}\left( A_{\pm
}^{\ast }\right) \alpha _{x}\left( A_{\pm }\right) \right)
\label{limit kac1}
\end{equation}%
for any translation-invariant state $\rho \in E_{1}$, where we have from (%
\ref{fourier transform}) that%
\begin{equation*}
\hat{f}_{\pm }\left( 0\right) \doteq \int_{\mathbb{R}^{d}}f\left( x\right) 
\mathrm{d}^{d}x\geq 0\ .
\end{equation*}%
Recall that $f_{-},f_{+}$ are assumed to be positive definite, i.e., the
Fourier\ transforms $\hat{f}_{-},\hat{f}_{+}$ of $f_{-},f_{+}$,
respectively, are positive functions on $\mathbb{R}^{d}$. Comparing (\ref%
{energy density0})--(\ref{sdssdd01}) and (\ref{sdssdd0})--(\ref{sdssddbis})
in light of (\ref{limit kac1}), it demonstrates that the parameters $\eta
_{-},\eta _{+}\in \mathbb{R}_{0}^{+}$ of the mean-field model to be taken in
the limit $\gamma _{\pm }\rightarrow 0^{+}$ of the short-range one are the
first Born approximation to the scattering length of the interparticle
potentials 
\begin{equation*}
\eta _{\pm }=\hat{f}_{\pm }(0)\in \mathbb{R}_{0}^{+}\ ,
\end{equation*}%
as expected of course. This is rigorously proven by Theorem \ref{propuni
copy(11)}, which, in the example presented here, refers to the following
statement:

\begin{theorem}[Conventional mean-field models]
\label{Kacthm1}\mbox{ }\newline
Fix an arbitrary reflection-symmetric\ finitely supported real-valued
function $\varepsilon $ on $\mathbb{Z}^{d}$. Let $f_{-},f_{+}\in
C_{0}^{2d}\left( \mathbb{R}^{d},\mathbb{R}\right) $ be reflection-symmetric,
positive definite functions on $\mathbb{R}^{d}$ with $\hat{f}_{-}(\gamma
^{-1}k)\leq \hat{f}_{-}\left( k\right) $ for $k\in \mathbb{R}^{d}$. Fix an
inverse temperature $\beta \in (0,\infty )$.

\begin{enumerate}
\item[i.)] Convergence of infinite volume pressures: 
\begin{equation*}
\lim_{\gamma _{+}\rightarrow 0^{+}}\lim_{\gamma _{-}\rightarrow
0^{+}}P_{\beta }\left( \gamma _{-},\gamma _{+}\right) =P_{\beta }^{\sharp
}\left( \hat{f}_{-}(0),\hat{f}_{+}(0)\right) \ .
\end{equation*}

\item[ii.)] Convergence of equilibrium states: For any $\gamma _{+}\in
\left( 0,1\right) $, take any weak$^{\ast }$ accumulation point $\omega
_{\gamma _{+}}$ of any net $(\omega _{\gamma _{-},\gamma _{+}})_{\gamma
_{-}\in \left( 0,1\right) }\subseteq \mathit{\Omega }_{\beta }(\gamma
_{-},\gamma _{+})$ as $\gamma _{-}\rightarrow 0^{+}$. Pick any weak$^{\ast }$
accumulation point $\omega $ of the net $(\omega _{\gamma _{+}})_{\gamma
_{+}\in \left( 0,1\right) }$, as $\gamma _{+}\rightarrow 0^{+}$. Then, 
\begin{equation*}
\omega _{\gamma _{-},\gamma _{+}}\underset{\text{weak}^{\ast },\gamma
_{-}\rightarrow 0^{+}}{\rightarrow }\omega _{\gamma _{+}}\underset{\text{weak%
}^{\ast },\gamma _{+}\rightarrow 0^{+}}{\rightarrow }\omega \in \mathit{%
\Omega }_{\beta }^{\sharp }(\hat{f}_{-}(0),\hat{f}_{+}(0))\ .
\end{equation*}
\end{enumerate}
\end{theorem}

Theorem \ref{Kacthm1} uses the particular order for the limits: first $%
\gamma _{-}\rightarrow 0^{+}$ and then $\gamma _{+}\rightarrow 0^{+}$,
meaning that the attractive range has to be much larger than the repulsive
one. In the opposite case, Theorem \ref{propuni copy(12)bis} shows that one
gets non-conventional pressures:\ 

\begin{theorem}[Non-conventional mean-field models]
\label{Kacthm1 copy(1)}\mbox{ }\newline
Fix an arbitrary reflection-symmetric finitely supported real-valued
function $\varepsilon $ on $\mathbb{Z}^{d}$. Let $f_{-},f_{+}\in
C_{0}^{2d}\left( \mathbb{R}^{d},\mathbb{R}\right) $ be reflection-symmetric,
positive definite functions on $\mathbb{R}^{d}$ with $\hat{f}_{-}(\gamma
^{-1}k)\leq \hat{f}_{-}\left( k\right) $ for $k\in \mathbb{R}^{d}$. Fix an
inverse temperature $\beta \in (0,\infty )$.

\begin{enumerate}
\item[i.)] Convergence of infinite-volume pressures: 
\begin{equation*}
\lim_{\gamma _{-}\rightarrow 0^{+}}\lim_{\gamma _{+}\rightarrow
0^{+}}P_{\beta }\left( \gamma _{-},\gamma _{+}\right) =P_{\beta }^{\flat
}\left( \hat{f}_{-}(0),\hat{f}_{+}(0)\right) \ .
\end{equation*}

\item[ii.)] Convergence of equilibrium states: For any $\gamma _{-}\in
\left( 0,1\right) $, take any weak$^{\ast }$ accumulation point $\omega
_{\gamma _{-}}$ of any net $(\omega _{\gamma _{-},\gamma _{+}})_{\gamma
_{+}\in \left( 0,1\right) }\subseteq \mathit{\Omega }_{\beta }(\gamma
_{-},\gamma _{+})$ as $\gamma _{+}\rightarrow 0^{+}$. Pick any weak$^{\ast }$
accumulation point $\omega $ of the net $(\omega _{\gamma _{-}})_{\gamma
_{-}\in \left( 0,1\right) }$, as $\gamma _{-}\rightarrow 0^{+}$. Then, 
\begin{equation*}
\omega _{\gamma _{-},\gamma _{+}}\underset{\text{weak}^{\ast },\gamma
_{+}\rightarrow 0^{+}}{\rightarrow }\omega _{\gamma _{-}}\underset{\text{weak%
}^{\ast },\gamma _{-}\rightarrow 0^{+}}{\rightarrow }\omega \in \mathit{%
\Omega }_{\beta }^{\flat }(\hat{f}_{-}(0),\hat{f}_{+}(0))\ .
\end{equation*}
\end{enumerate}
\end{theorem}

Recall\ that the $\sup $ and the $\inf $ in Equation (\ref{pressure bis}) do
not commute in general and we generally have 
\begin{equation*}
P_{\beta }^{\sharp }\left( \eta _{-},\eta _{+}\right) \neq P_{\beta }^{\flat
}\left( \eta _{-},\eta _{+}\right) \ .
\end{equation*}%
Hence, depending upon how the double limit $\gamma _{\pm }\rightarrow 0^{+}$
of the short-range model is taken, one can get an effective long-range
system that is different from the one described by the \emph{conventional}
mean-field model, which is the thermodynamic limit of the finite-volume
system described from local Hamiltonians (\ref{MF}). See also Theorem \ref%
{propuni copy(12)} showing that the Kac limit of pressures can attain 
\textbf{all} the values of the interval 
\begin{equation*}
\left[ P_{\beta }^{\sharp }(\hat{f}_{-}(0),\hat{f}_{+}(0)),P_{\beta }^{\flat
}(\hat{f}_{-}(0),\hat{f}_{+}(0))\right] \ ,
\end{equation*}%
by a convenient choice of the sequences $(\gamma _{+,n})_{n\in \mathbb{N}}$
and $(\gamma _{-,n})_{n\in \mathbb{N}}$.

As explained in Section \ref{section mixed} in the general case, the results
referring to the above examples are highly non-trivial: As expected, any Kac
or van der Waals limit leads to mean-field pressures and equilibrium states.
However, the limit mean-field model is not necessarily what one
traditionally guesses when one mixes repulsive and attractive long-range
components. In fact, it strongly depends upon the hierarchy of ranges
between attractive and repulsive interparticle forces. For instance, if the
range of repulsive forces is much larger than the range of the attractive
ones, then in the Kac limit for these forces one may get a mean-field model
that is unconventional. See Theorem \ref{Kacthm1 copy(1)}.

The observation that models in the Kac (or van der Waals) limit are not
equivalent to their expected mean-field approximation has already been
observed in the past, but in a different context. In 1987, de Smedt and
Zagrebnov studied a Bose gas in the continuum and in presence of two-body
interactions with positive Fourier transforms and whose range is tuned by
the parameter $\gamma =\gamma _{+}\in (0,1)$, exactly as in (\ref{SR}) or as
in (\ref{SRbis}) for $\gamma =\gamma _{+}$ and $\gamma _{-}=0$. To this
model, they added a rescaled external potential $V_{L}$ defined by $%
V_{L}(x)\doteq V(x/L)$, $V\in C^{\infty }\left( \mathbb{R}^{d},\mathbb{R}%
\right) $, where $L\in \mathbb{R}^{+}$ parametrizes the $d$-dimensional cube 
$\left[ 0,L\right] ^{d}$ where the Bose gas is enclosed. Indeed, the
Bose-Einstein condensation of the perfect Bose gas (i.e., without any Kac
limit) was previously shown by van den Berg and Lewis to be very sensitive
to scaled external fields, even in low dimensions and even if the rescaled
external potential looks at first glance rather negligible (being almost
constant) for large $L\gg 1$. For instance, a generalized Bose-Einstein
condensation of type II or III can be provoked \cite{vandenBerg} with such
rescaled external potentials. Using this setting, de Smedt and Zagrebnov 
\cite{Smedt-Zagrebnov} show that a discrepancy between the model in the Kac
limit and the expected mean-field gas, above a critical density (normally
marking the appearance of a Bose-Einstein condensation). In this case \cite%
{Smedt-Zagrebnov}, there is \textbf{no} combination of repulsive and
attractive interactions, but instead a perturbation with a (weak) scaled
external field. The latter refers to the scaled-external-field perturbation
method (used in \cite{Smedt-Zagrebnov} to the Kac limit), which has been
used in many situations.

\section{Appendix\label{Appendix}}

\subsection{Elementary Technical Results\label{Appendix copy(1)}}

In this section, we give a sequence of useful technical results, mainly on
summations of real-valued functions of the $d$-dimensional vector space $%
\mathbb{R}^{d}$, as in the first lemma:\ 

\begin{lemma}[Integral test for series estimates -- I]
\label{inttest}\mbox{ }\newline
Let $f$ be an arbitrary function $\mathbb{R}^{d}\rightarrow \mathbb{R}$ and $%
g:\mathbb{R}_{0}^{+}\rightarrow \mathbb{R}_{0}^{+}$ a monotonically
decreasing function such that, for any $x\in \mathbb{R}^{d}$,%
\begin{equation}
\left\vert f\left( x\right) \right\vert \leq g\left( \left\vert x\right\vert
\right) \qquad \text{and}\qquad \int_{0}^{\infty }g\left( r\right) r^{d-1}%
\mathrm{d}r<\infty \ .  \label{assumption}
\end{equation}%
Then, it follows that%
\begin{equation*}
\sum_{z\in \mathbb{Z}^{d}}\left\vert f\left( z\right) \right\vert \leq
g\left( 0\right) +\sum_{n=1}^{d}{\binom{d}{n}}\frac{2\pi ^{\frac{n}{2}}}{%
\Gamma (\frac{n}{2})}\int_{0}^{\infty }g\left( r\right) r^{n-1}\mathrm{d}%
r<\infty \ .
\end{equation*}
\end{lemma}

\begin{proof}
Note first that monotonic functions $\mathbb{R}_{0}^{+}\rightarrow \mathbb{R}%
_{0}^{+}$ are Borel measurable and thus $\int_{0}^{\infty }g\left( r\right)
r^{d-1}\mathrm{d}r$ is well-defined as the integral of a positive measurable
function, with respect to the Lebesgue integral. Fix all parameters of the
lemma. For any subset $\mathcal{I}\in 2^{\left\{ 1,\dots ,d\right\} }$ with $%
\left\vert \mathcal{I}\right\vert =n$ and $\mathcal{I}=\{j_{1},\dots
,j_{n}\} $ so that $j_{k}<j_{l}$ for $k<l$, we define the mapping $\xi _{%
\mathcal{I}}:\mathbb{Z}^{n}\rightarrow \mathbb{Z}^{d}$ by 
\begin{equation}
\xi _{\mathcal{I}}\left( z\right) \doteq \left( y_{1},\dots ,y_{d}\right)
,\qquad z=\left( z_{1},\dots ,z_{n}\right) \in \mathbb{Z}^{n}\ ,
\label{mappin idiot}
\end{equation}%
with $y_{j_{k}}\doteq z_{k}$ for any $k\in \left\{ 1,\dots ,n\right\} $ and $%
y_{j}=0$ otherwise. Then, the following equality holds true: 
\begin{equation}
\sum_{z\in \mathbb{Z}^{d}}\left\vert f\left( z\right) \right\vert
=\left\vert f\left( 0\right) \right\vert +\sum_{n=1}^{d}\sum_{\mathcal{I}\in
2^{\left\{ 1,\dots ,d\right\} }:\left\vert \mathcal{I}\right\vert =n}\
\sum_{z\in \mathbb{Z}^{n}\backslash \left\{ 0\right\} }\left\vert f\circ \xi
_{\mathcal{I}}\left( z\right) \right\vert \ .  \label{sdsdsdsdsdsd}
\end{equation}%
For any $n\in \left\{ 1,\dots ,d\right\} $, observe from (\ref{assumption})
that, obviously, 
\begin{equation*}
\left\vert f\circ \xi _{\mathcal{I}}\left( z\right) \right\vert \leq g\left(
\left\vert z\right\vert \right) \ ,\qquad z\in \mathbb{Z}^{n}\backslash
\left\{ 0\right\} \ ,
\end{equation*}%
and, since $g$ is monotonically decreasing, one obtains that 
\begin{equation}
\left\vert f\circ \xi _{\mathcal{I}}\left( z\right) \right\vert \leq g\left(
\left\vert z\right\vert \right) \leq g\left( \left\vert x\right\vert \right)
\ ,\qquad z\in \mathbb{Z}^{n}\backslash \left\{ 0\right\} ,\ x\in \mathcal{A}%
_{z}\subseteq \mathbb{R}^{n}\ ,  \label{cbncbnbvnvnvn}
\end{equation}%
by defining the subset 
\begin{equation*}
\mathcal{A}_{\left( z_{1},\dots ,z_{d}\right) }\doteq \mathcal{A}%
_{z_{1}}\times \cdots \times \mathcal{A}_{z_{n}}\qquad \text{with}\qquad 
\mathcal{A}_{z_{i}}\doteq \left\{ 
\begin{array}{ccc}
\left( z_{i}-1,z_{i}\right] & \text{if} & z_{i}>0\ . \\ 
\left[ z_{i},z_{i}+1\right) & \text{if} & z_{i}<0\ .%
\end{array}%
\right.
\end{equation*}%
By integrating (\ref{cbncbnbvnvnvn}) over $\mathcal{A}_{z}$, we deduce the
inequality 
\begin{equation*}
\left\vert f\circ \xi _{\mathcal{I}}\left( z\right) \right\vert \leq \int_{%
\mathcal{A}_{z}}g\left( \left\vert x\right\vert \right) \mathrm{d}%
x_{1}\cdots \mathrm{d}x_{n}\ ,\qquad z\in \mathbb{Z}^{n}\backslash \left\{
0\right\} \ .
\end{equation*}%
We incorporate this upper estimate in (\ref{sdsdsdsdsdsd}) to arrive at%
\begin{eqnarray*}
\sum_{z\in \mathbb{Z}^{d}}\left\vert f\left( z\right) \right\vert &\leq
&g\left( 0\right) +\sum_{n=1}^{d}\sum_{\mathcal{I}\in 2^{\left\{ 1,\dots
,d\right\} }:\left\vert \mathcal{I}\right\vert =n}\ \sum_{z\in \mathbb{Z}%
^{n}\backslash \left\{ 0\right\} }\int_{\mathcal{A}_{z}}g\left( \left\vert
x\right\vert \right) \mathrm{d}x_{1}\cdots \mathrm{d}x_{n} \\
&=&g\left( 0\right) +\sum_{n=1}^{d}\binom{d}{n}\int_{\mathbb{R}^{n}}g\left(
\left\vert x\right\vert \right) \mathrm{d}x_{1}\cdots \mathrm{d}x_{n}\ ,
\end{eqnarray*}%
from which the assertion follows.
\end{proof}

\begin{corollary}[Integral test for series estimates -- II]
\label{bound}\mbox{ }\newline
Under the conditions of Lemma \ref{inttest}, for any $\gamma \in \mathbb{R}%
^{+}$ and $a\in \mathbb{Z}^{d}$, the series%
\begin{equation*}
\sum_{z\in \mathbb{Z}^{d}}\gamma ^{d}f\left( \gamma z+a\right)
\end{equation*}%
is absolutely convergent. If $\gamma <1$ then 
\begin{equation*}
\sum_{z\in \mathbb{Z}^{d}}\left\vert \gamma ^{d}f\left( \gamma z+a\right)
\right\vert \leq M_{g}\ ,
\end{equation*}%
where $M_{g}$ is a constant that only depends upon $g$ and is additive and
homogeneous with respect to $g$.
\end{corollary}

\begin{proof}
Fix all parameters of the corollary. First, observe that, for any $\gamma
\in \mathbb{R}^{+}$ and $a\in \mathbb{Z}^{d}$, there is $b^{(a,\gamma )}\in 
\mathbb{R}^{d}$ such that $|b^{(a,\gamma )}|\leq \sqrt{d}/2$ and%
\begin{equation}
\sum_{z\in \mathbb{Z}^{d}}\gamma ^{d}f\left( \gamma z+a\right) =\sum_{z\in 
\mathbb{Z}^{d}}\gamma ^{d}f\left( \gamma (z+b^{(a,\gamma )})\right) \ .
\label{rtrtrtrtrtr}
\end{equation}%
Therefore, fix now $\gamma \in \mathbb{R}^{+}$ and $b\in \mathbb{R}^{d}$
such that $\left\vert b\right\vert \leq \sqrt{d}/2$. Define the functions%
\begin{equation}
h\left( x\right) \doteq \gamma ^{d}f\left( \gamma (x+b)\right) \ ,\quad x\in 
\mathbb{R}^{d},\quad \text{and}\quad m\left( r\right) \doteq \left\{ 
\begin{array}{lll}
\gamma ^{d}g(0) & \text{if} & 0\leq r\leq |b| \\ 
\gamma ^{d}g\left( \gamma \left\vert r-\left\vert b\right\vert \right\vert
\right) & \text{if} & r>|b|%
\end{array}%
\right. \ .  \label{sdsdsdsdsdsdsdsdsd}
\end{equation}%
Using the reverse triangle inequality and the monotonicity of the function $%
g $, we deduce from (\ref{assumption}) that%
\begin{equation*}
\left\vert h\left( x\right) \right\vert \leq \gamma ^{d}g\left( \gamma
\left\vert x+b\right\vert \right) \leq m\left( \left\vert x\right\vert
\right) \leq \gamma ^{d}g\left( 0\right) \ ,\qquad x\in \mathbb{R}^{d}\ ,
\end{equation*}%
as well as, for any $n\in \left\{ 1,\dots ,d\right\} $,%
\begin{align}
\int_{0}^{\infty }m\left( r\right) r^{n-1}\mathrm{d}r& \leq \gamma
^{d}\left( \left\vert b\right\vert ^{n}g\left( 0\right) +\int_{\left\vert
b\right\vert }^{\infty }g(\gamma \left( r-\left\vert b\right\vert \right)
)r^{n-1}\mathrm{d}r\right)  \notag \\
& =\gamma ^{d}\left( \left\vert b\right\vert ^{n}g\left( 0\right) +\gamma
^{-1}\int_{0}^{\infty }g(u)\left( \frac{u}{\gamma }+\left\vert b\right\vert
\right) ^{n-1}\mathrm{d}u\right)  \notag \\
& =\gamma ^{d}\left\vert b\right\vert ^{n}g\left( 0\right) +\sum_{k=0}^{n-1}{%
\binom{n-1}{k}}\left\vert b\right\vert ^{n-\left( k+1\right) }\gamma
^{d-\left( k+1\right) }\int_{0}^{\infty }g(u)u^{k}\mathrm{d}u  \notag \\
& \leq \gamma ^{d}\left( \frac{\sqrt{d}}{2}\right) ^{n}g\left( 0\right)
+\sum_{k=0}^{n-1}{\binom{n-1}{k}}\gamma ^{d-(k+1)}\left( \frac{\sqrt{d}}{2}%
\right) ^{n-(k+1)}\int_{0}^{\infty }g(u)u^{k}\mathrm{d}u<\infty \ .
\label{jhkhkjhk}
\end{align}%
We can thus invoke Lemma \ref{inttest} with the functions given by (\ref%
{sdsdsdsdsdsdsdsdsd}) to deduce that, for any $\gamma \in \mathbb{R}^{+}$
and $b\in \mathbb{R}^{d}$ so that $\left\vert b\right\vert \leq \sqrt{d}/2$,%
\begin{equation*}
\sum_{z\in \mathbb{Z}^{d}}\left\vert \gamma ^{d}f\left( \gamma \left(
z+b\right) \right) \right\vert \leq \gamma ^{d}g\left( 0\right)
+\sum_{n=1}^{d}{\binom{d}{n}}\frac{2\pi ^{\frac{n}{2}}}{\Gamma (\frac{n}{2})}%
\int_{0}^{\infty }m\left( r\right) r^{n-1}\mathrm{d}r<\infty \ .
\end{equation*}%
If $\gamma \in (0,1)$ then one checks from (\ref{jhkhkjhk}) that, for any $%
b\in \mathbb{R}^{d}$ so that $\left\vert b\right\vert \leq \sqrt{d}/2$,%
\begin{equation*}
\sum_{z\in \mathbb{Z}^{d}}\left\vert \gamma ^{d}f\left( \gamma \left(
z+b\right) \right) \right\vert \leq M_{g}
\end{equation*}%
with%
\begin{equation*}
M_{g}\doteq g\left( 0\right) +\sum_{n=1}^{d}{\binom{d}{n}}\frac{2\pi ^{\frac{%
n}{2}}}{\Gamma (\frac{n}{2})}\left( \left( \frac{\sqrt{d}}{2}\right)
^{n}g\left( 0\right) +\sum_{k=0}^{n-1}{\binom{n-1}{k}}\left( \frac{\sqrt{d}}{%
2}\right) ^{n-(k+1)}\int_{0}^{\infty }g(u)u^{k}\mathrm{d}u\right) \ .
\end{equation*}%
By Equation (\ref{rtrtrtrtrtr}), the corollary then follows.
\end{proof}

We recall now the so-called \emph{Poisson summation formula}. It is a
well-known identity in Fourier analysis. Its precise formulation, as we need
it here, is given in the next proposition and relates the summation of a
rescaled, continuously differentiable and integrable function $f$ to a
summation of the Fourier transform $\hat{f}$ of this function.

\begin{proposition}[Poisson summation formula]
\label{poissonsum}\mbox{ }\newline
Let $f\in C^{1}\left( \mathbb{R}^{d},\mathbb{C}\right) $ be a continuously
differentiable function satisfying 
\begin{equation}
\sup_{k\in \left\{ 0,\dots ,d\right\} }\sup_{x\in \mathbb{R}^{d}}\left\vert
\left( 1+\left\vert x\right\vert \right) ^{d+\varepsilon }\partial
_{x_{k}}f\left( x\right) \right\vert <\infty  \label{boundeds}
\end{equation}%
for some $\varepsilon \in \mathbb{R}^{+}$, with the convention $\partial
_{x_{0}}f=f$. Then, for any $\gamma \in \mathbb{R}^{+}$, 
\begin{equation*}
\sum_{z\in \mathbb{Z}^{d}}\gamma ^{d}f(\gamma a+\gamma z)=\sum_{z\in \mathbb{%
Z}^{d}}\hat{f}\left( 2\pi \gamma ^{-1}z\right) \mathrm{e}^{2\pi iz\cdot a}\ .
\end{equation*}
\end{proposition}

\begin{proof}
For the reader's convenience, we shortly give the proof of this well-known
identity. Fix all parameters of the proposition. By Inequality (\ref%
{boundeds}) and Lemma \ref{inttest}, one checks that all the series 
\begin{equation*}
\sum_{z\in \mathbb{Z}^{d}}\partial _{x_{k}}f\left( x+\gamma z\right) \
,\qquad k\in \left\{ 0,\dots ,d\right\} \ ,
\end{equation*}%
converge uniformly with respect to $x\in \mathbb{R}^{d}$. So, we can define
a continuously differentiable function $g\in C^{1}\left( \mathbb{R}^{d},%
\mathbb{C}\right) $ by the series%
\begin{equation}
g\left( x\right) =\sum_{z\in \mathbb{Z}^{d}}\gamma ^{d}f\left( x+\gamma
z\right) \ ,\qquad x\in \mathbb{R}^{d}\ .  \label{sdssd}
\end{equation}%
As previously used in (\ref{rtrtrtrtrtr}), this function is $\gamma $%
-periodic in the sense that, for any vector of the form $y=\gamma z\in 
\mathbb{R}^{d}$ for some $z\in \mathbb{Z}^{d}$, one has $g(x+y)=g(x)$ for
any $x\in \mathbb{R}^{d}$. Thus, this function can be written as a Fourier
series: 
\begin{equation}
g\left( x\right) =\sum_{z\in \mathbb{Z}^{d}}u_{z}\mathrm{e}^{2\pi \gamma
^{-1}iz\cdot x},\qquad x\in \mathbb{R}^{d}\ ,  \label{sdsdsdghghgh}
\end{equation}%
where, for any $z\in \mathbb{Z}^{d}$, 
\begin{equation}
u_{z}\doteq \gamma ^{-d}\int_{\Gamma }g\left( x\right) \mathrm{e}^{-2\pi
\gamma ^{-1}iz\cdot x}\mathrm{d}x\qquad \text{with}\qquad \Gamma \doteq %
\left[ -\frac{\gamma }{2},\frac{\gamma }{2}\right) ^{d}\ .  \label{fourseri}
\end{equation}%
Additionally, the Fourier coefficients $u_{z}$, $z\in \mathbb{Z}^{d}$, are
absolutely summable, for $g$ is continuously differentiable. Now, from (\ref%
{sdssd}) and (\ref{fourseri}) we compute that, for any $z\in \mathbb{Z}^{d}$%
, 
\begin{equation*}
u_{z}\doteq \sum_{y\in \mathbb{Z}^{d}}\int_{\Gamma }f\left( x+\gamma
y\right) \mathrm{e}^{-2\pi \gamma ^{-1}iz\cdot x}\mathrm{d}x=\int_{\mathbb{R}%
^{d}}f\left( u\right) \mathrm{e}^{-2\pi \gamma ^{-1}iz\cdot u}\mathrm{d}u=%
\hat{f}\left( 2\pi \gamma ^{-1}z\right) \ ,
\end{equation*}%
which together with (\ref{sdssd})--(\ref{sdsdsdghghgh}) for $x=\gamma a$
yields the proposition.
\end{proof}

We conclude the series of elementary results by giving two estimates on
summations of Fourier transforms of integrable functions.

\begin{lemma}[Summations of Fourier transforms of functions -- I]
\label{propsumlim}\mbox{ }\newline
For any $\varepsilon \in \mathbb{R}^{+}$, there is a constant $%
M_{\varepsilon }\in \mathbb{R}^{+}$ such that, for any $f\in \mathfrak{D}%
_{\varepsilon }$, 
\begin{equation*}
\sum_{k\in \mathbb{Z}^{d}\backslash \left\{ 0\right\} }|\hat{f}\left( \gamma
^{-1}k\right) |\leq \gamma ^{2}M_{\varepsilon }\left\Vert f\right\Vert _{%
\mathfrak{D}_{\varepsilon }}\ ,\qquad \gamma \in (0,1)\ .
\end{equation*}
\end{lemma}

\begin{proof}
Fix all parameters of the lemma. For any $k=(k_{1},\dots ,k_{d})\in \mathbb{Z%
}^{d}$ such that $k_{j_{1}},\dots ,k_{j_{n}}\neq 0$ for some subset $%
\mathcal{I}=\{j_{1},\dots ,j_{n}\}\subseteq \left\{ 1,\dots ,d\right\} $
with $\left\vert \mathcal{I}\right\vert =n$, one has that%
\begin{equation}
F\left( f\right) \left( k\right) \equiv \hat{f}\left( k\right) \doteq \int_{%
\mathbb{R}^{d}}f\left( x\right) \mathrm{e}^{-ik\cdot x}\mathrm{d}x=\frac{%
(-1)^{n}}{k_{j_{1}}^{2}\cdots k_{j_{n}}^{2}}F\left( \frac{\partial ^{2n}f}{%
\partial x_{j_{1}}^{2}\cdots \partial x_{j_{n}}^{2}}\right) \left( k\right)
\ ,  \label{sdssdd}
\end{equation}%
where the Fourier transform 
\begin{equation*}
F\left( \frac{\partial ^{2n}f}{\partial x_{j_{1}}^{2}\cdots \partial
x_{j_{n}}^{2}}\right)
\end{equation*}%
is well-defined and bounded, because all derivatives of $f$ of order up to $%
2d$ belong to $L^{1}(\mathbb{R}^{d})$, by assumption. Define 
\begin{equation}
C_{f}\doteq \left( 2d\right) !\left\Vert f\right\Vert _{\mathfrak{D}%
_{\varepsilon }}\int_{\mathbb{R}^{d}}\left( 1+\left\vert x\right\vert
\right) ^{-(d+\varepsilon )}\mathrm{d}^{d}x<\infty  \label{sdssdd2}
\end{equation}%
and note that%
\begin{equation*}
\max_{n\in \{1,\dots ,d\}}\max_{\{j_{1},\dots ,j_{n}\}\in 2^{\left\{ 1,\dots
,d\right\} }}\left\Vert F\left( \frac{\partial ^{2n}f}{\partial
x_{j_{1}}^{2}\cdots \partial x_{j_{n}}^{2}}\right) \right\Vert _{\infty
}\leq C_{f}\ .
\end{equation*}%
Similar to (\ref{sdsdsdsdsdsd}), observe that 
\begin{equation*}
\sum_{k\in \mathbb{Z}^{d}\backslash \left\{ 0\right\} }|\hat{f}\left( \gamma
^{-1}k\right) )|=\sum_{n=1}^{d}\sum_{\mathcal{I}\in 2^{\left\{ 1,\dots
,d\right\} }:\left\vert \mathcal{I}\right\vert =n}\ \sum_{k\in \mathbb{Z}%
^{n}\backslash \left\{ 0\right\} }|\hat{f}\circ \xi _{\mathcal{I}}\left(
\gamma ^{-1}k\right) |\ ,
\end{equation*}%
where $\xi _{\mathcal{I}}:\mathbb{Z}^{n}\rightarrow \mathbb{Z}^{d}$ is the
mapping defined by (\ref{mappin idiot}) for any subset $\mathcal{I}\in
2^{\left\{ 1,\dots ,d\right\} }$. Then, using (\ref{sdssdd}) and (\ref%
{sdssdd2}), we bound the last equality to get that, for any $\gamma \in 
\mathbb{R}^{+}$,%
\begin{eqnarray}
\sum_{k\in \mathbb{Z}^{d}\backslash \left\{ 0\right\} }|\hat{f}\left( \gamma
^{-1}k\right) )| &\leq &C_{f}\sum_{n=1}^{d}\sum_{\mathcal{I}\in 2^{\left\{
1,\dots ,d\right\} }:\left\vert \mathcal{I}\right\vert =n}\gamma
^{2n}\sum_{k=\left( k_{1},\ldots ,k_{n}\right) \in \mathbb{Z}^{n}\backslash
\left\{ 0\right\} }\frac{1}{k_{1}^{2}\cdots k_{n}^{2}}  \notag \\
&=&C_{f}\sum_{n=1}^{d}\binom{d}{n}\left( \gamma ^{2}\sum_{q\in \mathbb{Z}%
\backslash \left\{ 0\right\} }\frac{1}{q^{2}}\right) ^{n}=C_{f}\sum_{n=1}^{d}%
\binom{d}{n}\left( \frac{\gamma ^{2}\pi ^{2}}{3}\right) ^{n}\ .
\label{fgfgg}
\end{eqnarray}%
In particular, for any $\gamma \in (0,1)$, we obtain that 
\begin{equation*}
\sum_{k\in \mathbb{Z}^{d}\backslash \left\{ 0\right\} }|\hat{f}\left( \gamma
^{-1}k\right) )|\leq \gamma ^{2}M_{\varepsilon }\left\Vert f\right\Vert _{%
\mathfrak{D}_{\varepsilon }}\ ,
\end{equation*}%
where $M_{\varepsilon }\in \mathbb{R}^{+}$\ is the constant defined by 
\begin{equation*}
M_{\varepsilon }\doteq \left( 2d\right) !\int_{\mathbb{R}^{d}}\left(
1+\left\vert x\right\vert \right) ^{-(d+\varepsilon )}\mathrm{d}%
^{d}x\sum_{n=1}^{d}\binom{d}{n}\left( \frac{\pi ^{2}}{3}\right) ^{n}\ .
\end{equation*}
\end{proof}

\begin{lemma}[Summations of Fourier transforms of functions -- II]
\label{colsumlim}\mbox{ }\newline
For any $\varepsilon \in \mathbb{R}^{+}$, there is a constant $%
M_{\varepsilon }\in \mathbb{R}^{+}$ such that, for any $f\in \mathfrak{D}%
_{\varepsilon }$,%
\begin{equation*}
\sum_{z\in \mathbb{Z}^{d}\backslash \{0\}}|\hat{f}(\gamma ^{-1}(2\pi
z-\theta ))|\leq \gamma ^{2}M_{\varepsilon }\left\Vert f\right\Vert _{%
\mathfrak{D}_{\varepsilon }}\ ,\qquad \gamma \in (0,1),\ \theta \in \left(
-\pi ,\pi \right] ^{d}\ .
\end{equation*}
\end{lemma}

\begin{proof}
The proof is very similar to the one of Lemma \ref{propsumlim}. For
instance, in the same way one gets (\ref{fgfgg}), one checks that, for any $%
\gamma \in (0,1)$,%
\begin{equation*}
\sum_{z\in \mathbb{Z}^{d}\backslash \left\{ 0\right\} }|\hat{f}(\gamma
^{-1}(2\pi z-\theta ))|\leq \gamma ^{2}C_{f}\sum_{n=1}^{d}{\binom{d}{n}}%
\left( \sum_{q\in \mathbb{Z}\backslash \left\{ 0\right\} }\frac{1}{(2\pi
\left\vert q\right\vert -\pi )^{2}}\right) ^{n}
\end{equation*}%
with $C_{f}$ $\in \mathbb{R}^{+}$ being the constant (\ref{sdssdd2}).
\end{proof}

We conclude this subsection with a lemma, which is reminiscent of Dini's
theorem:

\begin{lemma}[Limiting variational problem for increasing functionals]
\label{colsumlim copy(1)}\mbox{ }\newline
Let $(f_{n})_{n\in \mathbb{N}}$ be a sequence of lower semicontinuous
functionals on a compact space $\mathcal{X}$ converging pointwise to some
functional $f$ on $\mathcal{X}$, such that, for any $\eta \in \mathbb{R}^{+}$%
, there exists $N\in \mathbb{N}$ satisfying $f_{m}\geq f_{n}-\eta $ for all
natural numbers $m,n\geq N$, $m\geq n$. Then,%
\begin{equation*}
\lim_{n\rightarrow \infty }\inf f_{n}\left( \mathcal{X}\right) =\inf f\left( 
\mathcal{X}\right) >-\infty \ .
\end{equation*}
\end{lemma}

\begin{proof}
Note that $\inf f_{n}\left( \mathcal{X}\right) >-\infty $ because $f_{n}$ is
lower semicontinuous on a compact set. Additionally, for any $\eta \in 
\mathbb{R}^{+}$, there is $N\in \mathbb{N}$ such that, for any natural
number $n\geq N$, $f\geq f_{n}-\eta $, which in turn implies that 
\begin{equation}
\inf f\left( \mathcal{X}\right) \geq \inf f_{n}\left( \mathcal{X}\right)
-\eta \geq \inf f_{N}\left( \mathcal{X}\right) -2\eta >-\infty \ .
\label{sdasddasdasd}
\end{equation}%
Thus the sequence $(\inf f_{n}\left( \mathcal{X}\right) )_{n\in \mathbb{N}}$
of real numbers is bounded and it suffices to prove that it has only one
accumulation point to prove its convergence. Take any subsequence $%
(f_{n_{k}})_{k\in \mathbb{N}}$ such that $(\inf f_{n_{k}}\left( \mathcal{X}%
\right) )_{k\in \mathbb{N}}$ converges to some accumulation point of the
whole sequence. Given $\eta \in \mathbb{R}^{+}$, by taking again a
subsequence of $(f_{n_{k}})_{k\in \mathbb{N}}$, we can assume\ without loss
of generality that 
\begin{equation*}
f_{n_{l}}\geq f_{n_{k}}-\frac{\eta }{2^{k}}
\end{equation*}%
for all natural numbers $l\geq k$. For any $l\in \left\{ 2,3,\ldots ,\infty
\right\} $, define the preimage 
\begin{equation*}
E_{l}\doteq f_{n_{l}}^{-1}\left( \mathcal{I}_{l}\right)
\end{equation*}%
of the open interval 
\begin{equation*}
\mathcal{I}_{l}\doteq \left( \inf f\left( \mathcal{X}\right)
-\sum_{j=1}^{l-1}\frac{\eta }{2^{j}},\infty \right) \ .
\end{equation*}%
Then, it follows that $E_{l}\subseteq E_{l+1}$, and since $(f_{n})_{n\in 
\mathbb{N}}$ converges pointwise to $f$, one has 
\begin{equation*}
\mathcal{X}=\bigcup_{j=1}^{\infty }E_{j}\ .
\end{equation*}%
Since $f_{n}$ is lower semicontinuous for any $n\in \mathbb{N}$, $%
(E_{j})_{j\in \mathbb{N}}$ is an open cover for $\mathcal{X}$, and by the
compactness of $\mathcal{X}$, there exists $q\in \mathbb{N}$ such that 
\begin{equation*}
\mathcal{X}=\bigcup_{j=1}^{q}E_{j}=E_{q}.
\end{equation*}%
Hence, for\ any $\eta \in \mathbb{R}^{+}$, there is some $q\in \mathbb{N}$
so that, for any natural number $j\geq q$, 
\begin{equation*}
\inf f_{n_{j}}\left( \mathcal{X}\right) \geq \inf f\left( \mathcal{X}\right)
-\eta \ .
\end{equation*}%
Combined with (\ref{sdasddasdasd}), we deduce that%
\begin{equation*}
\inf f\left( \mathcal{X}\right) +\eta \geq \lim_{j\rightarrow \infty }\inf
f_{n_{j}}\left( \mathcal{X}\right) \geq \inf f\left( \mathcal{X}\right)
-\eta \ ,
\end{equation*}%
for all subsequences $(f_{n_{k}})_{k\in \mathbb{N}}$ such that $(\inf
f_{n_{k}}\left( \mathcal{X}\right) )_{k\in \mathbb{N}}$ converges to some
accumulation point. Since $\eta \in \mathbb{R}^{+}$ is arbitrary, the
assertion follows.
\end{proof}

\subsection{Estimates on Kac Interactions}

In this section we give important estimates on the Kac interactions of
Definition \ref{definition Kac interaction}. We start by showing in the next
lemma that they are locally Lipschitz continuous:

\begin{lemma}[Locally Lipschitz continuity of the Kac function]
\label{lemma Kac norm}\mbox{ }\newline
\emph{(i)} For any fixed $\gamma \in (0,1)$, the range of the Kac function $%
\mathcal{K}_{\gamma }$ of Definition \ref{definition Kac interaction} is a
subspace of $\mathcal{W}_{1}^{\mathbb{R}}$ and, for every $\varepsilon \in 
\mathbb{R}^{+}$, there is a constant $D_{\varepsilon }\in \mathbb{R}^{+}$
such that 
\begin{equation*}
\left\Vert \mathcal{K}_{\gamma }\left( \Phi ,f\right) -\mathcal{K}_{\gamma
}\left( \Psi ,g\right) \right\Vert _{\mathcal{W}_{1}}\leq D_{\varepsilon
}\left( \left\Vert f\right\Vert _{\mathfrak{D}_{\varepsilon }}\left\Vert
\Phi -\Psi \right\Vert _{\mathcal{W}_{1}}\left( \left\Vert \Phi \right\Vert
_{\mathcal{W}_{1}}+\left\Vert \Psi \right\Vert _{\mathcal{W}_{1}}\right)
+\left\Vert f-g\right\Vert _{\mathfrak{D}_{\varepsilon }}\left\Vert \Psi
\right\Vert _{\mathcal{W}_{1}}^{2}\right)
\end{equation*}%
for any $\Phi ,\Psi \in \mathcal{W}_{1}$ and $f,g\in \mathfrak{D}%
_{\varepsilon }$. In particular, the mapping $\mathcal{K}_{\gamma }:\mathcal{%
W}_{1}\times \mathfrak{D}_{0}\rightarrow \mathcal{W}_{1}^{\mathbb{R}}$ is
locally Lipschitz continuous.\newline
\emph{(ii)} For any fixed $\varepsilon \in \mathbb{R}^{+}$, there is a
constant $\tilde{D}_{\varepsilon }\in \mathbb{R}^{+}$ such that, for any $%
f\in \mathfrak{D}_{\varepsilon }$ and $\Phi \in \mathcal{W}_{1}$,%
\begin{equation*}
\left\Vert \mathcal{K}_{\gamma _{1}}\left( \Phi ,f\right) -\mathcal{K}%
_{\gamma _{2}}\left( \Phi ,f\right) \right\Vert _{\mathcal{W}_{1}}\leq 
\tilde{D}_{\varepsilon }\left\Vert f\right\Vert _{\mathfrak{D}_{\varepsilon
}}\left\Vert \Phi \right\Vert _{\mathcal{W}_{1}}^{2}\left\vert \ln \frac{%
\gamma _{2}}{\gamma _{1}}\right\vert \ ,\qquad \gamma _{1},\gamma _{2}\in
\left( 0,1\right) \ .
\end{equation*}
\end{lemma}

\begin{proof}
\underline{(i):} In view of Corollary \ref{bound}, for any $\varepsilon \in 
\mathbb{R}^{+}$, there is a constant $M_{\varepsilon }$ such that, for any $%
f\in \mathfrak{D}_{\varepsilon }$, $\gamma \in (0,1)$ and $a\in \mathfrak{L}$%
, 
\begin{equation}
\sum_{z\in \mathfrak{L}}\gamma ^{d}\left\vert f\left( \gamma z+a\right)
\right\vert \leq M_{\varepsilon }\left\Vert f\right\Vert _{\mathfrak{D}%
_{\varepsilon }}\ .  \label{estimate 1}
\end{equation}%
In fact, choose $g(r)=\Vert f\Vert _{\mathfrak{D}_{\varepsilon
}}(1+r)^{-(d+\varepsilon )}$ in Corollary \ref{bound}, observing from (\ref%
{norm definition function kac}) that, for any $\varepsilon \in \mathbb{R}%
^{+} $ and $f\in \mathfrak{D}_{\varepsilon }$,%
\begin{equation*}
\sup_{x\in \mathbb{R}^{d}}\left\vert \left( 1+\left\vert x\right\vert
\right) ^{d+\varepsilon }f\left( x\right) \right\vert \leq \left\Vert
f\right\Vert _{\mathfrak{D}_{\varepsilon }}<\infty \ .
\end{equation*}%
Take now an interaction $\Phi \in \mathcal{W}_{1}$ and a function $f\in 
\mathfrak{D}_{\varepsilon }$ for a fixed $\varepsilon \in \mathbb{R}^{+}$.
Then, using the norm (\ref{iteration0}), the fact that $f$ is
reflection-symmetric and (\ref{estimate 1}), we compute from Definition \ref%
{definition Kac interaction} that%
\begin{eqnarray}
\left\Vert \mathcal{K}_{\gamma }\left( \Phi ,f\right) \right\Vert _{\mathcal{%
W}_{1}} &\leq &\sum_{\mathcal{Z}_{1},\mathcal{Z}_{2}\in \mathcal{P}_{\mathrm{%
f}}:\mathcal{Z}_{1}\cup \mathcal{Z}_{2}\supseteq \{0\}}\frac{\left\Vert \Phi
_{\mathcal{Z}_{1}}\right\Vert _{\mathcal{U}}\left\Vert \Phi _{\mathcal{Z}%
_{2}}\right\Vert _{\mathcal{U}}}{\left\vert \mathcal{Z}_{1}\right\vert
+\left\vert \mathcal{Z}_{2}\right\vert }\sum_{x\in \mathcal{Z}_{1},y\in 
\mathcal{Z}_{2}}\frac{\gamma ^{d}\left\vert f\left( \gamma \left( x-y\right)
\right) \right\vert }{\left\vert \mathcal{Z}_{1}\right\vert \left\vert 
\mathcal{Z}_{2}\right\vert }  \notag \\
&\leq &2\sum_{\mathcal{Z}_{1},\mathcal{Z}_{2}\in \mathcal{P}_{\mathrm{f}}:%
\mathcal{Z}_{1}\supseteq \{0\}}\frac{\left\Vert \Phi _{\mathcal{Z}%
_{1}}\right\Vert _{\mathcal{U}}\left\Vert \Phi _{\mathcal{Z}_{2}}\right\Vert
_{\mathcal{U}}}{\left\vert \mathcal{Z}_{1}\right\vert +\left\vert \mathcal{Z}%
_{2}\right\vert }\sum_{x\in \mathcal{Z}_{1},y\in \mathcal{Z}_{2}}\frac{%
\gamma ^{d}\left\vert f\left( \gamma \left( x-y\right) \right) \right\vert }{%
\left\vert \mathcal{Z}_{1}\right\vert \left\vert \mathcal{Z}_{2}\right\vert }
\label{fefef} \\
&\leq &2\sum_{\mathcal{Z}_{1}\in \mathcal{P}_{\mathrm{f}}:\mathcal{Z}%
_{1}\supseteq \{0\}}\frac{\left\Vert \Phi _{\mathcal{Z}_{1}}\right\Vert _{%
\mathcal{U}}}{\left\vert \mathcal{Z}_{1}\right\vert }\sum_{\mathcal{Z}%
_{2}\in \mathcal{P}_{\mathrm{f}}:\mathcal{Z}_{2}\supseteq \{0\}}\frac{%
\left\Vert \Phi _{\mathcal{Z}_{2}}\right\Vert _{\mathcal{U}}}{\left\vert 
\mathcal{Z}_{2}\right\vert }  \notag \\
&&\times \frac{1}{\left\vert \mathcal{Z}_{2}\right\vert \left\vert \mathcal{Z%
}_{1}\right\vert }\sum_{x\in \mathcal{Z}_{1},y\in \mathcal{Z}_{2}}\sum_{z\in 
\mathfrak{L}}\gamma ^{d}\left\vert f\left( \gamma \left( x-y-z\right)
\right) \right\vert  \label{fefefefefe} \\
&\leq &2M_{\varepsilon }\left\Vert f\right\Vert _{\mathfrak{D}_{\varepsilon
}}\left\Vert \Phi \right\Vert _{\mathcal{W}_{1}}^{2}\ .  \label{fefef2}
\end{eqnarray}%
In particular, the Kac function $\mathcal{K}_{\gamma }$ maps $\mathcal{W}%
_{1}\times \mathfrak{D}_{0}$ to $\mathcal{W}_{1}^{\mathbb{R}}$. Similarly,
for any $\Phi ,\Psi \in \mathcal{W}_{1}$ and $f\in \mathfrak{D}_{0}$, from
the inequality 
\begin{align*}
\left\Vert \mathcal{K}_{\gamma }\left( \Phi ,f\right) -\mathcal{K}_{\gamma
}\left( \Psi ,f\right) \right\Vert _{\mathcal{W}_{1}}& \leq \sum_{\mathcal{Z}%
_{1},\mathcal{Z}_{2}\in \mathcal{P}_{\mathrm{f}}:\mathcal{Z}_{1}\cup 
\mathcal{Z}_{2}\supseteq \{0\}}\frac{\left\Vert \Phi _{\mathcal{Z}_{1}}-\Psi
_{\mathcal{Z}_{1}}\right\Vert _{\mathcal{U}}\left\Vert \Phi _{\mathcal{Z}%
_{2}}\right\Vert _{\mathcal{U}}}{\left\vert \mathcal{Z}_{1}\right\vert
+\left\vert \mathcal{Z}_{2}\right\vert }\sum_{x\in \mathcal{Z}_{1},y\in 
\mathcal{Z}_{2}}\frac{\gamma ^{d}f\left( \gamma \left( x-y\right) \right) }{%
\left\vert \mathcal{Z}_{1}\right\vert \left\vert \mathcal{Z}_{2}\right\vert }
\\
& +\sum_{\mathcal{Z}_{1},\mathcal{Z}_{2}\in \mathcal{P}_{\mathrm{f}}:%
\mathcal{Z}_{1}\cup \mathcal{Z}_{2}\supseteq \{0\}}\frac{\left\Vert \Psi _{%
\mathcal{Z}_{1}}\right\Vert _{\mathcal{U}}\left\Vert \Phi _{\mathcal{Z}%
_{2}}-\Psi _{\mathcal{Z}_{2}}\right\Vert _{\mathcal{U}}}{\left\vert \mathcal{%
Z}_{1}\right\vert +\left\vert \mathcal{Z}_{2}\right\vert }\sum_{x\in 
\mathcal{Z}_{1},y\in \mathcal{Z}_{2}}\frac{\gamma ^{d}f\left( \gamma \left(
x-y\right) \right) }{\left\vert \mathcal{Z}_{1}\right\vert \left\vert 
\mathcal{Z}_{2}\right\vert }
\end{align*}%
and the same arguments used to prove (\ref{fefef2}), one gets that 
\begin{equation}
\left\Vert \mathcal{K}_{\gamma }\left( \Phi ,f\right) -\mathcal{K}_{\gamma
}\left( \Psi ,f\right) \right\Vert _{\mathcal{W}_{1}}\leq 2M_{\varepsilon
}\left\Vert f\right\Vert _{\mathfrak{D}_{\varepsilon }}\left\Vert \Phi -\Psi
\right\Vert _{\mathcal{W}_{1}}\left( \left\Vert \Phi \right\Vert _{\mathcal{W%
}_{1}}+\left\Vert \Psi \right\Vert _{\mathcal{W}_{1}}\right) \ .
\label{bounded}
\end{equation}%
Since the mapping $f\mapsto \mathcal{K}_{\gamma }\left( \Phi ,f\right) $ is
linear for any fixed $\gamma \in (0,1)$ and $\Phi \in \mathcal{W}_{1}$, the
first bound stated in the lemma then follows. This is done by using (\ref%
{bounded}) and the identity%
\begin{equation*}
\mathcal{K}_{\gamma }\left( \Phi ,f\right) -\mathcal{K}_{\gamma }\left( \Psi
,g\right) =\mathcal{K}_{\gamma }\left( \Phi ,f\right) -\mathcal{K}_{\gamma
}\left( \Psi ,f\right) +\mathcal{K}_{\gamma }\left( \Psi ,f-g\right)
\end{equation*}%
for any $\Phi ,\Psi \in \mathcal{W}_{1}$, $\varepsilon \in \mathbb{R}^{+}$
and $f,g\in \mathfrak{D}_{\varepsilon }$.

\noindent \underline{(ii):} By Definition \ref{definition Kac interaction},
for any $\varepsilon \in \mathbb{R}^{+}$, there is a constant $\tilde{D}%
_{\varepsilon }\in \mathbb{R}^{+}$ such that, for any $f\in \mathfrak{D}%
_{\varepsilon }$, $\Phi \in \mathcal{W}_{1}$ and $\gamma _{1},\gamma _{2}\in
\left( 0,1\right) $,%
\begin{eqnarray*}
&&\left\Vert \mathcal{K}_{\gamma _{1}}\left( \Phi ,f\right) -\mathcal{K}%
_{\gamma _{2}}\left( \Phi ,f\right) \right\Vert _{\mathcal{W}_{1}} \\
&\leq &2\sum_{\mathcal{Z}_{1}\in \mathcal{P}_{\mathrm{f}}:\mathcal{Z}%
_{1}\supseteq \{0\}}\frac{\left\Vert \Phi _{\mathcal{Z}_{1}}\right\Vert _{%
\mathcal{U}}}{\left\vert \mathcal{Z}_{1}\right\vert }\sum_{\mathcal{Z}%
_{2}\in \mathcal{P}_{\mathrm{f}}:\mathcal{Z}_{2}\supseteq \{0\}}\frac{%
\left\Vert \Phi _{\mathcal{Z}_{2}}\right\Vert _{\mathcal{U}}}{\left\vert 
\mathcal{Z}_{2}\right\vert }\frac{1}{\left\vert \mathcal{Z}_{2}\right\vert
\left\vert \mathcal{Z}_{1}\right\vert } \\
&&\times \sum_{x\in \mathcal{Z}_{1},y\in \mathcal{Z}_{2}}\sum_{z\in 
\mathfrak{L}}\int_{\gamma _{1}}^{\gamma _{2}}\gamma ^{d-1}\left\vert f\left(
\gamma \left( x-y-z\right) \right) d+\nabla f\left( \gamma \left(
x-y-z\right) \right) \cdot \left( \gamma \left( x-y-z\right) \right)
\right\vert \mathrm{d}\gamma \\
&\leq &\tilde{D}_{\varepsilon }\left\Vert f\right\Vert _{\mathfrak{D}%
_{\varepsilon }}\left\Vert \Phi \right\Vert _{\mathcal{W}_{1}}^{2}\left\vert
\ln \frac{\gamma _{2}}{\gamma _{1}}\right\vert \ ,
\end{eqnarray*}%
similar to Equations (\ref{fefef})--(\ref{fefefefefe}), observing that%
\begin{equation*}
\left\vert f\left( x\right) d+\nabla f\left( x\right) \cdot x\right\vert
\leq d\Vert f\Vert _{\mathfrak{D}_{\varepsilon }}(1+\left\vert x\right\vert
)^{-(d+\varepsilon )},\qquad x\in \mathbb{R}^{d}.
\end{equation*}
\end{proof}

Note that the bound (\ref{fefef2}) on the norm of the Kac interaction $%
\mathcal{K}_{\gamma }\left( \Phi ,f\right) $, referring to Lemma \ref{lemma
Kac norm} (i) can be refined, yielding the inequality 
\begin{equation}
\left\Vert \mathcal{K}_{\gamma }\left( \Phi ,f\right) \right\Vert _{\mathcal{%
W}_{1}}+\mathfrak{F}_{\gamma }\left( \Phi ,f\right) \leq 2M_{\varepsilon
}\left\Vert f\right\Vert _{\mathfrak{D}_{\varepsilon }}\left\Vert \Phi
\right\Vert _{\mathcal{W}_{1}}^{2}\ ,  \label{fefeffefef}
\end{equation}%
where, for any $\gamma \in (0,1)$, $\mathfrak{F}_{\gamma }$ is the mapping
from $\mathcal{W}_{1}\times \mathfrak{D}_{0}$ to $\mathbb{R}_{0}^{+}$
defined by%
\begin{equation}
\mathfrak{F}_{\gamma }\left( \Phi ,f\right) \doteq \sum_{\mathcal{Z}_{1},%
\mathcal{Z}_{2}\in \mathcal{P}_{\mathrm{f}}:\mathcal{Z}_{1}\cap \mathcal{Z}%
_{2}\supseteq \{0\}}\frac{\left\Vert \Phi _{\mathcal{Z}_{1}}\right\Vert _{%
\mathcal{U}}\left\Vert \Phi _{\mathcal{Z}_{2}}\right\Vert _{\mathcal{U}}}{%
\left\vert \mathcal{Z}_{1}\right\vert \left\vert \mathcal{Z}_{2}\right\vert }%
\sum_{x\in \mathcal{Z}_{1},y\in \mathcal{Z}_{2}}\frac{\gamma ^{d}\left\vert
f\left( \gamma \left( x-y\right) \right) \right\vert }{\left\vert \mathcal{Z}%
_{1}\right\vert +\left\vert \mathcal{Z}_{2}\right\vert }\ ,\quad \Phi \in 
\mathcal{W}_{1},\ f\in \mathfrak{D}_{0}\ .  \label{technical function}
\end{equation}%
To prove (\ref{fefeffefef}), it suffices to refine Inequality (\ref{fefef})
in an obvious way. However, such a correction to the bound (\ref{fefef2})
pointwise vanishes in the Kac limit $\gamma \rightarrow 0^{+}$:

\begin{lemma}[Vanishing of $\mathfrak{F}_{\protect\gamma }$ in the Kac limit]

\label{propuni}\mbox{ }\newline
\emph{(i)} For any $\varepsilon \in \mathbb{R}^{+}$, there is a constant $%
M_{\varepsilon }\in \mathbb{R}^{+}$ such that, for any $\gamma \in (0,1)$, $%
\Phi ,\Psi \in \mathcal{W}_{1}$, and $f,g\in \mathfrak{D}_{\varepsilon }$, 
\begin{eqnarray*}
\left\vert \mathfrak{F}_{\gamma }\left( \Phi ,f\right) -\mathfrak{F}_{\gamma
}\left( \Psi ,g\right) \right\vert &\leq &M_{\varepsilon }\left\Vert \Phi
-\Psi \right\Vert _{\mathcal{W}_{1}}\left\Vert f\right\Vert _{\mathfrak{D}%
_{\varepsilon }}\left( \left\Vert \Phi -\Psi \right\Vert _{\mathcal{W}%
_{1}}+2\max \left\{ \left\Vert \Psi \right\Vert _{\mathcal{W}%
_{1}},\left\Vert \Phi \right\Vert _{\mathcal{W}_{1}}\right\} \right) \\
&&+3M_{\varepsilon }\left\Vert f-g\right\Vert _{\mathfrak{D}_{\varepsilon
}}\left\Vert \Psi \right\Vert _{\mathcal{W}_{1}}^{2}\ .
\end{eqnarray*}%
\emph{(ii)} For any $\Phi \in \mathcal{W}_{1}$ and$\ f\in \mathfrak{D}_{0}$, 
\begin{equation*}
\lim_{\gamma \rightarrow 0^{+}}\mathfrak{F}_{\gamma }\left( \Phi ,f\right)
=0\ .
\end{equation*}
\end{lemma}

\begin{proof}
Let\ $\varepsilon \in \mathbb{R}^{+}$ and $f\in \mathfrak{D}_{\varepsilon }$%
. For any $\gamma \in (0,1)$ and $\Phi ,\Psi \in \mathcal{W}_{1}$, from (\ref%
{estimate 1}), (\ref{technical function}) and the triangle inequality, we
arrive at%
\begin{align}
\left\vert \mathfrak{F}_{\gamma }\left( \Phi ,f\right) -\mathfrak{F}_{\gamma
}\left( \Psi ,f\right) \right\vert & \leq 2\sum_{\mathcal{Z}_{1},\mathcal{Z}%
_{2}\in \mathcal{P}_{\mathrm{f}}:\mathcal{Z}_{1}\cap \mathcal{Z}%
_{2}\supseteq \{0\}}\frac{\left\Vert \Psi _{\mathcal{Z}_{1}}\right\Vert _{%
\mathcal{U}}\left\Vert \Phi _{\mathcal{Z}_{2}}-\Psi _{\mathcal{Z}%
_{2}}\right\Vert _{\mathcal{U}}}{\left\vert \mathcal{Z}_{1}\right\vert
\left\vert \mathcal{Z}_{2}\right\vert }\sum_{x\in \mathcal{Z}_{1},y\in 
\mathcal{Z}_{2}}\frac{\gamma ^{d}\left\vert f\left( \gamma \left( x-y\right)
\right) \right\vert }{\left\vert \mathcal{Z}_{1}\right\vert +\left\vert 
\mathcal{Z}_{2}\right\vert }  \notag \\
& +\sum_{\mathcal{Z}_{1},\mathcal{Z}_{2}\in \mathcal{P}_{\mathrm{f}}:%
\mathcal{Z}_{1}\cap \mathcal{Z}_{2}\supseteq \{0\}}\frac{\left\Vert \Phi _{%
\mathcal{Z}_{1}}-\Psi _{\mathcal{Z}_{1}}\right\Vert _{\mathcal{U}}\left\Vert
\Phi _{\mathcal{Z}_{2}}-\Psi _{\mathcal{Z}_{2}}\right\Vert _{\mathcal{U}}}{%
\left\vert \mathcal{Z}_{1}\right\vert \left\vert \mathcal{Z}_{2}\right\vert }%
\sum_{x\in \mathcal{Z}_{1},y\in \mathcal{Z}_{2}}\frac{\gamma ^{d}\left\vert
f\left( \gamma \left( x-y\right) \right) \right\vert }{\left\vert \mathcal{Z}%
_{1}\right\vert +\left\vert \mathcal{Z}_{2}\right\vert }  \notag \\
& \leq M_{\varepsilon }\left\Vert f\right\Vert _{\mathfrak{D}_{\varepsilon
}}\left\Vert \Phi -\Psi \right\Vert _{\mathcal{W}_{1}}\left( \left\Vert \Phi
-\Psi \right\Vert _{\mathcal{W}_{1}}+2\max \left\{ \left\Vert \Psi
\right\Vert _{\mathcal{W}_{1}},\left\Vert \Phi \right\Vert _{\mathcal{W}%
_{1}}\right\} \right) \ .  \label{equicontinuous1}
\end{align}%
Using the reverse triangle inequality, observe that%
\begin{equation*}
\left\vert \mathfrak{F}_{\gamma }\left( \Phi ,f\right) -\mathfrak{F}_{\gamma
}\left( \Psi ,g\right) \right\vert \leq \left\vert \mathfrak{F}_{\gamma
}\left( \Phi ,f\right) -\mathfrak{F}_{\gamma }\left( \Psi ,f\right)
\right\vert +\left\vert \mathfrak{F}_{\gamma }\left( \Psi ,f-g\right)
\right\vert
\end{equation*}%
for any $\Phi ,\Psi \in \mathcal{W}_{1}$, $\varepsilon \in \mathbb{R}^{+}$
and $f,g\in \mathfrak{D}_{\varepsilon }$. By combining this estimate with (%
\ref{equicontinuous1}), we obtain Assertion (i). Finally, note that, for
finite-range interactions (see (\ref{W0})), we clearly have 
\begin{equation}
\lim_{\gamma \rightarrow 0^{+}}\mathfrak{F}_{\gamma }\left( \Phi ,f\right)
=0\ ,\qquad \Phi \in \mathcal{W}_{0}\subseteq \mathcal{W}_{1}\ ,
\label{limit finite range}
\end{equation}%
since in this case the sum in (\ref{technical function}) is finite.
Therefore, by density of $\mathcal{W}_{0}\subseteq \mathcal{W}_{1}$ and the
equicontinuity of $(\mathfrak{F}_{\gamma }\left( \cdot ,f\right) )_{\gamma
\in (0,1)}$ at fixed $f\in \mathfrak{D}_{0}$ (Assertion (i)), (\ref{limit
finite range}) extends to all interactions $\Phi \in \mathcal{W}_{1}$%
.\bigskip
\end{proof}

\noindent \textit{Acknowledgments:} This work is supported by CNPq
(309723/2020-5) as well as by the Basque Government through the grant
IT1615-22 and the BERC 2022-2025 program, by the COST Action CA18232
financed by the European Cooperation in Science and Technology (COST), and
by the Ministry of Science and Innovation via the grant PID2020-112948GB-I00
funded by MCIN/AEI/10.13039/501100011033 and by \textquotedblleft ERDF A way
of making Europe\textquotedblright . We thank Domingos Marchetti for
valuable discussions and hints, as well as the referee for pointing out
additional references, like \cite{Smedt-Zagrebnov}.

\noindent \textbf{Jean-Bernard Bru} \newline
Departamento de Matem\'{a}ticas and EHU Quantum Center\newline
Facultad de Ciencia y Tecnolog\'{\i}a\newline
Universidad del Pa\'{\i}s Vasco / Euskal Herriko Unibertsitatea, UPV/EHU%
\newline
Apartado 644, 48080 Bilbao \medskip \newline
BCAM - Basque Center for Applied Mathematics\newline
Mazarredo, 14. \newline
48009 Bilbao\medskip \newline
IKERBASQUE, Basque Foundation for Science\newline
48011, Bilbao\medskip \newline

\noindent \textbf{Walter de Siqueira Pedra} \newline
Departamento de Matem\'{a}tica\newline
Instituto de Ci\^{e}ncias Matem\'{a}ticas e da Computa\c{c}\~{a}o\newline
Universidade de S\~{a}o Paulo\newline
Avenida Trabalhador S\~{a}o Carlense, 400 \newline
13566-590 S\~{a}o Carlos - SP, Brazil\medskip\ \newline
BCAM - Basque Center for Applied Mathematics (\textit{As external scientific
member}) \newline
Mazarredo, 14. \newline
48009 Bilbao\medskip \newline

\noindent \textbf{K. Rodrigues Alves} \newline
Departamento de Matem\'{a}ticas\newline
Facultad de Ciencia y Tecnolog\'{\i}a\newline
Universidad del Pa\'{\i}s Vasco / Euskal Herriko Unibertsitatea, UPV/EHU%
\newline
Apartado 644, 48080 Bilbao \medskip \newline
BCAM - Basque Center for Applied Mathematics\newline
Mazarredo, 14. \newline
48009 Bilbao

\end{document}